\RequirePackage[running]{lineno}
\documentclass[aps,prl,nofootinbib,twocolumn,showpacs,superscriptaddress,letterpaper,amsmath,amsfonts,amssymb, preprintnumbers]{revtex4-1}
\pdfoutput=1 

\usepackage{graphicx}
\usepackage{dcolumn}
\usepackage{bm}
\usepackage{xspace}
\usepackage{siunitx}

\usepackage{rotating}

\usepackage[colorlinks,hyperindex,breaklinks]{hyperref}
\hypersetup{linkcolor=blue,citecolor=blue,filecolor=black,urlcolor=blue} 

\def\bulnu{\ensuremath{B \to X_u \, \ell^+\, \nu_{\ell}}\xspace}
\def\bclnu{\ensuremath{B \to X_c \, \ell^+\, \nu_{\ell}}\xspace}

\def\bpilnu{\ensuremath{B \to \pi \, \ell^+\,\nu_{\ell}}\xspace}

\def\brholnu{\ensuremath{B \to \rho \, \ell^+\,\nu_{\ell}}\xspace}
\def\bomegalnu{\ensuremath{B \to \omega \, \ell^+\,\nu_{\ell}}\xspace}
\def\betalnu{\ensuremath{B \to \eta \, \ell^+\,\nu_{\ell}}\xspace}
\def\betaplnu{\ensuremath{B \to \eta' \, \ell^+\,\nu_{\ell}}\xspace}
\def\bdlnu{\ensuremath{B \to D \, \ell^+\,\nu_{\ell}}\xspace}
\def\bdslnu{\ensuremath{B \to D^* \, \ell^+\,\nu_{\ell}}\xspace}
\def\bddslnu{\ensuremath{B \to D^{**} \, \ell^+\,\nu_{\ell}}\xspace}

\def\BFMC{\ensuremath{\Delta \mathcal{B}= 1.59 \times 10^{-3}}\xspace}

\newcolumntype{d}[1]{D{.}{.}{#1}}

\begin{document}

%\linenumbers

\title{Measurement of Differential Branching Fractions of Inclusive \bulnu Decays}

\author{L. Cao}
\email{cao@physik.uni-bonn.de}
\affiliation{University of Bonn, 53115 Bonn}\affiliation{Deutsches Elektronen--Synchrotron, 22607 Hamburg}

\author{W. Sutcliffe}
\affiliation{University of Bonn, 53115 Bonn}

\author{R. Van Tonder}
\affiliation{University of Bonn, 53115 Bonn}

\author{F. U.\ Bernlochner}
\email{florian.bernlochner@uni-bonn.de}
\affiliation{University of Bonn, 53115 Bonn}

%%% Paper:    B to Xu l nu
%%% Journal:  Physical Review Letters
%%% Contacts: L. Cao (cao@physik.uni-bonn.de)
%%%           F. Bernlochner (florian.bernlochner@cern.ch)
%%%           R. Van Tonder (vantonder@physik.uni-bonn.de)
%%%           W. Sutcliffe (wsut@uni-bonn.de)
%%% Non-responding authors or those who said NO are commented out.
%%% ====================================================================
%%% Click the RELOAD button on your web browser to see the updated file.
%%% ====================================================================
%%% Use \input{author} to insert this material into your latex file.
%%%%% Force institutions to appear in alphabetical order when typeset.
\noaffiliation
\affiliation{Department of Physics, University of the Basque Country UPV/EHU, 48080 Bilbao}
%%%\affiliation{Beihang University, Beijing 100191}
\affiliation{University of Bonn, 53115 Bonn}
\affiliation{Brookhaven National Laboratory, Upton, New York 11973}
\affiliation{Budker Institute of Nuclear Physics SB RAS, Novosibirsk 630090}
\affiliation{Faculty of Mathematics and Physics, Charles University, 121 16 Prague}
%%%\affiliation{Chiba University, Chiba 263-8522}
\affiliation{Chonnam National University, Gwangju 61186}
\affiliation{University of Cincinnati, Cincinnati, Ohio 45221}
\affiliation{Deutsches Elektronen--Synchrotron, 22607 Hamburg}
%%%\affiliation{Duke University, Durham, North Carolina 27708}
\affiliation{University of Florida, Gainesville, Florida 32611}
\affiliation{Department of Physics, Fu Jen Catholic University, Taipei 24205}
\affiliation{Key Laboratory of Nuclear Physics and Ion-beam Application (MOE) and Institute of Modern Physics, Fudan University, Shanghai 200443}
%%%\affiliation{Justus-Liebig-Universit\"at Gie\ss{}en, 35392 Gie\ss{}en}
\affiliation{Gifu University, Gifu 501-1193}
\affiliation{II. Physikalisches Institut, Georg-August-Universit\"at G\"ottingen, 37073 G\"ottingen}
\affiliation{SOKENDAI (The Graduate University for Advanced Studies), Hayama 240-0193}
\affiliation{Gyeongsang National University, Jinju 52828}
\affiliation{Department of Physics and Institute of Natural Sciences, Hanyang University, Seoul 04763}
\affiliation{University of Hawaii, Honolulu, Hawaii 96822}
\affiliation{High Energy Accelerator Research Organization (KEK), Tsukuba 305-0801}
\affiliation{J-PARC Branch, KEK Theory Center, High Energy Accelerator Research Organization (KEK), Tsukuba 305-0801}
\affiliation{National Research University Higher School of Economics, Moscow 101000}
\affiliation{Forschungszentrum J\"{u}lich, 52425 J\"{u}lich}
%%%\affiliation{Hiroshima Institute of Technology, Hiroshima 731-5193}
%%%\affiliation{Hiroshima University, Higashi-Hiroshima, Hiroshima 739-8530}
\affiliation{IKERBASQUE, Basque Foundation for Science, 48013 Bilbao}
%%%\affiliation{University of Illinois at Urbana-Champaign, Urbana, Illinois 61801}
\affiliation{Indian Institute of Science Education and Research Mohali, SAS Nagar, 140306}
\affiliation{Indian Institute of Technology Bhubaneswar, Satya Nagar 751007}
%%%\affiliation{Indian Institute of Technology Guwahati, Assam 781039}
\affiliation{Indian Institute of Technology Hyderabad, Telangana 502285}
\affiliation{Indian Institute of Technology Madras, Chennai 600036}
\affiliation{Indiana University, Bloomington, Indiana 47408}
\affiliation{Institute of High Energy Physics, Chinese Academy of Sciences, Beijing 100049}
\affiliation{Institute of High Energy Physics, Vienna 1050}
\affiliation{Institute for High Energy Physics, Protvino 142281}
%%%\affiliation{Institute of Mathematical Sciences, Chennai 600113}
\affiliation{INFN - Sezione di Napoli, I-80126 Napoli}
\affiliation{INFN - Sezione di Roma Tre, I-00146 Roma}
\affiliation{INFN - Sezione di Torino, I-10125 Torino}
\affiliation{Advanced Science Research Center, Japan Atomic Energy Agency, Naka 319-1195}
\affiliation{J. Stefan Institute, 1000 Ljubljana}
%%%\affiliation{Kanagawa University, Yokohama 221-8686}
\affiliation{Institut f\"ur Experimentelle Teilchenphysik, Karlsruher Institut f\"ur Technologie, 76131 Karlsruhe}
\affiliation{Kavli Institute for the Physics and Mathematics of the Universe (WPI), University of Tokyo, Kashiwa 277-8583}
\affiliation{Kennesaw State University, Kennesaw, Georgia 30144}
%%%\affiliation{King Abdulaziz City for Science and Technology, Riyadh 11442}
%%%\affiliation{Department of Physics, Faculty of Science, King Abdulaziz University, Jeddah 21589}
\affiliation{Kitasato University, Sagamihara 252-0373}
\affiliation{Korea Institute of Science and Technology Information, Daejeon 34141}
\affiliation{Korea University, Seoul 02841}
\affiliation{Kyoto Sangyo University, Kyoto 603-8555}
%%%\affiliation{Kyoto University, Kyoto 606-8502}
\affiliation{Kyungpook National University, Daegu 41566}
\affiliation{Universit\'{e} Paris-Saclay, CNRS/IN2P3, IJCLab, 91405 Orsay}
%%%\affiliation{\'Ecole Polytechnique F\'ed\'erale de Lausanne (EPFL), Lausanne 1015}
\affiliation{P.N. Lebedev Physical Institute of the Russian Academy of Sciences, Moscow 119991}
\affiliation{Liaoning Normal University, Dalian 116029}
\affiliation{Faculty of Mathematics and Physics, University of Ljubljana, 1000 Ljubljana}
\affiliation{Ludwig Maximilians University, 80539 Munich}
\affiliation{Luther College, Decorah, Iowa 52101}
\affiliation{Malaviya National Institute of Technology Jaipur, Jaipur 302017}
%%%\affiliation{University of Malaya, 50603 Kuala Lumpur}
\affiliation{Faculty of Chemistry and Chemical Engineering, University of Maribor, 2000 Maribor}
\affiliation{Max-Planck-Institut f\"ur Physik, 80805 M\"unchen}
\affiliation{School of Physics, University of Melbourne, Victoria 3010}
\affiliation{University of Mississippi, University, Mississippi 38677}
%%%\affiliation{University of Miyazaki, Miyazaki 889-2192}
%%%\affiliation{Moscow Physical Engineering Institute, Moscow 115409}
\affiliation{Graduate School of Science, Nagoya University, Nagoya 464-8602}
%%%\affiliation{Kobayashi-Maskawa Institute, Nagoya University, Nagoya 464-8602}
\affiliation{Universit\`{a} di Napoli Federico II, I-80126 Napoli}
%%%\affiliation{Nara University of Education, Nara 630-8528}
\affiliation{Nara Women's University, Nara 630-8506}
%%%\affiliation{National Central University, Chung-li 32054}
\affiliation{National United University, Miao Li 36003}
\affiliation{Department of Physics, National Taiwan University, Taipei 10617}
\affiliation{H. Niewodniczanski Institute of Nuclear Physics, Krakow 31-342}
\affiliation{Nippon Dental University, Niigata 951-8580}
\affiliation{Niigata University, Niigata 950-2181}
%%%\affiliation{University of Nova Gorica, 5000 Nova Gorica}
\affiliation{Novosibirsk State University, Novosibirsk 630090}
%%%\affiliation{Okinawa Institute of Science and Technology, Okinawa 904-0495}
\affiliation{Osaka City University, Osaka 558-8585}
%%%\affiliation{Osaka University, Osaka 565-0871}
\affiliation{Pacific Northwest National Laboratory, Richland, Washington 99352}
\affiliation{Panjab University, Chandigarh 160014}
\affiliation{Peking University, Beijing 100871}
\affiliation{University of Pittsburgh, Pittsburgh, Pennsylvania 15260}
\affiliation{Punjab Agricultural University, Ludhiana 141004}
%%%\affiliation{Research Center for Electron Photon Science, Tohoku University, Sendai 980-8578}
\affiliation{Research Center for Nuclear Physics, Osaka University, Osaka 567-0047}
\affiliation{Meson Science Laboratory, Cluster for Pioneering Research, RIKEN, Saitama 351-0198}
%%%\affiliation{Theoretical Research Division, Nishina Center, RIKEN, Saitama 351-0198}
%%%\affiliation{RIKEN BNL Research Center, Upton, New York 11973}
%%%\affiliation{Dipartimento di Matematica e Fisica, Universit\`{a} di Roma Tre, I-00146 Roma}
%%%\affiliation{Saga University, Saga 840-8502}
\affiliation{Department of Modern Physics and State Key Laboratory of Particle Detection and Electronics, University of Science and Technology of China, Hefei 230026}
\affiliation{Seoul National University, Seoul 08826}
\affiliation{Showa Pharmaceutical University, Tokyo 194-8543}
%%%\affiliation{Soochow University, Suzhou 215006}
\affiliation{Soongsil University, Seoul 06978}
%%%\affiliation{University of South Carolina, Columbia, South Carolina 29208}
%%%\affiliation{Stefan Meyer Institute for Subatomic Physics, Vienna 1090}
\affiliation{Sungkyunkwan University, Suwon 16419}
\affiliation{School of Physics, University of Sydney, New South Wales 2006}
%%%\affiliation{Department of Physics, Faculty of Science, University of Tabuk, Tabuk 71451}
\affiliation{Tata Institute of Fundamental Research, Mumbai 400005}
%%%\affiliation{Excellence Cluster Universe, Technische Universit\"at M\"unchen, 85748 Garching}
\affiliation{Department of Physics, Technische Universit\"at M\"unchen, 85748 Garching}
%%%\affiliation{School of Physics and Astronomy, Tel Aviv University, Tel Aviv 69978}
\affiliation{Toho University, Funabashi 274-8510}
\affiliation{Department of Physics, Tohoku University, Sendai 980-8578}
\affiliation{Earthquake Research Institute, University of Tokyo, Tokyo 113-0032}
\affiliation{Department of Physics, University of Tokyo, Tokyo 113-0033}
\affiliation{Tokyo Institute of Technology, Tokyo 152-8550}
\affiliation{Tokyo Metropolitan University, Tokyo 192-0397}
%%%\affiliation{Tokyo University of Agriculture and Technology, Tokyo 184-8588}
\affiliation{Utkal University, Bhubaneswar 751004}
\affiliation{Virginia Polytechnic Institute and State University, Blacksburg, Virginia 24061}
\affiliation{Wayne State University, Detroit, Michigan 48202}
\affiliation{Yamagata University, Yamagata 990-8560}
\affiliation{Yonsei University, Seoul 03722}
% \author{A.~Abdesselam}\affiliation{Department of Physics, Faculty of Science, University of Tabuk, Tabuk 71451} % Tabuk
  \author{I.~Adachi}\affiliation{High Energy Accelerator Research Organization (KEK), Tsukuba 305-0801}\affiliation{SOKENDAI (The Graduate University for Advanced Studies), Hayama 240-0193} % KEK
% \author{K.~Adamczyk}\affiliation{H. Niewodniczanski Institute of Nuclear Physics, Krakow 31-342} % Krakow
% \author{J.~K.~Ahn}\affiliation{Korea University, Seoul 02841} % Korea
  \author{H.~Aihara}\affiliation{Department of Physics, University of Tokyo, Tokyo 113-0033} % Tokyo
% \author{S.~Al~Said}\affiliation{Department of Physics, Faculty of Science, University of Tabuk, Tabuk 71451}\affiliation{Department of Physics, Faculty of Science, King Abdulaziz University, Jeddah 21589} % Tabuk
% \author{K.~Arinstein}\affiliation{Budker Institute of Nuclear Physics SB RAS, Novosibirsk 630090}\affiliation{Novosibirsk State University, Novosibirsk 630090} % BINP
% \author{Y.~Arita}\affiliation{Graduate School of Science, Nagoya University, Nagoya 464-8602} % Nagoya
  \author{D.~M.~Asner}\affiliation{Brookhaven National Laboratory, Upton, New York 11973} % BNL
% \author{H.~Atmacan}\affiliation{University of Cincinnati, Cincinnati, Ohio 45221} % Cincinnati
% \author{V.~Aulchenko}\affiliation{Budker Institute of Nuclear Physics SB RAS, Novosibirsk 630090}\affiliation{Novosibirsk State University, Novosibirsk 630090} % BINP
  \author{T.~Aushev}\affiliation{National Research University Higher School of Economics, Moscow 101000} % HSE
% \author{R.~Ayad}\affiliation{Department of Physics, Faculty of Science, University of Tabuk, Tabuk 71451} % Tabuk
% \author{T.~Aziz}\affiliation{Tata Institute of Fundamental Research, Mumbai 400005} % Tata
  \author{V.~Babu}\affiliation{Deutsches Elektronen--Synchrotron, 22607 Hamburg} % DESY
  \author{S.~Bahinipati}\affiliation{Indian Institute of Technology Bhubaneswar, Satya Nagar 751007} % IITB
% \author{A.~M.~Bakich}\affiliation{School of Physics, University of Sydney, New South Wales 2006} % Sydney
% \author{Y.~Ban}\affiliation{Peking University, Beijing 100871} % Peking
% \author{E.~Barberio}\affiliation{School of Physics, University of Melbourne, Victoria 3010} % Melbourne
% \author{M.~Barrett}\affiliation{High Energy Accelerator Research Organization (KEK), Tsukuba 305-0801} % KEK
% \author{M.~Bauer}\affiliation{Institut f\"ur Experimentelle Teilchenphysik, Karlsruher Institut f\"ur Technologie, 76131 Karlsruhe} % Karlsruhe
% ...
% Lebedev
% \author{V.~Zhulanov}\affiliation{Budker Institute of Nuclear Physics SB RAS, Novosibirsk 630090}\affiliation{Novosibirsk State University, Novosibirsk 630090} % BINP
% \author{T.~Zivko}\affiliation{J. Stefan Institute, 1000 Ljubljana} % Ljubljana
% \author{A.~Zupanc}\affiliation{Faculty of Mathematics and Physics, University of Ljubljana, 1000 Ljubljana}\affiliation{J. Stefan Institute, 1000 Ljubljana} % Ljubljana
% \author{N.~Zwahlen}\affiliation{\'Ecole Polytechnique F\'ed\'erale de Lausanne (EPFL), Lausanne 1015} % Lausanne
\collaboration{The Belle Collaboration}

\begin{abstract}
The first measurements of differential branching fractions of inclusive semileptonic \bulnu decays are performed using the full Belle data set of 711 fb$^{-1}$ of integrated luminosity at the $\Upsilon(4S)$ resonance and for $\ell = e, \mu$. With the availability of these measurements,  new avenues for future shape-function model-independent determinations of the Cabibbo-Kobayashi-Maskawa matrix element $|V_{ub}|$ can be pursued to gain new insights in the existing tension with respect to exclusive determinations. The differential branching fractions are reported as a function of the lepton energy, the four-momentum-transfer squared, light-cone momenta, the hadronic mass, and the hadronic mass squared. They are obtained by subtracting the backgrounds from semileptonic \bclnu decays and other processes, and corrected for resolution and acceptance effects.   
\end{abstract}


\preprint{ Belle Preprint 2021-15, KEK Preprint 2021-13}

\date{\today}

\maketitle

In this Letter we present the first measurements of the differential branching fractions of inclusive semileptonic \mbox{\bulnu} decays, obtained from analyzing the full Belle data set of 711 fb$^{-1}$ of integrated luminosity at the $\Upsilon(4S)$ resonance and for $\ell = e, \mu$. The measured distributions can be used for future studies of the nonperturbative decay dynamics of \bulnu transitions, and novel determinations of the $b$-quark mass $m_b$ and of the Cabibbo-Kobayashi-Maskawa (CKM) matrix element $|V_{ub}|$. The presented measurements use the same collision events that were analyzed in Ref.~\cite{Cao:2021xqf}. Therein, partial branching fractions of charmless semileptonic decays were reported using an analysis technique relying on the full reconstruction of the second $B$ meson of the $e^+ \, e^- \to \Upsilon(4S) \to B \bar B$ process. This approach allows for the direct reconstruction of the four momentum of the hadronic $X$ system of the \bulnu process and other kinematic quantities of interest. The analysis strategy of the presented measurements follows Ref.~\cite{Cao:2021xqf}, but more stringent selection criteria are applied to improve the resolution of key variables and further suppress backgrounds from \bclnu decays and other processes. Charge conjugation is implied throughout this Letter and \bulnu is defined as the average branching fraction of $B^+$ and $B^0$ meson decays.

Differential branching fractions are reported as a function of the lepton energy in the signal $B$ rest frame $E_\ell^B$, the invariant mass $M_X$ and mass squared $M_X^2$ of the hadronic $X$ system, the four-momentum-transfer squared $q^2 = \left( p_B - p_{X} \right)^2$ of the $B$ to the lepton and neutrino system, and the two light-cone momenta $P_\pm = \left( E^{B}_X \mp |\bold {p}^{B}_X|  \right)$ with $E_{X}^B$ and $\bold {p}_X^B$ in the signal $B$ rest frame. Measurements of these distributions are of great interest as they allow for the study of nonperturbative shape functions~\cite{Neubert:1993ch}. Shape functions describe the Fermi motion of the $b$ quark inside the $B$ meson, and enter in the calculation of the dynamics of \bulnu decays. Currently, properties of the leading-order $\Lambda_{\mathrm{QCD}}/m_b$ shape function can only be studied using the photon energy spectrum of $B \to X_s \, \gamma$ decays and moments of the lepton energy or hadronic invariant mass in charmed semileptonic $B$ decays~\cite{Gambino:2004qm,Bauer:2004ve,Benson:2004sg}. The modeling of both the leading and subleading shape functions introduce large theory uncertainties on predictions of the \bulnu\ decay rate, and hence on the determination of $|V_{ub}|$. With the presented differential branching fractions, we provide the necessary experimental input for future model-independent approaches, whose aim is to reduce this model dependence by directly measuring the shape function~\cite{Bernlochner:2020jlt,Gambino:2016fdy}. This will lead to more reliable determinations of $|V_{ub}|$ from inclusive processes and give new insights into the persistent tension with the values obtained from exclusive determinations~\cite{HFLAV:2019otj} of about 3 standard deviations.

We analyze \mbox{$(772 \pm 10) \times 10^6$} $B$ meson pairs recorded at the $\Upsilon(4S)$ resonance energy and $\SI{79}{fb^{-1}}$ of collision events recorded $\SI{60}{MeV}$ below the $\Upsilon(4S)$ peak, which were both recorded at the KEKB $e^+ e^-$ collider~\cite{KEKB} by the Belle detector. Belle is a large-solid-angle magnetic spectrometer and a detailed description of its subdetectors and performance can be found in Ref.~\citep{Abashian:2000cg}. Monte Carlo (MC) samples of $B$ meson decays and continuum processes ($e^+ e^- \to q \bar q$ with $q = u,d,s,c$) are simulated using the \texttt{EvtGen} generator~\citep{EvtGen} and a detailed description of all samples and models is given in Ref.~\cite{Cao:2021xqf}. The simulated samples are used for the background subtraction and to correct for detector resolution, selection, and acceptance effects. The sample sizes used correspond to approximately ten and five times, respectively, the Belle collision data for $B$ meson production and continuum processes. 

Semileptonic \bulnu decays are modeled as a mixture of specific exclusive modes and nonresonant contributions using a so-called "hybrid" approach~\cite{hybrid}, following closely the implementation of~\cite{Prim:2019gtj,markus_prim_2020_3965699}. In the hybrid approach, the triple differential rate of the inclusive and combined exclusive predictions are combined such that partial rates of the inclusive prediction are recovered. This is achieved by assigning three dimensional weights to the inclusive contribution as a function of the generator-level $q^2$, $E_\ell^B$, and $M_X$. For the inclusive contribution, we use two different calculations, i.e. the De Fazio and
Neubert (DFN) model~\cite{DeFazio:1999ptt} and the Bosch-Lange-Neubert-Paz (BLNP) model~\cite{Lange:2005yw}, and treat their difference as a systematic uncertainty. The simulated inclusive \bulnu events are hadronized with the JETSET algorithm~\cite{SJOSTRAND199474} into final states with two or more mesons. A summary of the used \bulnu branching fractions and decay models is given in Table~\ref{tab:MC}. Semileptonic \bclnu\ decays are dominated by \bdlnu\ and \bdslnu\ decays, which are simulated with form factor parametrizations discussed in Refs.~\cite{Boyd:1994tt,Grinstein:2017nlq,Bigi:2017njr} and values determined by Refs.~\cite{Glattauer:2015teq,Waheed:2018djm}. The remaining \bclnu\ decays are simulated as a mix of resonant and nonresonant modes, using Ref.~\cite{Bernlochner:2016bci} for the modeling of \bddslnu form factors. The known difference between inclusive and the sum of measured exclusive \bclnu is filled with $B \to D^{(*)} \, \eta  \, \ell^+  \nu_\ell$ decays. 
\bgroup
\def\arraystretch{1.2}
\begin{table}[b!]
\centering
\caption{Semileptonic \bulnu decays are modeled as a mixture of specific exclusive modes and nonresonant contributions.
The branching fractions are from the world averages from Ref.~\citep{pdg:2020} and the models and form factors (FFs) used are listed. We use natural units ($\hbar = c = 1$).
\label{tab:MC}}
\begin{tabular}{l|cc}
\hline \hline
 $\mathcal{B}$ & Value $B^+$ & Value $B^0$  \\
\hline
 \bpilnu$^{\mathrm{~a,e}}$ & $\left(7.8 \pm 0.3 \right) \times 10^{-5}$ & $\left(1.5 \pm 0.06 \right) \times 10^{-4}$ \\
  \betalnu$^{\mathrm{~b,e}}$  & $\left(3.9 \pm 0.5 \right)  \times 10^{-5}$ & $\cdots$ \\
  \betaplnu$^{\mathrm{~b,e}}$  & $\left(2.3 \pm 0.8 \right) \times 10^{-5}$ & $\cdots$ \\
  \bomegalnu$^{\mathrm{~c,e}}$  & $\left(1.2 \pm 0.1 \right)  \times 10^{-4}$ & $\cdots$ \\
  \brholnu$^{\mathrm{~c,e}}$  & $\left(1.6 \pm 0.1\right)  \times 10^{-4}$ & $\left(2.9 \pm 0.2\right)  \times 10^{-4}$ \\
 \bulnu$^{\mathrm{~d,e}}$~~  & $\left(2.2 \pm 0.3 \right)  \times 10^{-3}$ & $\left(2.0 \pm 0.3 \right)  \times 10^{-3}$ \\ 
\hline
\hline
\end{tabular}
\begin{footnotesize}
\def\arraystretch{1.0}
\vspace{1ex}
\begin{tabular}{ll}
$^{\mathrm{a}}$ & BCL FFs~\citep{Bourrely:2008za} from fit to LQCD~\citep{Lattice:2015tia} and Ref.~\citep{Amhis:2019ckw} .\\
$^{\mathrm{b}}$ & Pole FFs from LCSR~\citep{Duplancic:2015zna}.\\
$^{\mathrm{c}}$ & BSZ FFs fit~\citep{Bernlochner:2021rel} to LCSR~\citep{Bharucha:2012wy} and Refs.~\cite{Sibidanov:2013rkk,Lees:2012mq,delAmoSanchez:2010af}.\\
$^{\mathrm{d}}$ & DFN~\cite{DeFazio:1999ptt} ($m_{b}^{\text{KN}} = (4.66 \pm 0.04)\,\mathrm{GeV}$, $a^{\text{KN}} = 1.3 \pm 0.5$) \\
 & or BLNP model~\cite{Lange:2005yw} ($m_{b}^{\mathrm{SF}} = 4.61\,\mathrm{GeV}$, $\mu_{\pi}^{2\, \text{SF}} = 0.20 \,\mathrm{GeV}^2$) .\\
$^{\mathrm{e}}$ & Inclusive and exclusive decays are mixed using hybrid \\
& approach~\cite{hybrid} .\\
\end{tabular}
\vspace{-2mm}
\end{footnotesize}
\end{table}
\egroup

 \begin{figure}[t!]
  \includegraphics[width=0.42\textwidth]{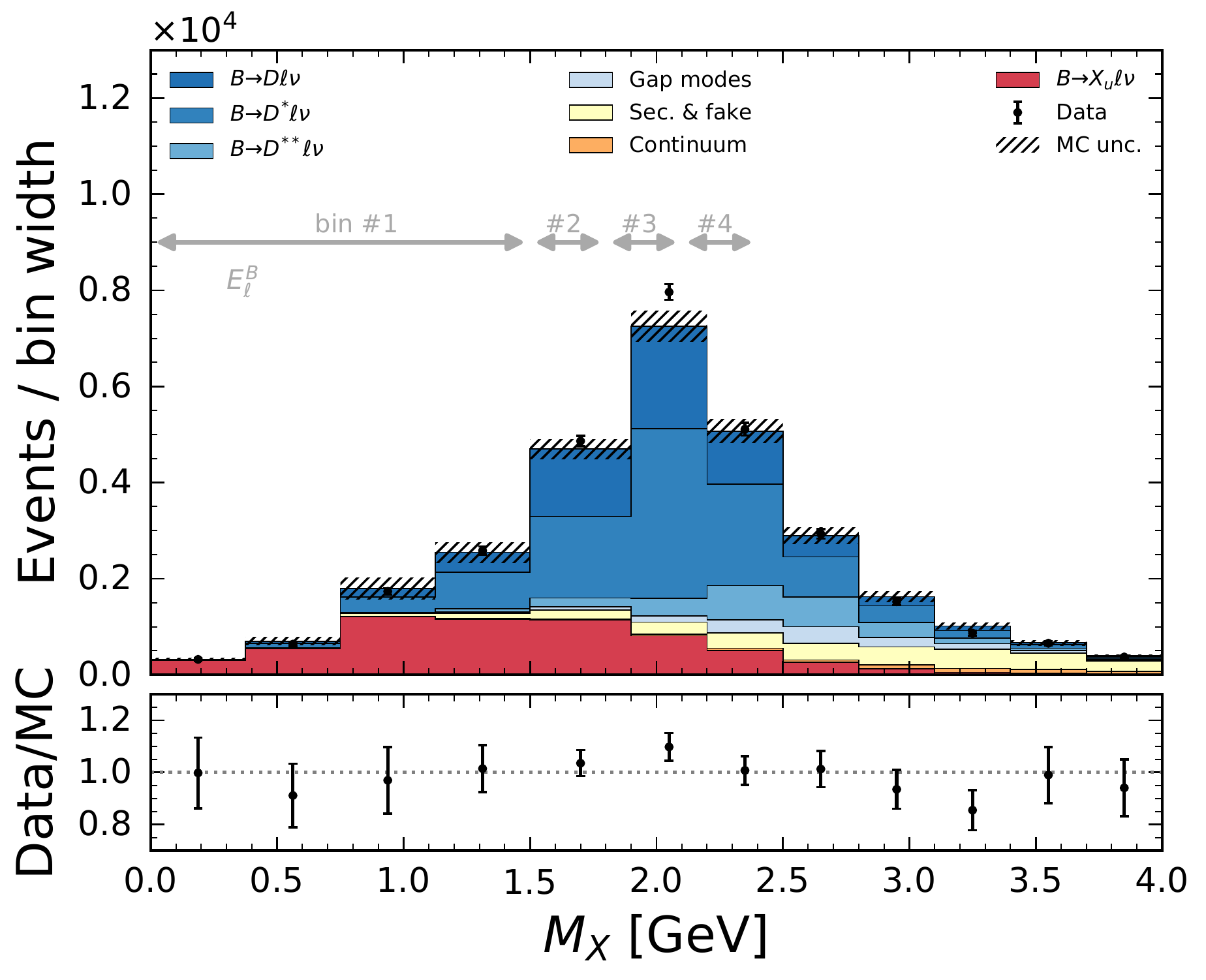} 
  \includegraphics[width=0.42\textwidth]{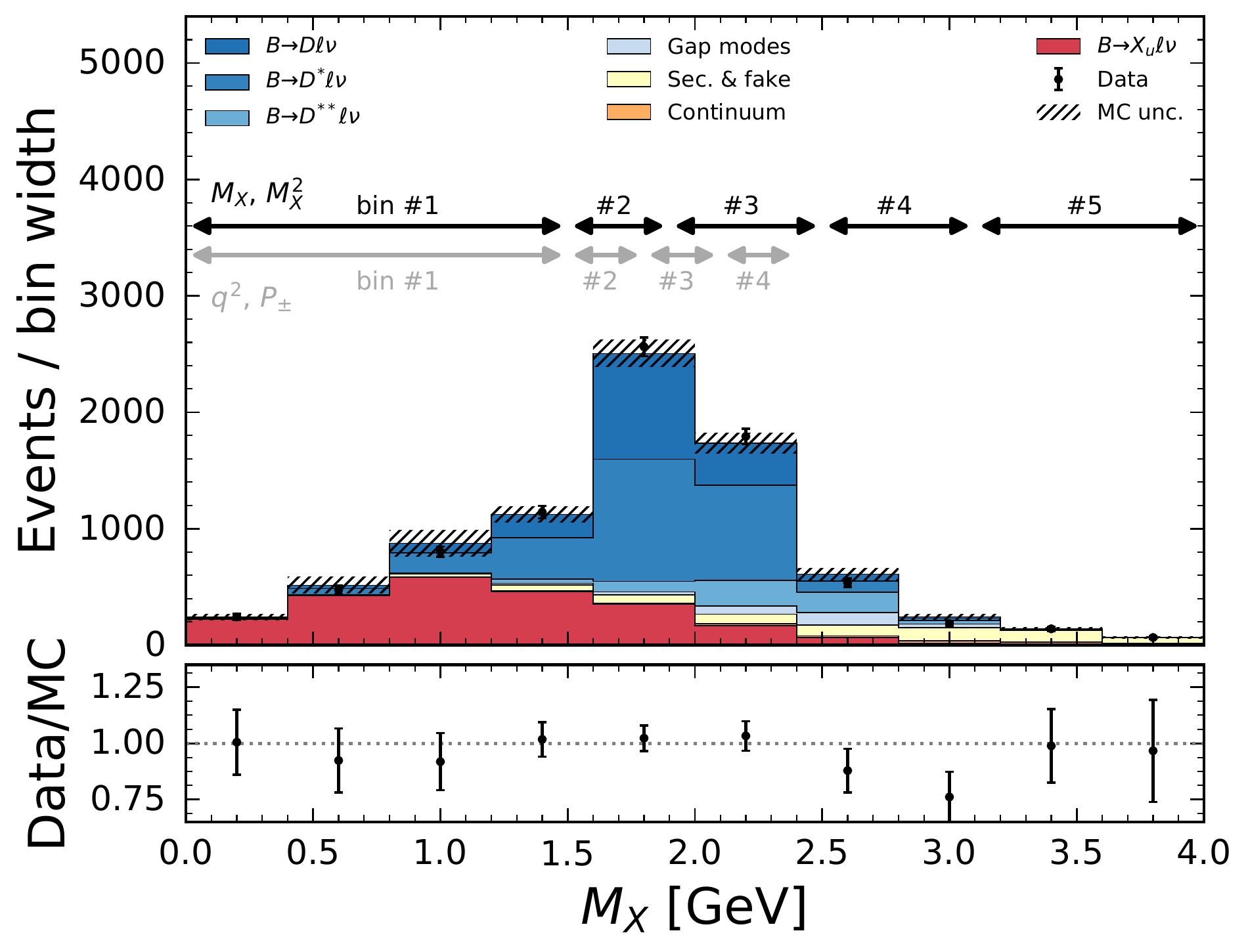} 
\caption{
 The reconstructed $M_X$ distributions after the BDT selection without (top) and with (bottom) the requirement of $\big|E_{\mathrm{miss}} - |\bold{p}_{\mathrm{miss}}|\big| < 0.1 \, \mathrm{GeV}$ are shown. The arrows indicate the coarse binning used in the background subtraction fit for the different variables. Removing the $M_X > 2.4 \, \mathrm{GeV}$ events improves the signal to background ratio for $E_\ell^B$, $q^2$, and $P_\pm$, but is not necessary for measurements of $M_X$ and $M_X^2$. 
 }
\label{fig:MXs}
\end{figure}

Collision events are reconstructed using the multivariate algorithm of Ref.~\cite{Feindt:2011mr}, in which one of the two $B$ mesons is fully reconstructed in hadronic final states (labeled as $B_{\mathrm{tag}}$). Signal candidates are reconstructed by identifying an electron or muon candidate with \mbox{$E_\ell^B = |\bold{p}_{\ell}^B| > 1 \, \mathrm{GeV}$} in the signal $B$ rest frame, and by reconstructing the hadronic $X$ system of the \bulnu semileptonic process using charged particles and neutral energy depositions of the collision event not used in the reconstruction of the $B_{\mathrm{tag}}$ candidate. The largest background after the reconstruction is from the CKM-favored \bclnu process, which possesses a very similar decay signature, completely dominating the selected candidate events. To identify \bulnu candidates, eleven distinguishing features are combined into a single discriminant using a multivariate classifier in the form of boosted decision trees (BDTs) using the implementation of Ref.~\cite{Chen:2016:XST:2939672.2939785}. The most discriminating variables are the reconstructed neutrino mass, $M_{\mathrm{miss}}^2$, the vertex fit probability of the $X \ell$ decay vertex, and the number of identified $K^\pm$ and $K_{S}^{0}$ in the $X$ system. To improve the resolution on the reconstructed variables or the signal to background ratio, additional selections are applied. For the measurements involving the hadronic $X$ system ($M_X$, $M_X^2$, $q^2$, $P_{\pm}$), we demand the missing energy, $E_{\mathrm{miss}}$, and the magnitude of the missing momentum $\big|\bold{p}_{\mathrm{miss}}\big|$ of the collision to be consistent with each other by requiring $\big|E_{\mathrm{miss}} - |\bold{p}_{\mathrm{miss}}|\big| < 0.1 \, \mathrm{GeV}$. This improves the resolution by 21\%-37\%, depending on the observable, and removes poorly reconstructed events. The signal efficiency after the BDT selection and this additional requirement is 8\% while rejecting 99.5\% of all \bclnu background events, as defined with respect to all selected signal or \bclnu events after successfully identifying a suitable $B_{\mathrm{tag}}$ candidate. To reduce the contamination of \bclnu and other backgrounds, for the measurements of $q^2$ and the light-cone momenta $P_{\pm}$, an additional requirement of $M_X < 2.4 \, \mathrm{GeV}$ is imposed: this selection, mostly targeting poorly understood high-mass $X_c$ states, removes in addition background from secondary leptons and reduces the \bclnu contamination by an additional 20\%. The reconstruction resolution of the lepton energy is excellent, thus no requirement on the missing energy and the magnitude of the missing momentum of the event is imposed, but to reduce background contributions we also require $M_X < 2.4 \, \mathrm{GeV}$. This results in a signal efficiency of 17\% and 99\% of \bclnu background events are rejected as defined with respect to all events after the $B_{\mathrm{tag}}$ selection. 

The differential branching fractions are extracted by subtracting the remaining background contributions from \bclnu and other sources in the measured distributions. This is implemented in a four-step procedure: first a binned likelihood fit to the $M_X$ distribution is carried out to estimate the number of background events. The $M_X$ fit takes the shape of signal and background from MC simulations and includes as nuisance parameters systematic effects that can impact the template shapes. To reduce the dependence on the precise modeling of the \bulnu process, a coarse binning is used. In particular, the resonance region ($M_X \in [0, 1.5] \, \mathrm{GeV}$) is described by a single bin. The analyzed hadronic invariant mass spectra with and without the selection on $\big|E_{\mathrm{miss}} - |\bold{p}_{\mathrm{miss}}|\big| < 0.1 \, \mathrm{GeV}$ and the used binning for the different fits are shown in Fig.~\ref{fig:MXs}. 

In the second step, the background is subtracted using the estimated normalization from the corresponding $M_X$ fit in the kinematic variable under study. The background shape is taken from MC simulation. The statistical uncertainty on the background-subtracted yields are determined using a bootstrapping procedure~\cite{booststrap,PhysRevD.39.274} to properly incorporate the correlation from the $M_X$ fit as the same data events are analyzed. The same method is used to determine the statistical correlations between all bins of all measured distributions. The systematic uncertainties associated with modeling the background shape and normalization are also propagated into the uncertainties of the estimated signal yields. In the third step, the signal yields are unfolded using the Singular Value Decomposition (SVD) algorithm from Ref.~\cite{Hocker:1995kb} with the implementation of Ref.~\cite{roounfold}. The regularization parameter of the unfolding method was carefully tuned with simulated samples to minimize the dependence on $m_b$, the shape function modeling, and the composition of the \bulnu signal. In the final step the unfolded yields are corrected for efficiency and acceptance effects to the partial phase space defined by $E_\ell^B > 1 \, \mathrm{GeV}$, also correcting for QED final-state radiation. The full analysis procedure was validated with independent MC samples and ensembles of pseudoexperiments and no biases of central values or uncertainties were observed.

\begin{figure}[t]
  \includegraphics[width=0.4\textwidth]{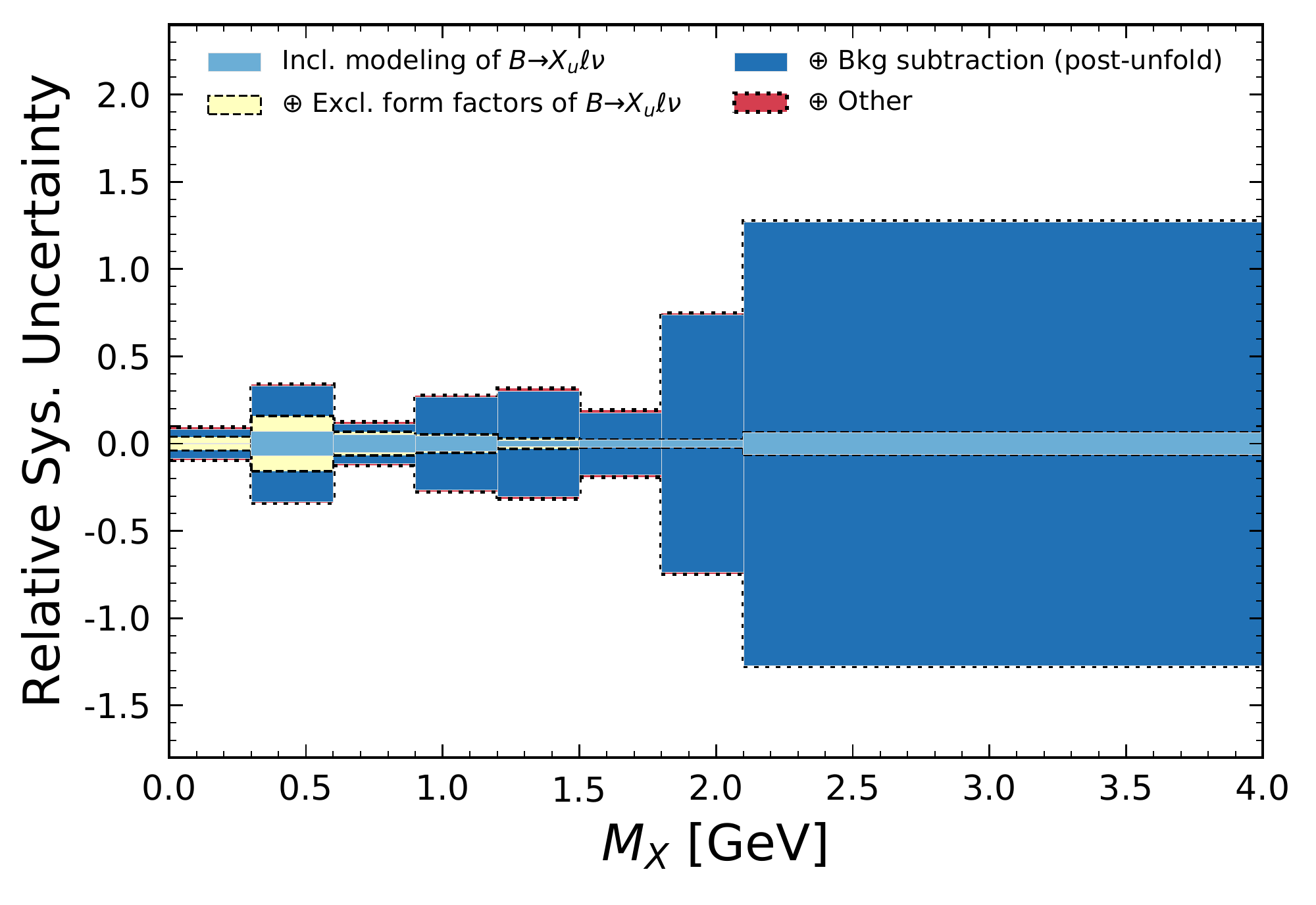} 
  \includegraphics[width=0.4\textwidth]{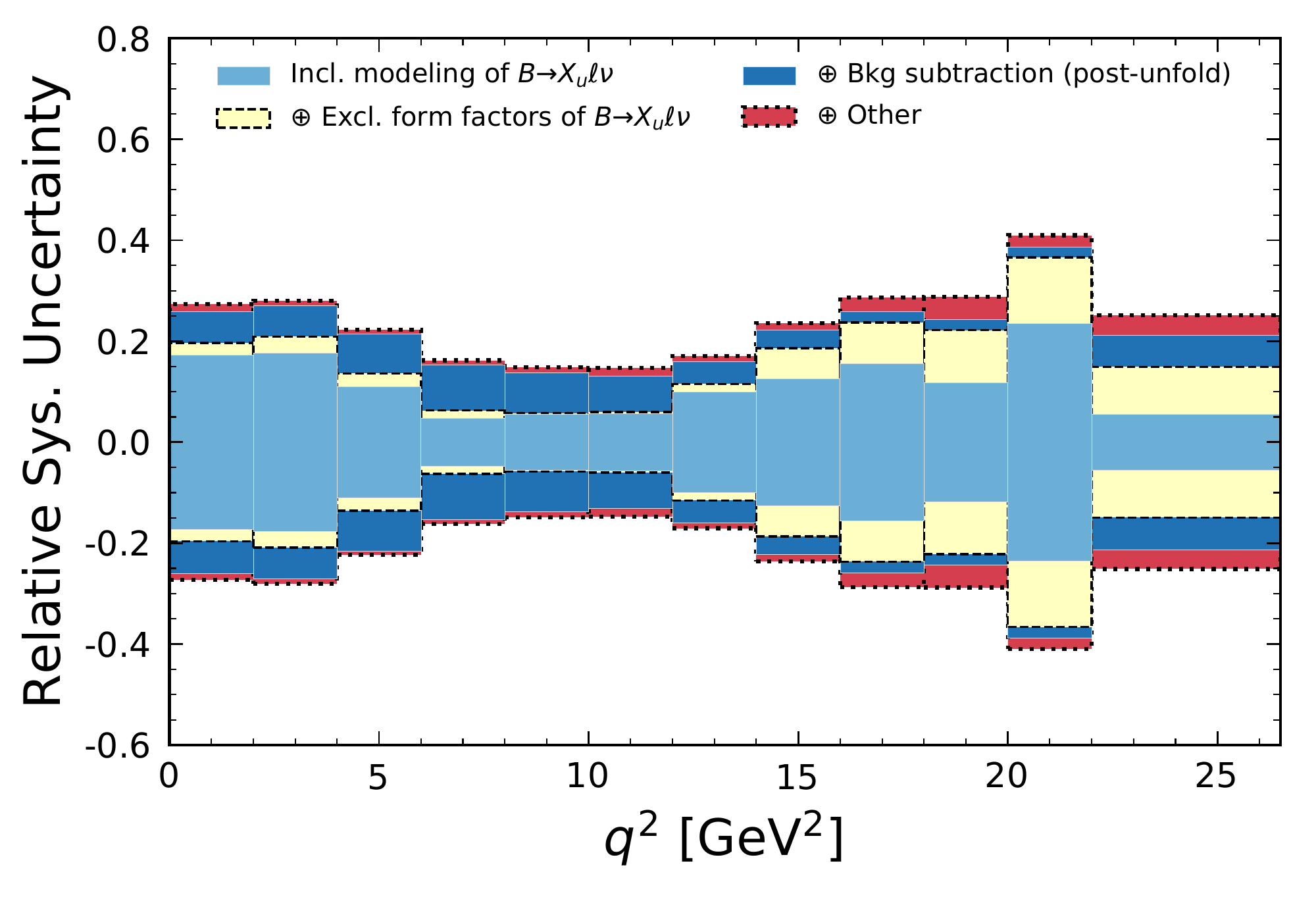} 
\caption{
The relative systematic uncertainties on the unfolded differential branching fraction as a function of $M_X$ and $q^2$ are shown. The different uncertainty sources are color coded. 
 }
\label{fig:syst_examples}
\end{figure}

Systematic uncertainties from the background subtraction, the modeling of the detector response for \bulnu, and uncertainties entering the total normalization are consistently propagated through the background subtraction, unfolding, and efficiency correction procedure. For the background subtraction we evaluate \bulnu and \bclnu modeling (FFs, nonperturbative parameters and composition) and detector related systematic uncertainties. The largest systematic uncertainties are typically from the assumptions entering the modeling of the \bulnu signal composition, but depending on the region of phase space also the background subtraction uncertainty can be a dominant source of uncertainty. Figure~\ref{fig:syst_examples} shows the relative uncertainties on the unfolded differential branching fractions as a function of $M_X$ and $q^2$. The total systematic uncertainties range from 9 to 130\% in relative error, and the background uncertainty is the dominant source of error in regions of phase space that are enriched in \bclnu (e.g. above $M_X \approx m_{D^0} = 1.86 \, \mathrm{GeV}$). The exclusive \bulnu modeling errors only contribute significantly in the resonance region at low $M_X$ or high $q^2$. The full systematic and statistical correlations between all measured distributions are determined to allow for a future simultaneous analysis of all measured distributions, and are provided with the full systematic uncertainties of all measured distributions in Supplemental Material, Ref.~\cite{supplemental}. 

 \begin{figure*}[th!]
  \includegraphics[width=0.4\textwidth]{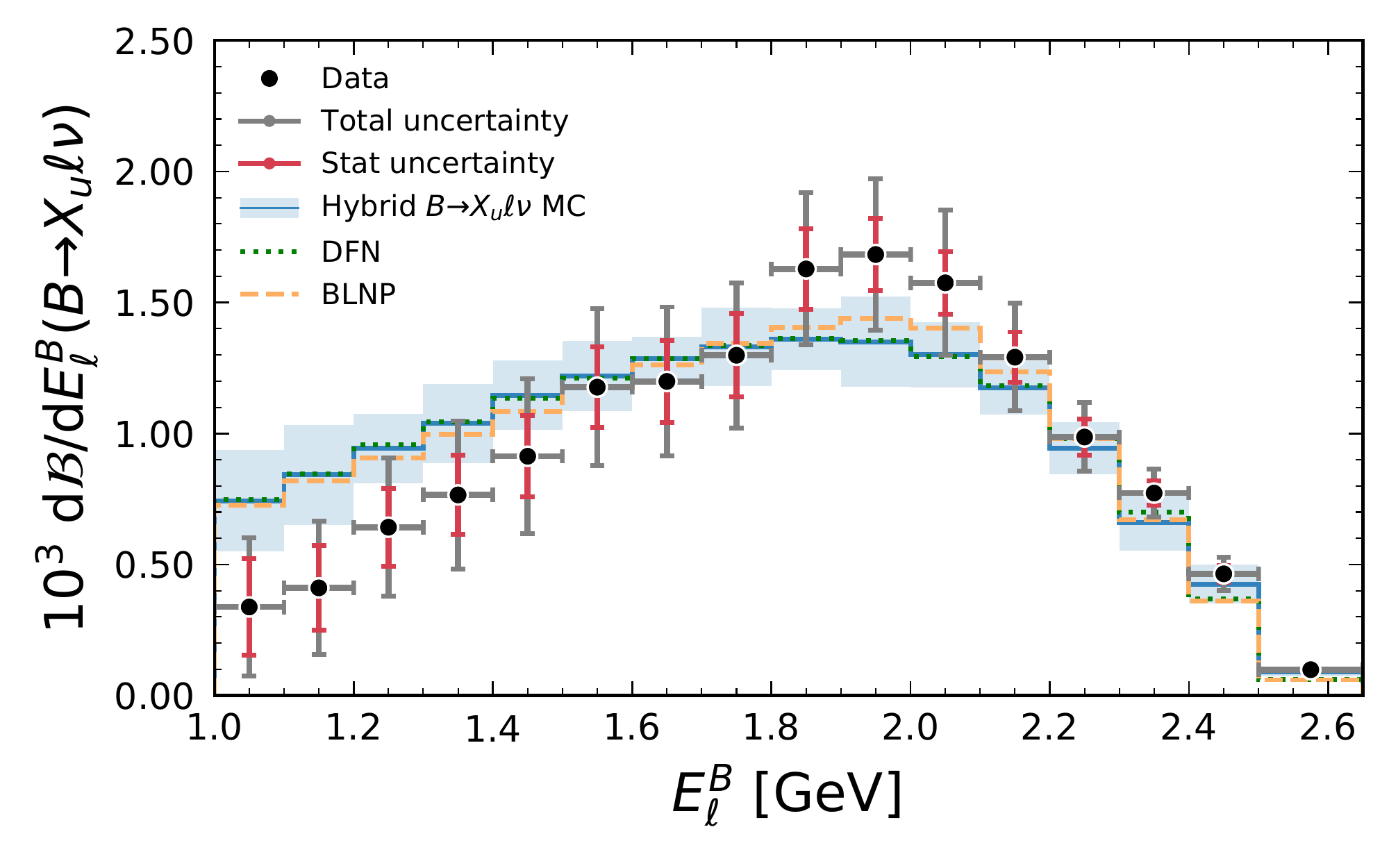} 
  \includegraphics[width=0.4\textwidth]{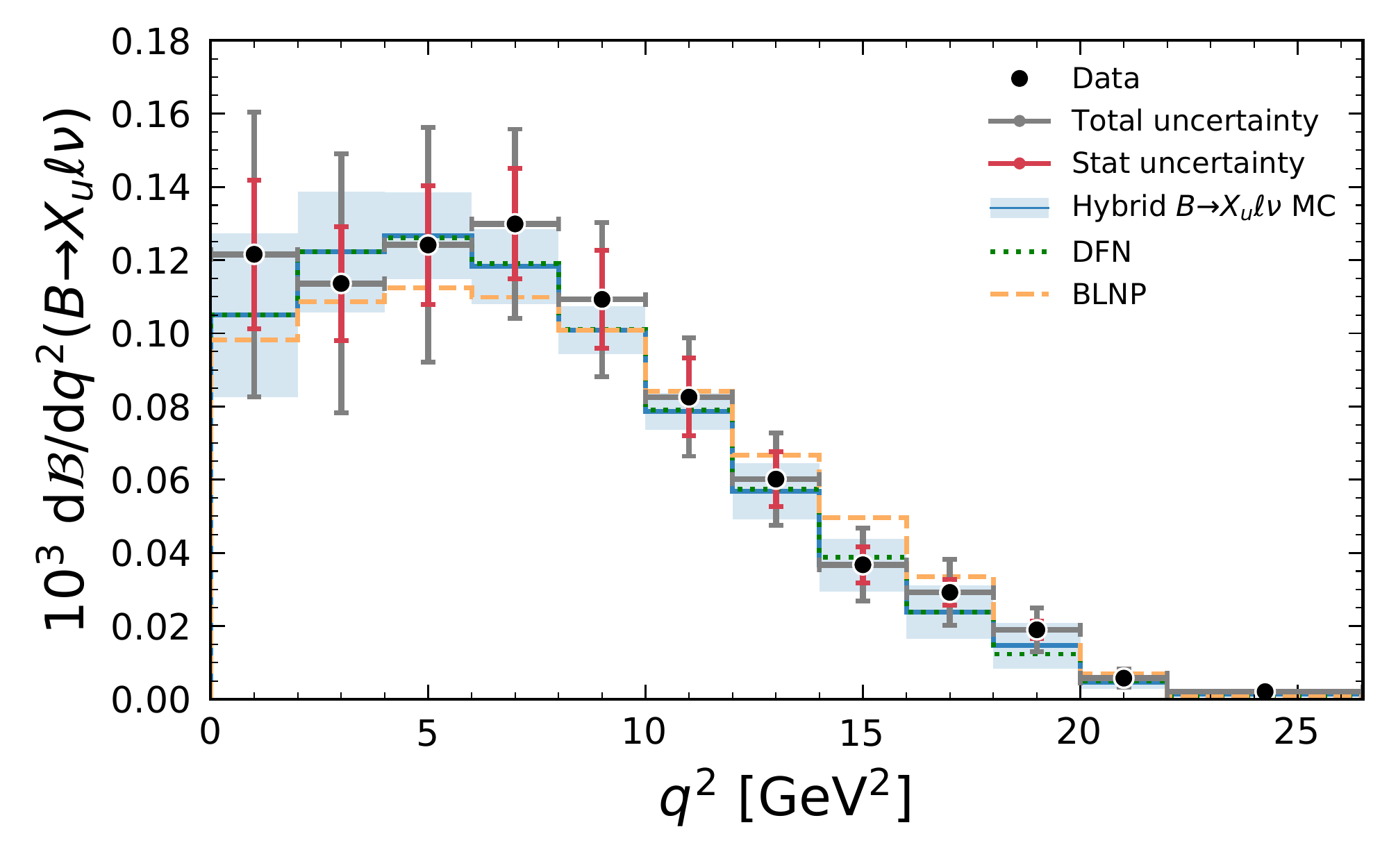} 
  \includegraphics[width=0.4\textwidth]{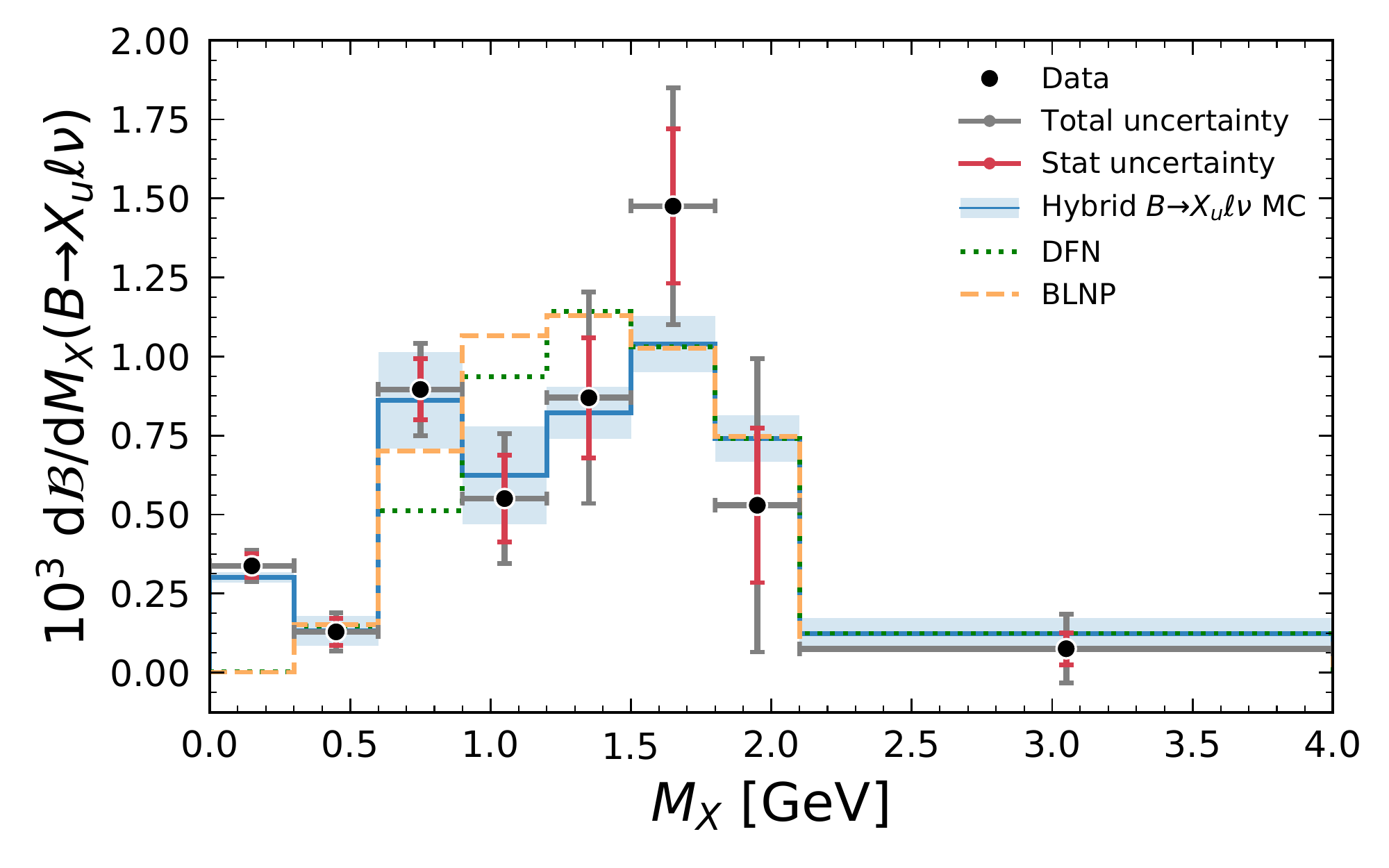} 
  \includegraphics[width=0.4\textwidth]{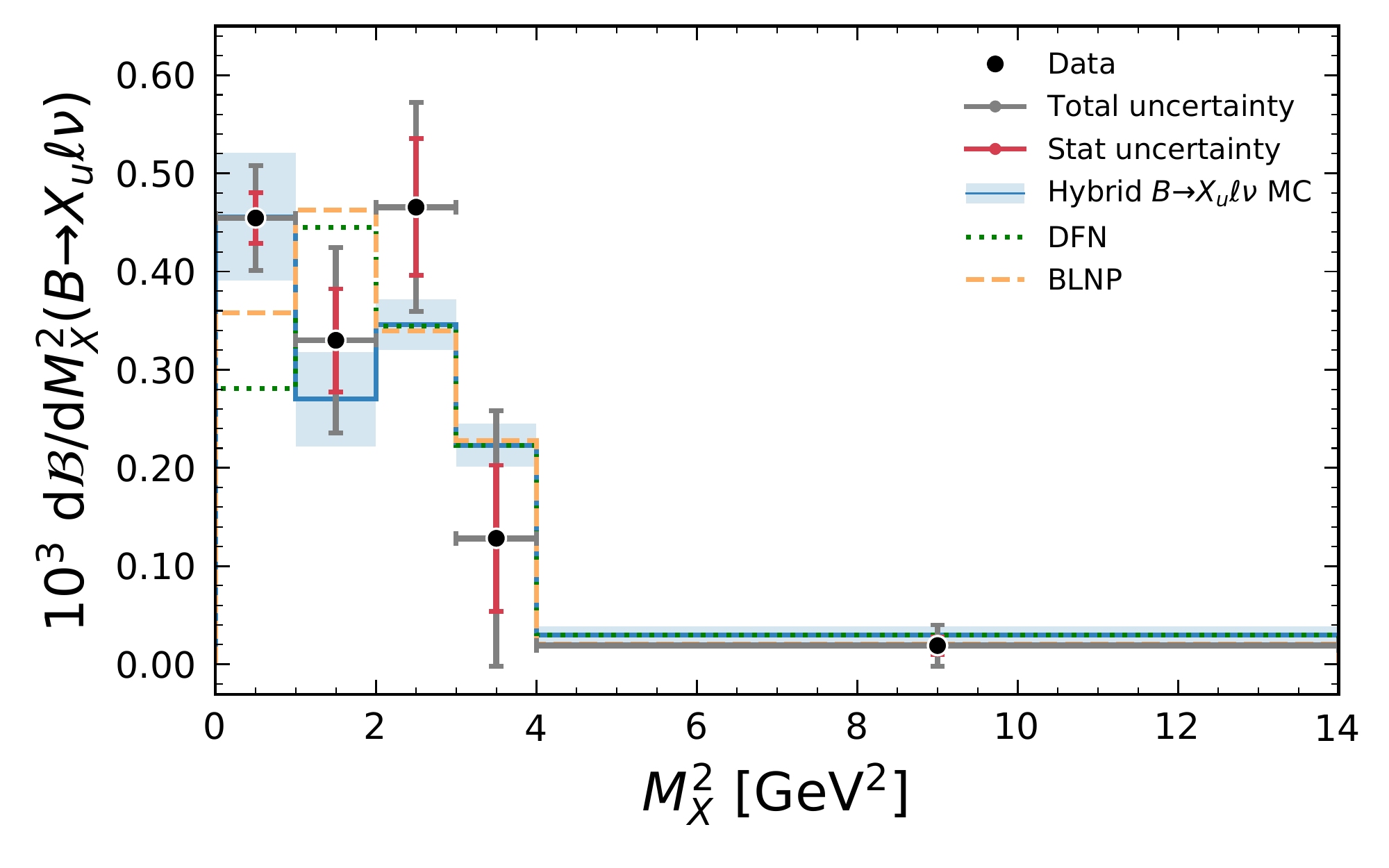} 
  \includegraphics[width=0.4\textwidth]{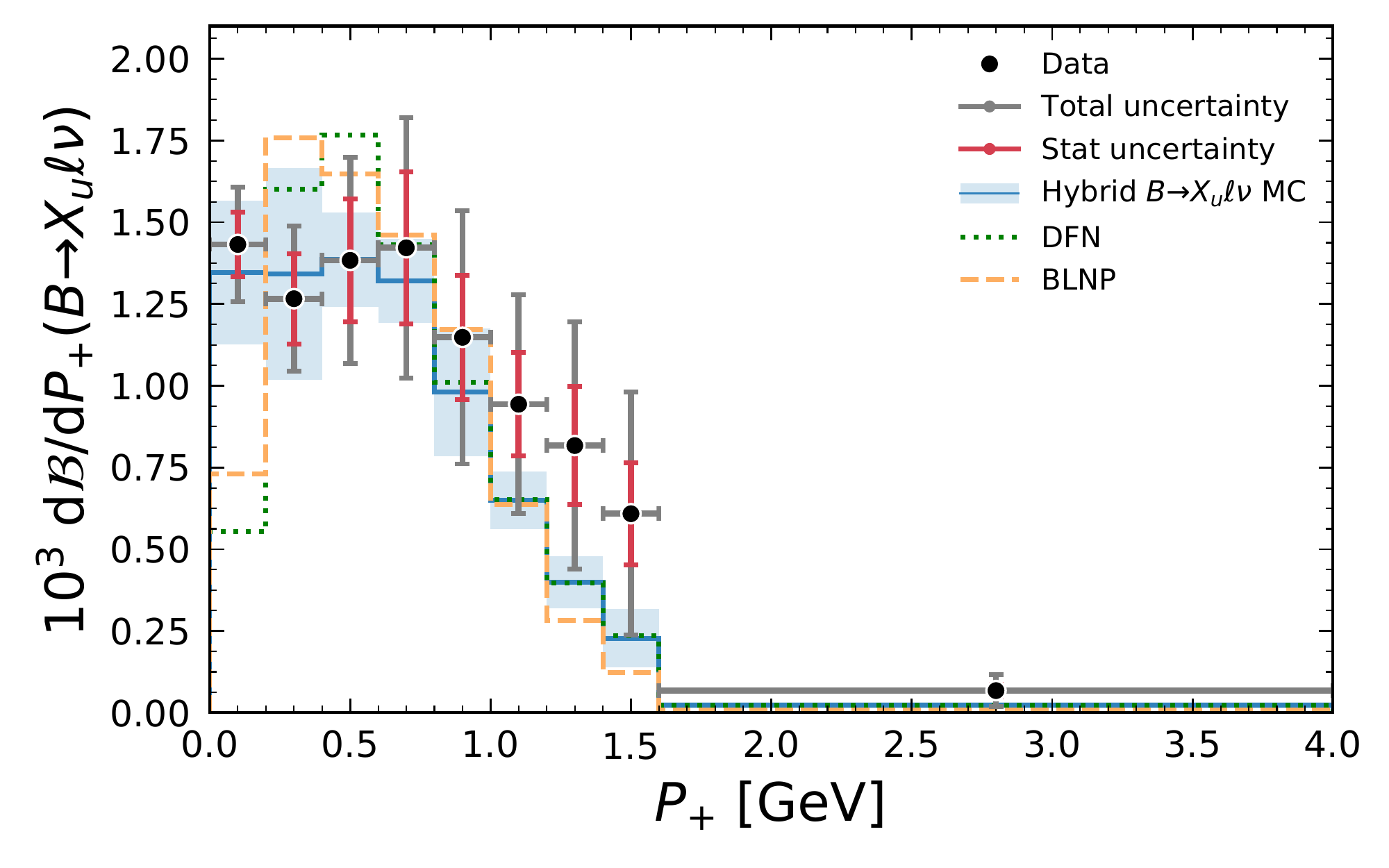} 
  \includegraphics[width=0.4\textwidth]{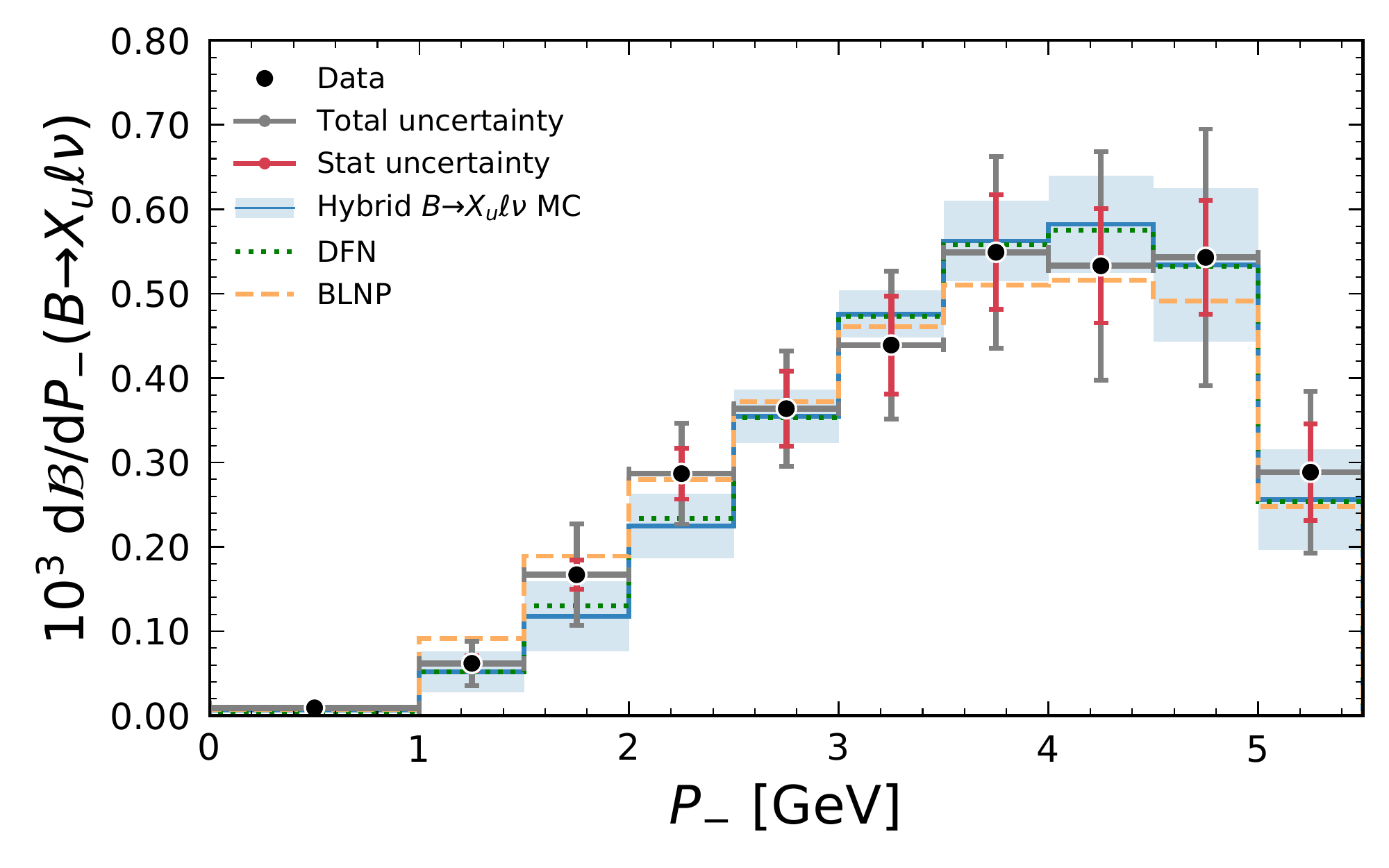} 
\caption{
The measured differential \bulnu branching fractions are shown: the lepton energy in the $B$ rest frame ($E_\ell^B$), the four-momentum-transfer squared of the $B$ to the $X_u$ system [$q^2 = \left( p_B - p_X \right)^2$], the invariant hadronic mass and mass squared of the $X_u$ system ($M_X$, $M_X^2$), and the light-cone momenta of the hadronic $X_u$ system [$P_\pm = (E_X^B \mp |\bold{p}_X^B|)$]. The hybrid MC prediction and two inclusive calculations are also shown and scaled to \BFMC.
 }
\label{fig:dBF}
\end{figure*}
 
The measured differential branching fractions as a function of $E_\ell^B$, $q^2$, $M_X$, $M_X^2$, $P_{-}$, and $P_+$ are shown in Fig.~\ref{fig:dBF} and the numerical values with full correlations can be found in Supplemental Material, Ref.~\cite{supplemental}. The distributions are compared to the \bulnu hybrid MC and the fully inclusive DFN~\cite{DeFazio:1999ptt} and BLNP~\cite{Lange:2005yw} predictions with model parameters listed in Table~\ref{tab:MC}. All predictions are scaled to match the \bulnu partial branching fraction ($\Delta \mathcal{B}$) with $E_\ell^B > 1 \, \mathrm{GeV}$ of \BFMC from Ref.~\cite{Cao:2021xqf}. The uncertainty band of the hybrid prediction includes variations on the composition, form factors, and the inclusive modeling, whose central value is based on the DFN prediction but includes the difference to BLNP as an additional uncertainty. The agreement between the measured and predicted distributions is fair overall, with differences occurring for the fully inclusive predictions in the resonance region of, e.g., low $M_X$, and near the end point of $q^2$ and $E_\ell^B$. There the hybrid MC describes the \bulnu process more adequately due to the explicit inclusion of resonant contributions. The largest discrepancy is observed in $E_\ell^B$, but the data points in the range of $E_\ell^B \in [1 - 1.8] \, \mathrm{GeV}$ exhibit strong correlations and are only weakly correlated or anticorrelated with the other bins of the spectrum. To quantify the agreement with the three displayed predictions we carry out a $\chi^2$ test using the experimental covariance only. We find a good $\chi^{2}$ of $13.5$ for the measured $E_\ell^B$ spectrum and the hybrid prediction with 16 degrees of freedom. Similarly we find for the DFN and BLNP predictions $\chi^2$ values of 16.2 and 16.5, respectively. 
 
In conclusion, this Letter presents the first measurements of differential branching fractions of inclusive semileptonic \bulnu decays as a function of $E_\ell^B$, $q^2$, $M_X$, $M_X^2$, $P_{-}$, and $P_+$ (a first preliminary measurement of the shape of the spectrum of $M_X^2$ was presented in Ref.~\cite{Tackmann:2008qa} and Ref.~\cite{BaBar:2016rxh} reported a differential branching fraction measurement as a function $E_e^B$, but without providing the full experimental uncertainties). The measurements use the full Belle data set of 711 fb$^{-1}$ of integrated luminosity at the $\Upsilon(4S)$ resonance and for $\ell = e, \mu$ in which one of the two $B$ mesons was fully reconstructed in hadronic modes. The differential branching fractions are obtained by subtracting \bclnu and other backgrounds with the normalization determined by a fit to the $M_X$ distribution of the hadronic $X$ system. The resulting distributions are corrected for detector resolution and efficiency effects and unfolded to the phase space of the lepton energy of $E_\ell^B > 1 \, \mathrm{GeV}$ in the rest frame of the signal $B$ meson. The measurements are, depending on the region of phase space, statistically or systematically limited, and show fair agreement to hybrid and inclusive predictions of \bulnu decays. The measured distributions are sensitive to the shape function governing the nonperturbative dynamics of the $b \to u$ transition and will allow future direct determinations of the shape function and $|V_{ub}|$, as proposed by Refs.~\cite{Bernlochner:2020jlt,Gambino:2016fdy}. These novel analyses will provide new insights into the persistent tensions on the value of $|V_{ub}|$ from inclusive and exclusive determinations~\cite{HFLAV:2019otj}.

We thank Kerstin Tackmann, Frank Tackmann, Zoltan Ligeti, and Dean Robinson for discussions about the content of this manuscript. 
L. C., W. S., R. vT., and F. B. were supported by the German Research Foundation (DFG) Emmy-Noether Grant No. BE 6075/1-1.  L.C. was also supported by the Helmholtz W2/W3-116 grant.

We thank the KEKB group for the excellent operation of the
accelerator; the KEK cryogenics group for the efficient
operation of the solenoid; and the KEK computer group, and the Pacific Northwest National
Laboratory (PNNL) Environmental Molecular Sciences Laboratory (EMSL)
computing group for strong computing support; and the National
Institute of Informatics, and Science Information NETwork 5 (SINET5) for
valuable network support.  We acknowledge support from
the Ministry of Education, Culture, Sports, Science, and
Technology (MEXT) of Japan, the Japan Society for the 
Promotion of Science (JSPS), and the Tau-Lepton Physics 
Research Center of Nagoya University; 
the Australian Research Council including grants
DP180102629, % Sevior
DP170102389, % Varvell
DP170102204, % Yabsley
DP150103061, % Urquijo
FT130100303; % Urquijo;
Austrian Federal Ministry of Education, Science and Research (FWF) and
FWF Austrian Science Fund No.~P~31361-N36;
the National Natural Science Foundation of China under Contracts
No.~11435013,  %Zhen-An Liu
No.~11475187,  %Chang-Zheng Yuan
No.~11521505,  %Chang-Zheng Yuan
No.~11575017,  %Cheng-Ping Shen
No.~11675166,  %Wen-Biao Yan
No.~11705209;  %Yi-Ming Li
Key Research Program of Frontier Sciences, Chinese Academy of Sciences (CAS), Grant No.~QYZDJ-SSW-SLH011; % Chang-Zheng Yuan
the  CAS Center for Excellence in Particle Physics (CCEPP); %Chang-Zheng Yuan,
the Shanghai Pujiang Program under Grant No.~18PJ1401000;  %Tao Luo
the Shanghai Science and Technology Committee (STCSM) under Grant No.~19ZR1403000; %Xiaolong Wang
the Ministry of Education, Youth and Sports of the Czech
Republic under Contract No.~LTT17020;
Horizon 2020 ERC Advanced Grant No.~884719 and ERC Starting Grant No.~947006 ``InterLeptons'' (European Union);
the Carl Zeiss Foundation, the Deutsche Forschungsgemeinschaft, the
Excellence Cluster Universe, and the VolkswagenStiftung;
the Department of Atomic Energy (Project Identification No. RTI 4002) and the Department of Science and Technology of India; 
the Istituto Nazionale di Fisica Nucleare of Italy; 
National Research Foundation (NRF) of Korea Grant
Nos.~2016R1\-D1A1B\-01010135, 2016R1\-D1A1B\-02012900, 2018R1\-A2B\-3003643,
2018R1\-A6A1A\-06024970, 2018R1\-D1A1B\-07047294, 2019K1\-A3A7A\-09033840,
2019R1\-I1A3A\-01058933;
Radiation Science Research Institute, Foreign Large-size Research Facility Application Supporting project, the Global Science Experimental Data Hub Center of the Korea Institute of Science and Technology Information and KREONET/GLORIAD;
the Polish Ministry of Science and Higher Education and 
the National Science Center;
the Ministry of Science and Higher Education of the Russian Federation, Agreement 14.W03.31.0026, % from 15.02.2018
and the HSE University Basic Research Program, Moscow; % from 15.04.2021
University of Tabuk research grants
S-1440-0321, S-0256-1438, and S-0280-1439 (Saudi Arabia);
the Slovenian Research Agency Grant Nos. J1-9124 and P1-0135;
Ikerbasque, Basque Foundation for Science, Spain;
the Swiss National Science Foundation; 
the Ministry of Education and the Ministry of Science and Technology of Taiwan;
and the United States Department of Energy and the National Science Foundation.

\bibliographystyle{apsrev4-1}
\bibliography{BtoXulnu}

\newpage
\onecolumngrid
\section*{Supplemental Material} 
\setcounter{page}{1}
\subsection{HEPDATA and Forward-Folding}

The results will be made fully available in HEPData (https://www.hepdata.net), including the background-subtracted yields, migration matrices, and efficiency curves. This will allow interested parties to also forward-fold \bulnu theory predictions and directly compare such with the background-subtracted Belle data. In addition, the first to third moments of each differential spectrum are provided. 

\subsection{Systematic Uncertainties}

\begin{figure*}[th!]
  \includegraphics[width=0.45\textwidth]{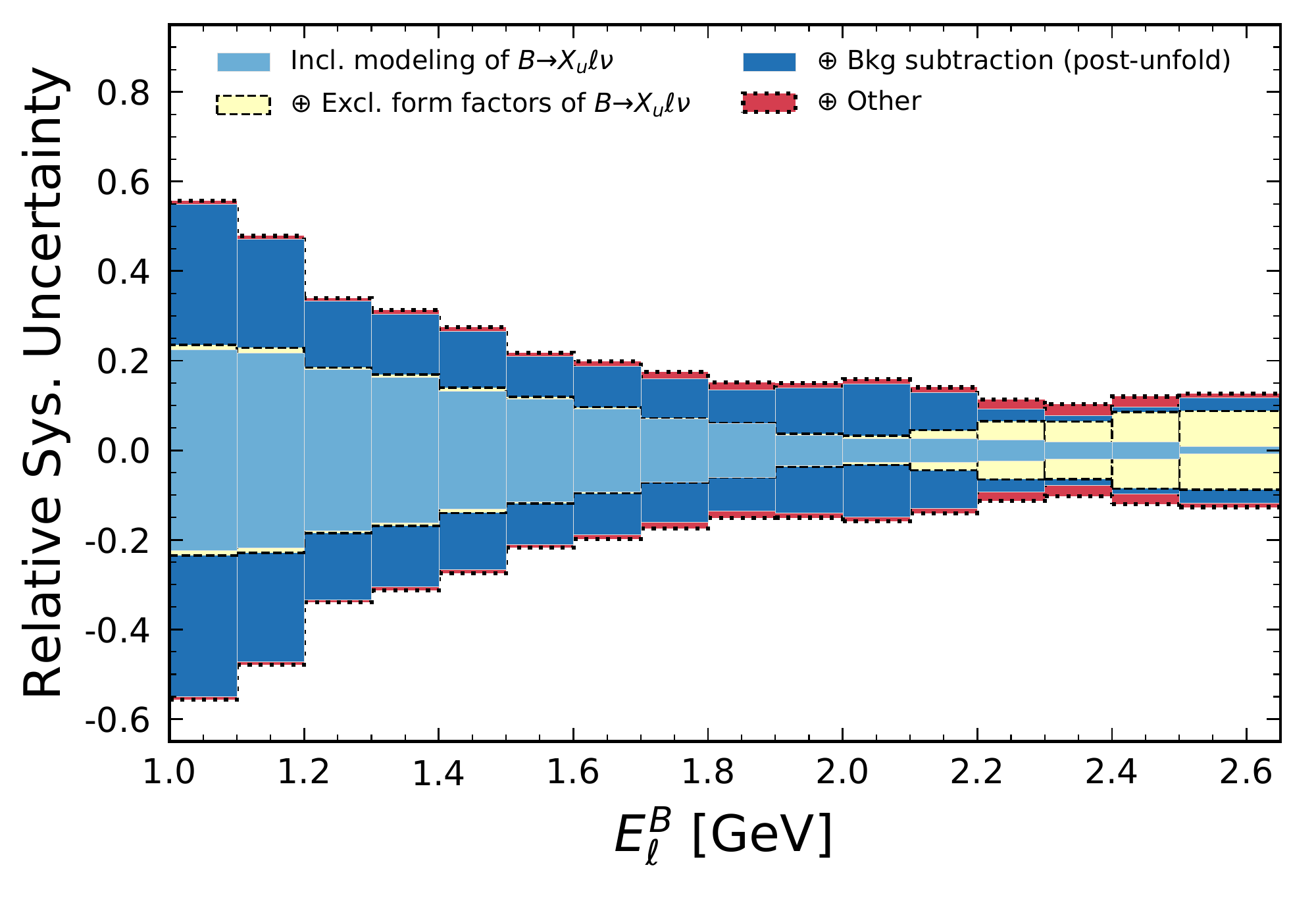} 
  \includegraphics[width=0.45\textwidth]{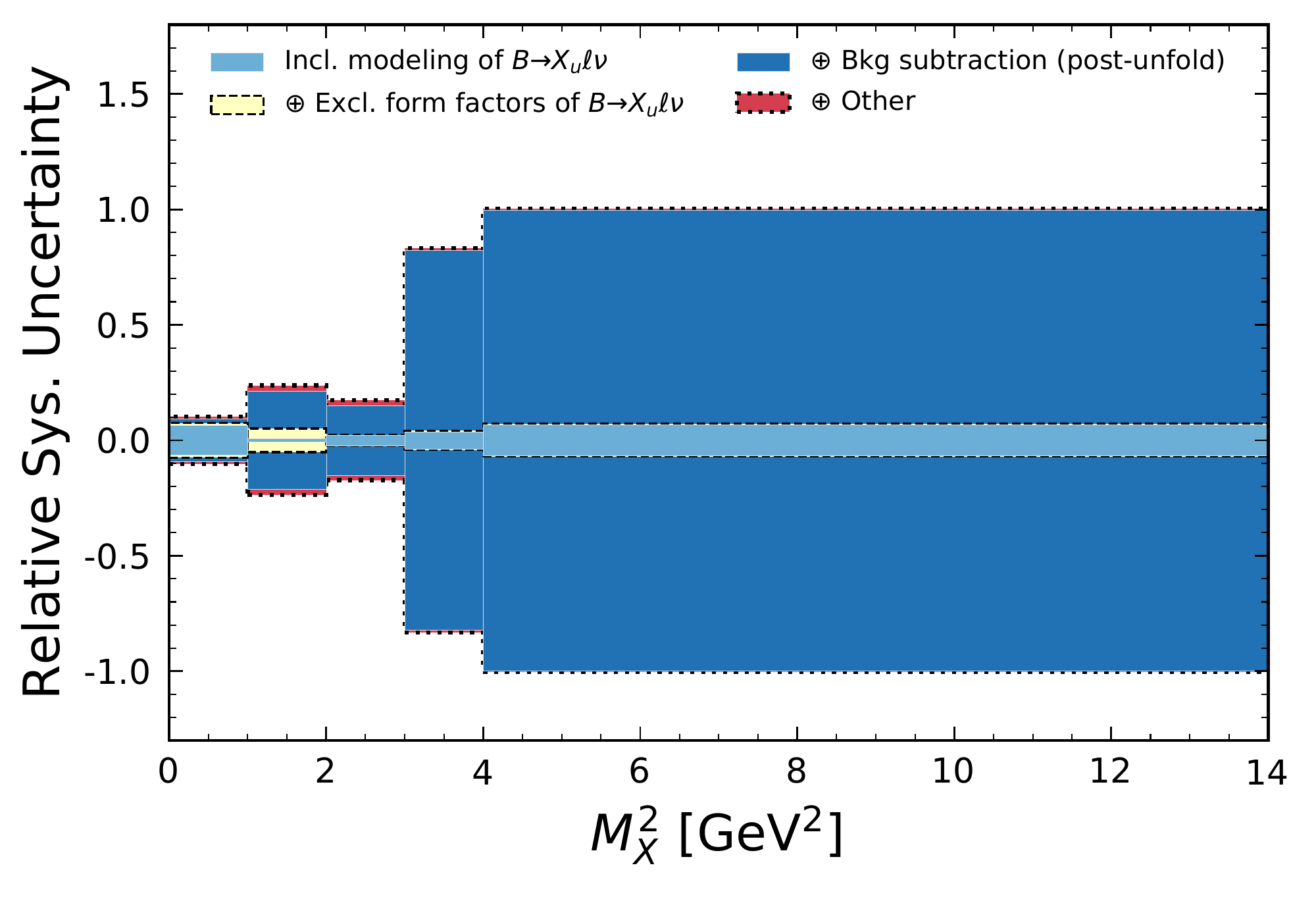} 
  \includegraphics[width=0.45\textwidth]{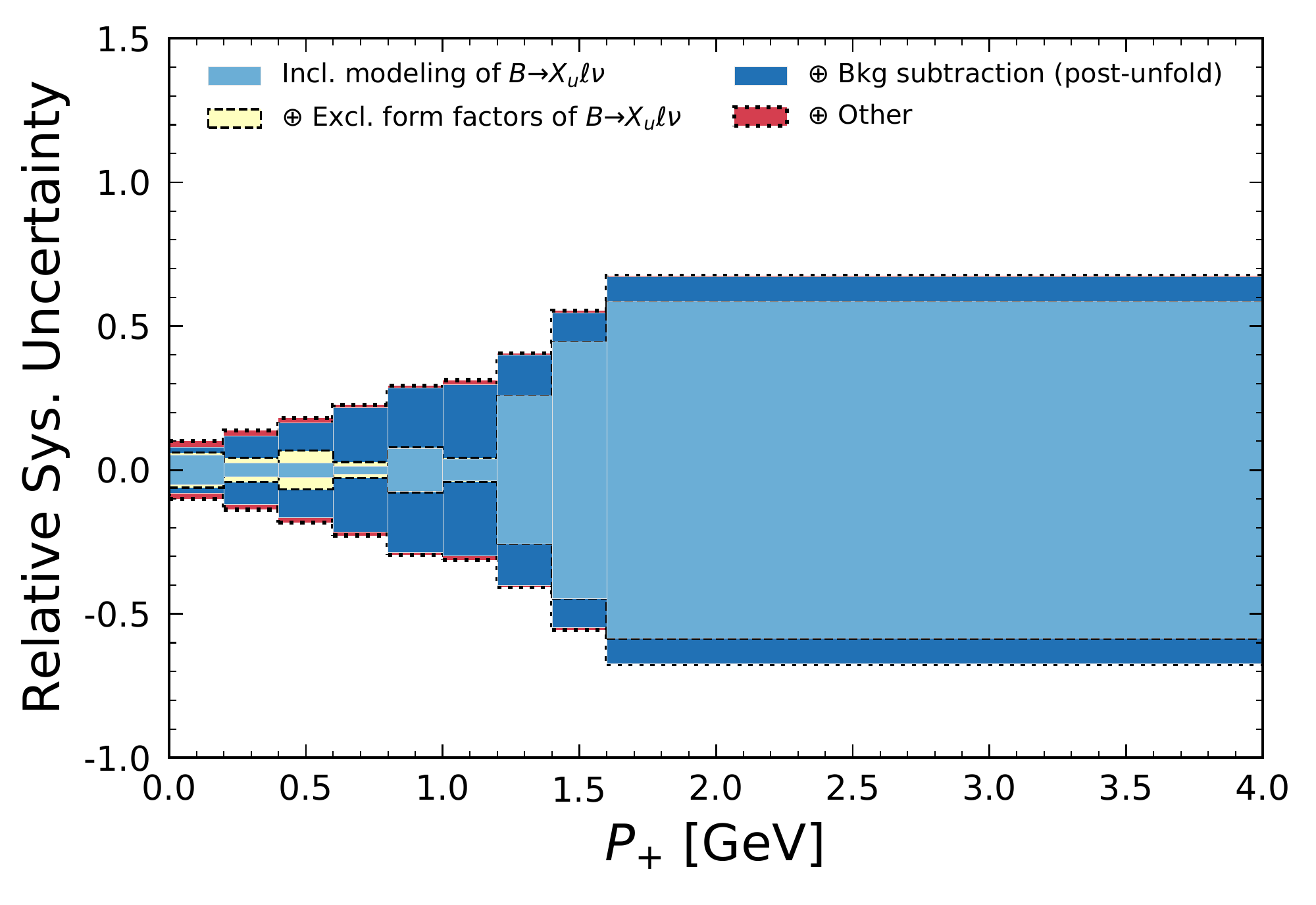} 
  \includegraphics[width=0.45\textwidth]{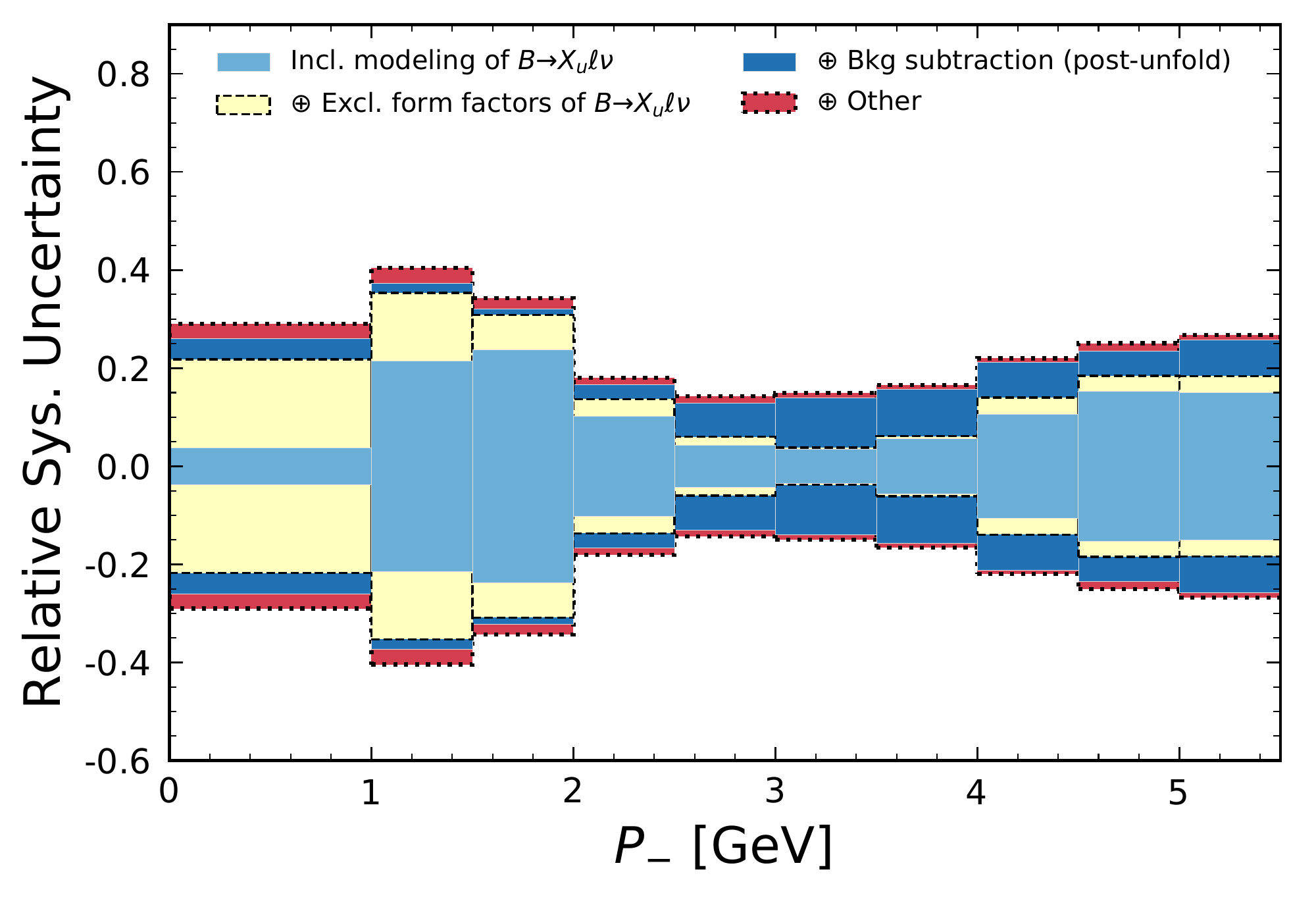} 
\caption{
The systematic uncertainties on the differential \bulnu branching fractions as a function of $E_\ell^B$, $M_X^2$, and $P_\pm$ are shown. They are separated into systematic uncertainties associated to the modeling of the inclusive and exclusive parts of the \bulnu signal, the background subtraction, and other sources. 
 }
\label{fig:syst_dBF}
\end{figure*}

Figure~\ref{fig:syst_dBF} displays the systematic uncertainties of $E_\ell^B$, $M_X^2$, $P_\pm$ shown as relative errors with respect to the measured differential branching fraction. The low $E_\ell^B$ region is dominated by uncertainties from the modeling of the inclusive \bulnu and the background subtraction uncertainties. The endpoint of the $E_\ell^B$  spectrum is dominated by contributions from resonant \bulnu decays. The systematic uncertainties for large $M_X^2$ are fully dominated by the background subtraction error. The systematic errors for $P_-$ show three distinct regions: the intermediate region is dominated by the background subtraction uncertainties, whereas the low and high $P_-$ regions are dominated by exclusive and inclusive \bulnu modeling uncertainties. The uncertainties at large $P_+$ values are fully dominated by the modeling of the inclusive parts of the \bulnu hybrid. 

Systematic uncertainties are consistently propagated through the entire analysis procedure, including the unfolding and efficiency correction. Tables~\ref{tab:data-dbf-sys-mx}-\ref{tab:data-dbf-sys-pminus} provide a full summary of all the considered systematic uncertainties. A brief summary on the most significant uncertainties follows (note that they are evaluated in the same manner as described in Ref.~\cite{Cao:2021xqf}):

\begin{itemize}
 \item[-] The uncertainty on the tracking efficiency is evaluated by assigning an error of 0.35\% per charged track on the \bulnu signal side. 
 \item[-] The tagging calibration uncertainties are evaluated by producing different sets of calibration factors, which take into account the correlation structure from common systematic uncertainties and that individual channels and ranges of the output classifier are statistically independent. The uncertainty on the calibration factors is about 3.6\% and only a negligible dependence on the studied kinematic distributions is observed. 
 \item[-] The uncertainties on the composition of the used \bulnu MC is evaluated by variations of the \bpilnu, \brholnu, \bomegalnu, \betalnu, \betaplnu branching fractions and form factors. Semileptonic \bpilnu\ decays are simulated using the Bourrely-Caprini-Lellouch (BCL) parametrization~\citep{Bourrely:2008za} with form factor central values and uncertainties from the global fit carried out by Ref.~\citep{Lattice:2015tia}. The processes of \brholnu\ and \bomegalnu\ are modeled using the BCL form factor parametrization. We use the fit of Ref.~\cite{Bernlochner:2021rel}, that combines the measurements of Refs.~\cite{Sibidanov:2013rkk,Lees:2012mq,delAmoSanchez:2010af} with the light-cone sum rule predictions of Ref.~\cite{Bharucha:2012wy} to determine a set of form factor central values and uncertainties. The processes of \betalnu and \betaplnu are modeled using the LCSR calculation of Ref.~\cite{Duplancic:2015zna}. The uncertainty on non-resonant \bulnu contributions in the used hybrid model is estimated by changing the underlying model from that of DFN~\cite{DeFazio:1999ptt} to that of BLNP~\cite{Lange:2005yw}. In addition, the uncertainty on the used DFN parameters $m_b^{\text{KN}} = (4.66 \pm 0.04) \, \mathrm{GeV}$ and $a^{\text{KN}} = 1.3 \pm 0.5$ from Ref.~\cite{Buchmuller:2005zv} are incorporated. For each of these variations, new hybrid weights are calculated to propagate the uncertainties into the full analysis procedure in a consistent way. 
 \item[-] The uncertainties of $X_u$ fragmentation into $s \bar s$ quark pairs is evaluated by variations of the corresponding JETSET parameter $\gamma_s$~\cite{SJOSTRAND199474}. We vary the $s \bar s$ production probability within $\gamma_s = 0.30 \pm 0.09$, with an uncertainty covering the range of the direct measurements of Refs.~\cite{Althoff:1984iz,Bartel:1983qp} of $\gamma_s = 0.35 \pm 0.05$ and $\gamma_s = 0.27 \pm 0.06$. 
\item[-]  The $X_u$ system of the non-resonant signal component is hadronized by JETSET into final states with two or more pions. We assign an uncertainty on the multiplicity modeling by changing the pion multiplicity of non-resonant \bulnu\ to the distribution observed in data in the signal enriched region of $M_X < 1.7$ GeV, identical to the approach adopted in Ref.~\cite{Cao:2021xqf}.
\item[-] The uncertainties on the modeling of \bdlnu\, \bdslnu\, and \bddslnu\ are evaluated by variations of the BGL parameters and heavy quark form factors within their uncertainties. The \bdlnu\ decays are modeled using the Boyd-Grinstein-Lebed (BGL) parametrization~\cite{Boyd:1994tt} with form factor central values and uncertainties taken from the fit in Ref.~\cite{Glattauer:2015teq}. For \bdslnu\, we use the BGL implementation proposed by Refs.~\cite{Grinstein:2017nlq,Bigi:2017njr} with form factor central values and uncertainties from the fit to the measurement of Ref.~\cite{Waheed:2018djm}. Both backgrounds are normalized to the average branching fraction of Ref.~\cite{Amhis:2019ckw} assuming isospin symmetry. Semileptonic \bddslnu decays with $D^{**} = \{ D_0^*, D_1^*, D_1, D_2^* \}$ denoting the four orbitally excited charmed mesons are modeled using the heavy-quark-symmetry-based form factors proposed in Ref.~\cite{Bernlochner:2016bci}. In addition, the branching fraction uncertainties are included and the uncertainties on the \bclnu gap branching fractions are taken to be large enough to account for the difference between the sum of all exclusive branching fractions measured and the inclusive branching fraction measured.
\item[-] The impact on the efficiency of the lepton- and hadron-identification uncertainties is evaluated by producing replicas of the simulated samples with new corrections weights sampled from the measured corrections, that parametrize the difference between simulated and recorded collision event efficiencies on the identification efficiency as a function of the laboratory momentum and polar angle of the charged particle in question. 
\item[-] The slow pion and $K_S^0$ reconstruction efficiencies are also evaluated using replicas of simulated samples, by producing sets of new correction weights, that parametrize the difference in the reconstruction efficiency between simulated and recorded collision event efficiencies. 
\end{itemize}

\begin{table}[h]
\renewcommand\arraystretch{1.2}
\centering
\begin{tabular}{lllllllll}
\hline
\hline
$M_{X}$ [GeV]    & 0-0.3 & 0.3-0.6 & 0.6-0.9 & 0.9-1.2 & 1.2-1.5 & 1.5-1.8 & 1.8-2.1 & 2.1-4.0 \\
\hline
Tracking efficiency          &    0.55 &    0.56 &    0.82 &    0.86 &    0.95 &    1.05 &    1.15 &    1.19 \\
Tagging calibration            &    3.69 &    3.69 &    3.65 &    3.64 &    3.64 &    3.57 &    3.79 &    3.66 \\
Slow pion efficiency            &    0.00 &    0.07 &    0.04 &    0.05 &    0.04 &    0.04 &    0.06 &    0.04 \\
$K_{S}^{0}$           &    0.04 &    0.05 &    0.04 &    0.02 &    0.04 &    0.03 &    0.02 &    0.05 \\
$e$ID                &    0.72 &    0.83 &    0.74 &    0.69 &    0.73 &    0.74 &    0.94 &    1.22 \\
$\mu$ID         &    1.59 &    1.25 &    1.34 &    1.29 &    1.44 &    1.35 &    1.09 &    0.70 \\
$K/\pi$ ID           &    0.39 &    0.67 &    0.68 &    0.74 &    0.81 &    1.02 &    1.27 &    1.24 \\
$\mathcal{B}(B\to X_{u}\ell\nu)$          &    0.18 &    0.44 &    0.07 &    0.59 &    0.82 &    0.69 &    0.73 &    0.46 \\
$\mathcal{B}(B\to \pi  \ell\nu)$         &    0.42 &    0.45 &    0.45 &    0.14 &    0.05 &    0.04 &    0.05 &    0.05 \\
$\mathcal{B}(B\to \rho  \ell\nu)$        &    0.42 &    1.00 &    0.61 &    0.56 &    0.33 &    0.16 &    0.22 &    0.15 \\
$\mathcal{B}(B\to \omega  \ell\nu)$      &    0.42 &    0.39 &    0.65 &    0.12 &    0.11 &    0.06 &    0.11 &    0.10 \\
$\mathcal{B}(B\to \eta  \ell\nu)$        &    0.41 &    1.16 &    0.46 &    0.11 &    0.06 &    0.03 &    0.03 &    0.14 \\
$\mathcal{B}(B\to \eta^{\prime}  \ell\nu)$     &    0.42 &    0.39 &    0.46 &    0.24 &    0.30 &    0.03 &    0.14 &    0.11 \\
$B\to \pi  \ell\nu$ FF          &    0.98 &    3.08 &    1.52 &    0.53 &    1.05 &    0.37 &    0.36 &    0.38 \\
$B\to \rho  \ell\nu$ FF         &    2.77 &    8.54 &    3.96 &    2.94 &    1.65 &    0.59 &    0.83 &    0.89 \\
$B\to \omega  \ell\nu$ FF     &    2.40 &    9.71 &    1.10 &    0.90 &    1.41 &    0.70 &    0.65 &    1.32 \\
$B\to \eta  \ell\nu$ FF         &    0.71 &    3.58 &    0.09 &    0.09 &    0.51 &    0.28 &    0.27 &    0.07 \\
$B\to \eta^{\prime}  \ell\nu$ FF      &    0.69 &    3.65 &    0.16 &    0.27 &    0.48 &    0.29 &    0.32 &    0.15 \\
Hybrid model          &    0.21 &    5.86 &    5.08 &    4.01 &    0.50 &    1.97 &    2.02 &    6.13 \\
DFN parameters        &    0.18 &    3.66 &    1.01 &    1.38 &    1.64 &    0.87 &    0.50 &    1.35 \\
$\gamma_{s}$           &    0.47 &    4.17 &    2.36 &    3.98 &    3.08 &    4.10 &    9.31 &    3.60 \\
$\pi^{+}$ multiplicity modeling   &    0.57 &    0.42 &    0.45 &    4.15 &    7.98 &    4.78 &    3.98 &    2.34 \\
$N_{\mathit{B\bar{B}}}$              &    1.25 &    1.25 &    1.25 &    1.25 &    1.25 &    1.25 &    1.25 &    1.25 \\
Background subtraction &    5.97 &   26.93 &    8.23 &   25.15 &   29.65 &   16.80 &   73.36 &  126.64 \\
MC stat. (migration matrix)    &    4.04 &   11.22 &    3.54 &    6.85 &    4.30 &    4.71 &    6.85 &    8.22 \\

\hline
Total syst. uncertainty  &    9.36 &   33.77 &   12.32 &   27.56 &   31.62 &   19.21 &   74.55 &  127.23 \\
Total stat. uncertainty     &   11.11 &   32.64 &   10.77 &   24.99 &   21.88 &   16.54 &   46.24 &   66.76 \\
Total uncertainty    &   14.53 &   46.97 &   16.36 &   37.20 &   38.45 &   25.35 &   87.73 &  143.68 \\
\hline
\hline
\end{tabular}
\caption{The relative uncertainties (\%) of the measured differential branching fraction of $M_{X}$ are shown.}
    \label{tab:data-dbf-sys-mx}
\end{table}

\begin{table}[h]
\renewcommand\arraystretch{1.2}
\centering
\begin{tabular}{llllll}
\hline
\hline
$M_{X}^{2}$ [GeV$^{2}$]   & 0-1 & 1-2 & 2-3 & 3-4 & 4-14 \\
\hline
Tracking efficiency         &     0.69 &     0.96 &     1.01 &     1.20 &     1.16 \\
Tagging calibration          &     3.67 &     3.62 &     3.60 &     3.72 &     3.73 \\
Slow pion efficiency        &     0.03 &     0.05 &     0.05 &     0.05 &     0.03 \\
$K_{S}^{0}$       &     0.04 &     0.02 &     0.04 &     0.04 &     0.04 \\
$e$ID              &     0.74 &     0.70 &     0.74 &     0.96 &     1.12 \\
$\mu$ID         &     1.42 &     1.33 &     1.37 &     0.99 &     0.89 \\
$K/\pi$ ID          &     0.55 &     0.89 &     0.92 &     1.22 &     1.24 \\
$\mathcal{B}(B\to X_{u}\ell\nu)$          &     0.77 &     0.11 &     0.08 &     2.15 &     0.12 \\
$\mathcal{B}(B\to \pi  \ell\nu)$        &     0.67 &     0.43 &     0.07 &     0.23 &     0.04 \\
$\mathcal{B}(B\to \rho  \ell\nu)$        &     0.37 &     0.49 &     0.07 &     0.28 &     0.10 \\
$\mathcal{B}(B\to \omega  \ell\nu)$     &     0.45 &     0.07 &     0.03 &     0.16 &     0.05 \\
$\mathcal{B}(B\to \eta  \ell\nu)$       &     0.36 &     0.10 &     0.03 &     0.14 &     0.16 \\
$\mathcal{B}(B\to \eta^{\prime}  \ell\nu)$      &     0.56 &     0.47 &     0.18 &     0.09 &     0.06 \\
$B\to \pi  \ell\nu$ FF          &     1.24 &     1.87 &     0.09 &     1.08 &     0.95 \\
$B\to \rho  \ell\nu$ FF         &     2.14 &     3.03 &     0.41 &     1.03 &     1.01 \\
$B\to \omega  \ell\nu$ FF     &     2.88 &     3.10 &     0.38 &     1.08 &     1.08 \\
$B\to \eta  \ell\nu$ FF         &     0.88 &     1.18 &     0.02 &     0.54 &     0.36 \\
$B\to \eta^{\prime}  \ell\nu$ FF    &     0.91 &     1.30 &     0.03 &     0.53 &     0.49 \\
Hybrid model          &     6.25 &     0.37 &     1.95 &     3.16 &     6.71 \\
DFN parameters         &     1.25 &     0.56 &     1.18 &     1.90 &     1.13 \\
$\gamma_{s}$             &     0.90 &     3.76 &     2.63 &     9.53 &     7.82 \\
$\pi^{+}$ multiplicity   &     0.24 &     8.89 &     6.56 &     7.71 &     1.98 \\
$N_{\mathit{B\bar{B}}}$          &     1.25 &     1.25 &     1.25 &     1.25 &     1.25 \\
Background subtraction &     4.86 &    20.19 &    14.45 &    81.67 &    99.28 \\
MC stat. (migration matrix)    &     2.12 &     4.29 &     3.99 &     8.41 &     7.19 \\
\hline
Total syst. uncertainty      &    10.30 &    23.76 &    17.29 &    83.27 &   100.22 \\
Total stat. uncertainty     &     5.66 &    15.88 &    14.97 &    58.02 &    46.97 \\
Total uncertainty   &    11.75 &    28.58 &    22.88 &   101.49 &   110.68 \\
\hline
\hline
\end{tabular}
\caption{The relative uncertainties (\%) of the measured differential branching fraction of $M_{X}^{2}$ are shown.}
    \label{tab:data-dbf-sys-mx2}
\end{table}

\begin{sidewaystable}[h]
\renewcommand\arraystretch{1.2}
\centering
\begin{tabular}{lllllllllllll}
\hline
\hline
$q^{2}$ [GeV$^{2}$]    & 0-2 & 2-4 & 4-6 & 6-8 & 8-10 & 10-12  & 12-14 & 14-16 & 16-18 & 18-20 & 20-22 & 22-26.5 \\
\hline
Tracking efficiency                   &    0.93 &    0.95 &    0.94 &    0.90 &    0.89 &    0.87 &    0.82 &    0.75 &    0.72 &    0.67 &     0.55 &     0.57 \\
Tagging calibration                     &    3.58 &    3.69 &    3.71 &    3.69 &    3.65 &    3.63 &    3.63 &    3.64 &    3.64 &    3.73 &     3.86 &     3.76 \\
Slow pion efficiency                    &    0.02 &    0.03 &    0.04 &    0.03 &    0.04 &    0.04 &    0.04 &    0.04 &    0.04 &    0.06 &     0.09 &     0.07 \\
$K_{S}^{0}$                      &    0.03 &    0.03 &    0.04 &    0.03 &    0.03 &    0.04 &    0.04 &    0.04 &    0.03 &    0.03 &     0.04 &     0.04 \\
$e$ID                   &    0.79 &    0.77 &    0.77 &    0.75 &    0.74 &    0.66 &    0.69 &    0.83 &    0.87 &    0.83 &     0.82 &     0.79 \\
$\mu$ID                 &    1.39 &    1.26 &    1.24 &    1.32 &    1.34 &    1.46 &    1.45 &    1.39 &    1.33 &    1.52 &     1.64 &     1.47 \\
$K/\pi$ ID                  &    0.85 &    0.96 &    0.93 &    0.83 &    0.79 &    0.77 &    0.70 &    0.65 &    0.53 &    0.35 &     0.20 &     0.22 \\
$\mathcal{B}(B\to X_{u}\ell\nu)$                 &    1.26 &    1.01 &    0.75 &    0.54 &    0.42 &    0.01 &    0.56 &    1.06 &    2.01 &    1.47 &     0.64 &     0.45 \\
$\mathcal{B}(B\to \pi  \ell\nu)$                  &    1.21 &    0.95 &    0.77 &    0.59 &    0.27 &    0.36 &    1.25 &    2.93 &    5.10 &    6.53 &     5.64 &     5.65 \\
$\mathcal{B}(B\to \rho  \ell\nu)$                 &    1.14 &    0.96 &    0.89 &    0.64 &    0.39 &    0.59 &    1.37 &    3.06 &    5.13 &    6.55 &     5.62 &     5.64 \\
$\mathcal{B}(B\to \omega  \ell\nu)$               &    1.13 &    0.92 &    0.74 &    0.52 &    0.21 &    0.32 &    1.23 &    2.92 &    5.09 &    6.53 &     5.62 &     5.61 \\
$\mathcal{B}(B\to \eta  \ell\nu)$                  &    1.14 &    0.93 &    0.74 &    0.54 &    0.26 &    0.35 &    1.23 &    2.91 &    5.09 &    6.53 &     5.62 &     5.61 \\
$\mathcal{B}(B\to \eta^{\prime}  \ell\nu)$                 &    1.13 &    0.94 &    0.74 &    0.53 &    0.22 &    0.33 &    1.23 &    2.92 &    5.08 &    6.53 &     5.62 &     5.61 \\
$B\to \pi  \ell\nu$ FF                   &    2.60 &    3.00 &    2.38 &    1.25 &    0.38 &    0.83 &    2.12 &    4.39 &    3.83 &    2.07 &     5.51 &     2.63 \\
$B\to \rho  \ell\nu$ FF                &    5.05 &    5.44 &    4.29 &    2.36 &    1.29 &    1.18 &    2.46 &    7.53 &   10.45 &    7.01 &    15.88 &     8.74 \\
$B\to \omega  \ell\nu$ FF                 &    6.69 &    8.65 &    5.85 &    2.92 &    1.07 &    1.19 &    4.27 &    9.92 &   13.29 &   16.85 &    20.29 &     9.44 \\
$B\to \eta  \ell\nu$ FF                 &    2.35 &    2.53 &    1.67 &    0.85 &    0.34 &    0.35 &    1.41 &    2.76 &    2.80 &    2.76 &     6.99 &     3.00 \\
$B\to \eta^{\prime}  \ell\nu$ FF               &    2.42 &    2.62 &    1.74 &    0.92 &    0.38 &    0.39 &    1.34 &    2.78 &    3.20 &    2.90 &     6.53 &     3.30 \\
Hybrid model                  &   16.17 &   16.41 &    9.91 &    2.71 &    3.62 &    4.69 &    9.82 &   12.48 &   15.53 &   11.75 &    23.34 &     5.45 \\
DFN parameters                     &    6.01 &    6.43 &    4.78 &    3.86 &    4.12 &    3.19 &    2.03 &    1.44 &    1.10 &    1.01 &     3.26 &     1.01 \\
$\gamma_{s}$                      &    6.28 &    4.39 &    1.39 &    1.41 &    1.52 &    3.14 &    0.43 &    0.07 &    1.98 &    1.99 &     0.47 &     1.09 \\
$\pi^{+}$ multiplicity    &    2.74 &    3.08 &    2.85 &    2.18 &    3.18 &    4.00 &    3.09 &    0.10 &    0.14 &    0.07 &     1.41 &     0.42 \\
$N_{\mathit{B\bar{B}}}$              &    1.25 &    1.25 &    1.25 &    1.25 &    1.25 &    1.25 &    1.25 &    1.25 &    1.25 &    1.25 &     1.25 &     1.25 \\
Background subtraction &   16.66 &   16.89 &   16.52 &   13.79 &   12.22 &   11.30 &   10.67 &   11.54 &    9.77 &    9.14 &    10.18 &    12.96 \\
MC stat. (migration matrix) &    2.94 &    3.15 &    2.37 &    2.29 &    2.56 &    2.87 &    2.69 &    3.62 &    3.40 &    3.81 &     7.33 &     7.81 \\
\hline
Total syst. uncertainty        &   27.32 &   28.03 &   22.28 &   16.20 &   14.86 &   14.73 &   17.05 &   23.60 &   28.71 &   28.79 &    40.98 &    25.11 \\
Total stat. uncertainty              &   16.71 &   13.67 &   13.05 &   11.61 &   12.26 &   12.87 &   12.45 &   13.30 &   11.96 &   12.12 &    15.69 &    19.97 \\
Total uncertainty           &   32.02 &   31.19 &   25.82 &   19.93 &   19.26 &   19.56 &   21.11 &   27.09 &   31.10 &   31.23 &    43.88 &    32.08 \\
\hline
\hline
\end{tabular}
\caption{The relative uncertainties (\%) of the measured differential branching fraction of $q^{2}$ are shown.}
    \label{tab:data-dbf-sys-q2}
\end{sidewaystable}

\begin{sidewaystable}[h]
\renewcommand\arraystretch{1.2}
\centering
\begin{tabular}{lllllllllllllllll}
\hline
\hline
$E_{\ell}^{B}$ [GeV]    & 1.0-1.1 & 1.1-1.2 & 1.2-1.3 & 1.3-1.4 & 1.4-1.5 & 1.5-1.6  & 1.6-1.7 & 1.7-1.8 & 1.8-1.9 & 1.9-2.0 & 2.0-2.1 & 2.1-2.2 & 2.2-2.3 & 2.3-2.4 & 2.4-2.5 & 2.5-2.65 \\
\hline
Tracking efficiency                  &    0.39 &    0.39 &    0.39 &    0.39 &    0.39 &    0.38 &    0.39 &    0.39 &    0.39 &    0.39 &     0.38 &     0.38 &     0.38 &     0.38 &     0.38 &     0.38 \\
Tagging calibration                     &    3.75 &    3.71 &    3.67 &    3.65 &    3.63 &    3.67 &    3.69 &    3.68 &    3.69 &    3.71 &     3.67 &     3.66 &     3.69 &     3.67 &     3.73 &     3.74 \\
Slow pion efficiency                     &    0.04 &    0.04 &    0.04 &    0.04 &    0.04 &    0.05 &    0.06 &    0.06 &    0.07 &    0.07 &     0.07 &     0.08 &     0.08 &     0.08 &     0.07 &     0.06 \\
$K_{S}^{0}$                     &    0.04 &    0.04 &    0.04 &    0.04 &    0.04 &    0.04 &    0.04 &    0.04 &    0.04 &    0.04 &     0.04 &     0.04 &     0.04 &     0.04 &     0.03 &     0.04 \\
$e$ID                  &    1.34 &    1.20 &    0.93 &    0.79 &    0.71 &    0.70 &    0.69 &    0.70 &    0.72 &    0.71 &     0.73 &     0.76 &     0.78 &     0.76 &     0.78 &     0.79 \\
$\mu$ID                  &    1.15 &    1.17 &    1.29 &    1.35 &    1.36 &    1.37 &    1.37 &    1.36 &    1.38 &    1.41 &     1.43 &     1.46 &     1.43 &     1.52 &     1.53 &     1.43 \\
$K/\pi$ ID                 &    0.90 &    0.91 &    0.83 &    0.83 &    0.87 &    0.88 &    0.95 &    0.91 &    0.85 &    0.82 &     0.75 &     0.71 &     0.67 &     0.66 &     0.57 &     0.56 \\
$\mathcal{B}(B\to X_{u}\ell\nu)$                 &    1.12 &    0.81 &    1.12 &    0.82 &    0.44 &    1.01 &    0.78 &    0.57 &    0.74 &    0.52 &     0.52 &     0.49 &     0.04 &     1.31 &     1.18 &     0.16 \\
$\mathcal{B}(B\to \pi  \ell\nu)$                  &    0.69 &    0.66 &    0.64 &    0.60 &    0.55 &    0.45 &    0.35 &    0.27 &    0.23 &    0.29 &     0.41 &     0.71 &     1.22 &     1.97 &     2.14 &     0.85 \\
$\mathcal{B}(B\to \rho  \ell\nu)$                  &    0.65 &    0.62 &    0.66 &    0.67 &    0.69 &    0.55 &    0.42 &    0.33 &    0.32 &    0.32 &     0.51 &     0.75 &     1.26 &     2.08 &     2.19 &     0.86 \\
$\mathcal{B}(B\to \omega  \ell\nu)$               &    0.64 &    0.61 &    0.57 &    0.52 &    0.47 &    0.39 &    0.30 &    0.19 &    0.04 &    0.14 &     0.37 &     0.69 &     1.19 &     1.95 &     2.12 &     0.85 \\
$\mathcal{B}(B\to \eta  \ell\nu)$                 &    0.64 &    0.61 &    0.57 &    0.53 &    0.47 &    0.39 &    0.31 &    0.21 &    0.07 &    0.15 &     0.36 &     0.69 &     1.19 &     1.95 &     2.11 &     0.85 \\
$\mathcal{B}(B\to \eta^{\prime}  \ell\nu)$                 &    0.64 &    0.61 &    0.57 &    0.52 &    0.46 &    0.40 &    0.32 &    0.22 &    0.10 &    0.18 &     0.38 &     0.69 &     1.19 &     1.95 &     2.11 &     0.85 \\
$B\to \pi  \ell\nu$ FF                 &    1.61 &    1.53 &    0.94 &    1.26 &    1.39 &    1.05 &    0.72 &    0.30 &    0.16 &    0.28 &     0.48 &     0.82 &     1.20 &     0.92 &     1.94 &     2.61 \\
$B\to \rho  \ell\nu$ FF                   &    3.62 &    3.56 &    2.16 &    2.63 &    2.78 &    1.86 &    1.52 &    0.86 &    0.40 &    0.57 &     1.25 &     2.02 &     3.36 &     3.19 &     4.01 &     4.55 \\
$B\to \omega  \ell\nu$ FF                 &    5.29 &    5.52 &    3.01 &    3.38 &    3.32 &    1.94 &    1.73 &    1.00 &    0.58 &    1.20 &     1.38 &     2.65 &     4.57 &     4.97 &     6.55 &     6.29 \\
$B\to \eta  \ell\nu$ FF                  &    1.69 &    1.62 &    0.92 &    1.10 &    1.17 &    0.77 &    0.61 &    0.28 &    0.12 &    0.21 &     0.37 &     0.83 &     1.31 &     1.06 &     1.79 &     2.09 \\
$B\to \eta^{\prime}  \ell\nu$ FF               &    1.56 &    1.58 &    0.95 &    1.16 &    1.20 &    0.80 &    0.65 &    0.33 &    0.09 &    0.23 &     0.33 &     0.77 &     1.21 &     0.99 &     1.89 &     2.22 \\
Hybrid model                   &   21.68 &   21.05 &   17.09 &   15.34 &   12.37 &   10.28 &    8.24 &    5.84 &    4.90 &    1.21 &     0.72 &     1.68 &     1.93 &     0.89 &     1.29 &     0.43 \\
DFN parameters                     &    5.74 &    5.52 &    5.69 &    5.31 &    4.41 &    5.12 &    4.26 &    3.88 &    3.55 &    3.20 &     2.50 &     2.02 &     1.30 &     1.67 &     1.45 &     0.66 \\
$\gamma_{s}$                      &    7.06 &    6.26 &    3.45 &    5.13 &    4.76 &    0.87 &    1.64 &    4.12 &    3.92 &    1.72 &     1.50 &     0.62 &     3.52 &     1.34 &     2.59 &     1.14 \\
$\pi^{+}$ multiplicity          &    3.19 &    3.35 &    2.89 &    2.65 &    2.55 &    3.41 &    3.86 &    3.83 &    3.63 &    3.13 &     3.06 &     2.86 &     2.28 &     1.62 &     1.11 &     0.23 \\
$N_{\mathit{B\bar{B}}}$                       &    1.25 &    1.25 &    1.25 &    1.25 &    1.25 &    1.25 &    1.25 &    1.25 &    1.25 &    1.25 &     1.25 &     1.25 &     1.25 &     1.25 &     1.25 &     1.25 \\
Background subtraction    &   49.32 &   40.78 &   27.56 &   25.05 &   22.39 &   17.20 &   16.10 &   14.09 &   11.89 &   13.35 &    14.39 &    12.03 &     6.05 &     3.85 &     3.85 &     6.54 \\
MC stat. (migration matrix) &    5.92 &    5.53 &    3.03 &    3.17 &    3.00 &    1.54 &    1.91 &    2.28 &    1.74 &    1.62 &     1.38 &     1.82 &     2.56 &     2.19 &     2.26 &     4.22 \\
\hline
Total syst. uncertainty       &   55.70 &   47.86 &   33.93 &   31.26 &   27.48 &   21.73 &   19.82 &   17.48 &   15.15 &   15.05 &    15.83 &    14.06 &    11.34 &    10.26 &    12.01 &    12.70 \\
Total stat. uncertainty                 &   54.34 &   39.47 &   23.05 &   19.69 &   17.06 &   13.07 &   13.04 &   12.21 &    9.40 &    8.18 &     7.60 &     7.47 &     6.95 &     6.08 &     6.96 &    11.22 \\
Total uncertainty           &   77.82 &   62.03 &   41.02 &   36.94 &   32.34 &   25.36 &   23.73 &   21.32 &   17.82 &   17.13 &    17.56 &    15.92 &    13.30 &    11.92 &    13.88 &    16.94 \\
\hline
\hline
\end{tabular}
\caption{The relative uncertainties (\%) of the measured differential branching fraction of $E_{\ell}^{B}$ are shown.}
    \label{tab:data-dbf-sys-pl}
\end{sidewaystable}

\begin{sidewaystable}[h]
\renewcommand\arraystretch{1.2}
\centering
\begin{tabular}{llllllllll}
\hline
\hline
$P_{+}$ [GeV]    & 0-0.2 & 0.2-0.4 & 0.4-0.6 & 0.6-0.8 & 0.8-1.0  & 1.0-1.2 & 1.2-1.4 & 1.4-1.6 & 1.6-4.0 \\
\hline
Tracking efficiency                   &    0.65 &    0.87 &    0.95 &    1.03 &    1.09 &    1.11 &    1.09 &    1.06 &    1.05 \\
Tagging calibration                     &    3.69 &    3.59 &    3.69 &    3.64 &    3.63 &    3.64 &    3.68 &    3.66 &    3.65 \\
Slow pion efficiency                     &    0.01 &    0.06 &    0.04 &    0.04 &    0.05 &    0.05 &    0.04 &    0.04 &    0.04 \\
$K_{S}^{0}$                      &    0.04 &    0.03 &    0.04 &    0.03 &    0.03 &    0.03 &    0.03 &    0.04 &    0.04 \\
$e$ID                    &    0.70 &    0.79 &    0.77 &    0.74 &    0.77 &    0.87 &    0.91 &    0.89 &    0.89 \\
$\mu$ID                   &    1.48 &    1.29 &    1.29 &    1.35 &    1.34 &    1.24 &    1.22 &    1.21 &    1.19 \\
$K\pi$ ID                  &    0.56 &    0.74 &    0.85 &    0.94 &    1.06 &    1.08 &    1.02 &    0.96 &    0.93 \\
$\mathcal{B}(B\to X_{u}\ell\nu)$                &    0.67 &    0.66 &    1.45 &    1.22 &    0.31 &    0.59 &    0.42 &    1.28 &    1.62 \\
$\mathcal{B}(B\to \pi  \ell\nu)$                    &    1.84 &    1.70 &    0.49 &    0.08 &    0.05 &    0.04 &    0.04 &    0.09 &    0.12 \\
$\mathcal{B}(B\to \rho  \ell\nu)$                   &    1.69 &    2.03 &    0.52 &    0.39 &    0.14 &    0.07 &    0.09 &    0.06 &    0.06 \\
$\mathcal{B}(B\to \omega  \ell\nu)$                 &    1.73 &    1.68 &    0.43 &    0.19 &    0.02 &    0.04 &    0.03 &    0.08 &    0.10 \\
$\mathcal{B}(B\to \eta  \ell\nu)$                   &    1.67 &    1.68 &    0.42 &    0.08 &    0.01 &    0.04 &    0.04 &    0.04 &    0.05 \\
$\mathcal{B}(B\to \eta^{\prime}  \ell\nu)$                 &    1.68 &    1.68 &    0.41 &    0.12 &    0.05 &    0.02 &    0.03 &    0.04 &    0.05 \\
$B\to \pi  \ell\nu$ FF                  &    0.73 &    0.32 &    1.54 &    0.55 &    1.23 &    0.74 &    0.03 &    0.36 &    0.54 \\
$B\to \rho  \ell\nu$ FF                 &    2.40 &    2.19 &    5.04 &    2.14 &    1.33 &    1.32 &    0.28 &    0.58 &    0.93 \\
$B\to \omega  \ell\nu$ FF               &    1.96 &    2.66 &    3.29 &    0.75 &    1.40 &    0.87 &    0.12 &    0.67 &    0.93 \\
$B\to \eta  \ell\nu$ FF                 &    0.69 &    0.19 &    0.62 &    0.31 &    0.59 &    0.33 &    0.01 &    0.18 &    0.26 \\
$B\to \eta^{\prime}  \ell\nu$ FF                &    0.59 &    0.03 &    0.61 &    0.44 &    0.68 &    0.35 &    0.01 &    0.18 &    0.26 \\
Hybrid model                   &    4.74 &    2.04 &    1.68 &    0.84 &    7.24 &    3.10 &   25.53 &   43.81 &   55.28 \\

DFN parameters                     &    1.92 &    1.30 &    1.82 &    1.10 &    1.74 &    2.15 &    2.87 &    8.00 &   19.00 \\
$\gamma_{s}$                      &    1.69 &    2.79 &    3.23 &    2.81 &    4.60 &    8.56 &    5.83 &    7.64 &    4.77 \\
$\pi^{+}$ multiplicity           &    0.24 &    2.94 &    5.47 &    4.58 &    2.04 &    0.23 &    0.57 &    1.40 &    1.74 \\
$N_{\mathit{B\bar{B}}}$                          &    1.25 &    1.25 &    1.25 &    1.25 &    1.25 &    1.25 &    1.25 &    1.25 &    1.25 \\
Background subtraction    &    4.68 &   10.27 &   14.48 &   21.22 &   27.32 &   29.13 &   30.15 &   31.37 &   32.57 \\
MC stat. (migration matrix) &    2.42 &    4.01 &    3.60 &    3.06 &    3.33 &    4.07 &    4.95 &    5.24 &    5.53 \\
\hline
Total syst. uncertainty         &   10.08 &   13.78 &   18.19 &   22.73 &   29.39 &   31.24 &   40.59 &   55.47 &   67.51 \\
Total stat. uncertainty                 &    6.86 &   10.89 &   13.64 &   16.39 &   16.51 &   16.77 &   22.05 &   25.67 &   27.62 \\
Total uncertainty          &   12.20 &   17.56 &   22.74 &   28.03 &   33.71 &   35.46 &   46.19 &   61.12 &   72.94 \\
\hline
\hline
\end{tabular}
\caption{The relative uncertainties (\%) of the measured differential branching fraction of $P_{+}$ are shown.}
    \label{tab:data-dbf-sys-pplus}
\end{sidewaystable}

\begin{sidewaystable}[h]
\renewcommand\arraystretch{1.2}
\centering
\begin{tabular}{lllllllllll}
\hline
\hline
$P_{-}$ [GeV]    & 0-1.0 & 1.0-1.5 & 1.5-2.0 & 2.0-2.5 & 2.5-3.0  & 3.0-3.5 & 3.5-4.0 & 4.0-4.5 & 4.5-5.0 & 5.0-5.5\\
\hline
Tracking efficiency                   &    0.61 &    0.62 &    0.73 &    0.80 &    0.84 &    0.88 &    0.89 &    0.92 &    0.92 &    0.91 \\
Tagging calibration                     &    3.75 &    3.77 &    3.67 &    3.64 &    3.59 &    3.63 &    3.72 &    3.72 &    3.65 &    3.59 \\
Slow pion efficiency                    &    0.06 &    0.06 &    0.04 &    0.04 &    0.04 &    0.04 &    0.03 &    0.03 &    0.03 &    0.02 \\
$K_{S}^{0}$                      &    0.03 &    0.03 &    0.03 &    0.04 &    0.03 &    0.03 &    0.03 &    0.04 &    0.03 &    0.03 \\
$e$ID                    &    0.87 &    0.84 &    0.78 &    0.76 &    0.72 &    0.74 &    0.74 &    0.73 &    0.76 &    0.77 \\
$\mu$ID                   &    1.38 &    1.50 &    1.47 &    1.39 &    1.37 &    1.42 &    1.31 &    1.25 &    1.32 &    1.42 \\
$K\pi$ ID                  &    0.23 &    0.25 &    0.47 &    0.69 &    0.78 &    0.77 &    0.83 &    0.90 &    0.88 &    0.86 \\
$\mathcal{B}(B\to X_{u}\ell\nu)$                &    0.40 &    1.49 &    2.97 &    0.22 &    0.33 &    0.02 &    0.47 &    0.44 &    0.93 &    1.03 \\
$\mathcal{B}(B\to \pi  \ell\nu)$                    &    5.22 &    6.42 &    4.64 &    2.35 &    0.93 &    0.21 &    0.50 &    0.84 &    1.16 &    1.41 \\
$\mathcal{B}(B\to \rho  \ell\nu)$                   &    5.21 &    6.43 &    4.71 &    2.39 &    1.03 &    0.59 &    0.57 &    1.02 &    1.13 &    1.32 \\
$\mathcal{B}(B\to \omega  \ell\nu)$                 &    5.21 &    6.40 &    4.62 &    2.33 &    0.91 &    0.10 &    0.42 &    0.78 &    1.10 &    1.32 \\
$\mathcal{B}(B\to \eta  \ell\nu)$                   &    5.21 &    6.40 &    4.62 &    2.33 &    0.91 &    0.17 &    0.44 &    0.79 &    1.11 &    1.33 \\
$\mathcal{B}(B\to \eta^{\prime}  \ell\nu)$                 &    5.21 &    6.39 &    4.62 &    2.33 &    0.91 &    0.10 &    0.42 &    0.79 &    1.11 &    1.32 \\
$B\to \pi  \ell\nu$ FF                  &    3.65 &    4.49 &    5.03 &    2.13 &    1.59 &    0.58 &    0.68 &    2.55 &    2.88 &    2.75 \\
$B\to \rho  \ell\nu$ FF                 &   12.44 &   13.45 &   11.09 &    4.09 &    1.85 &    1.10 &    1.64 &    4.65 &    5.04 &    5.86 \\
$B\to \omega  \ell\nu$ FF               &   16.21 &   23.12 &   14.35 &    7.37 &    3.15 &    0.62 &    1.49 &    6.77 &    7.79 &    7.59 \\
$B\to \eta  \ell\nu$ FF                 &    3.44 &    4.69 &    3.99 &    1.87 &    1.02 &    0.23 &    0.39 &    1.94 &    2.35 &    2.65 \\
$B\to \eta^{\prime}  \ell\nu$ FF                &    4.25 &    5.23 &    4.28 &    1.79 &    0.95 &    0.24 &    0.39 &    1.99 &    2.45 &    2.74 \\
Hybrid model                   &    3.03 &   21.43 &   23.71 &    9.77 &    3.11 &    1.27 &    3.35 &    9.53 &   14.19 &   13.85 \\

DFN parameters                     &    2.17 &    1.59 &    1.12 &    3.11 &    2.90 &    3.27 &    4.57 &    4.77 &    5.76 &    5.68 \\
$\gamma_{s}$                     &    2.60 &    4.18 &    3.11 &    0.56 &    0.90 &    0.31 &    1.03 &    1.01 &    6.43 &    4.42 \\
$\pi^{+}$ multiplicity        &    0.27 &    0.51 &    0.03 &    1.76 &    3.66 &    3.16 &    2.67 &    3.22 &    3.10 &    2.52 \\
$N_{\mathit{B\bar{B}}}$                          &    1.25 &    1.25 &    1.25 &    1.25 &    1.25 &    1.25 &    1.25 &    1.25 &    1.25 &    1.25 \\
Background subtraction    &   13.11 &   10.18 &    8.43 &    9.12 &   11.20 &   13.27 &   14.29 &   15.73 &   14.29 &   17.77 \\
MC stat. (migration matrix) &    5.75 &    6.25 &    2.94 &    2.37 &    2.54 &    2.21 &    1.92 &    2.47 &    2.50 &    2.90 \\
\hline
Total syst. uncertainty        &   29.00 &   40.42 &   34.30 &   18.05 &   14.30 &   14.98 &   16.56 &   21.98 &   25.06 &   26.78 \\
Total stat. uncertainty                &   18.64 &   13.81 &   10.43 &   10.61 &   12.17 &   13.20 &   12.34 &   12.69 &   12.47 &   19.72 \\
Total uncertainty           &   34.48 &   42.71 &   35.84 &   20.94 &   18.78 &   19.97 &   20.65 &   25.39 &   27.99 &   33.26 \\
\hline
\hline
\end{tabular}
\caption{The relative uncertainties (\%) of the measured differential branching fraction of $P_{-}$ are shown.}
    \label{tab:data-dbf-sys-pminus}
\end{sidewaystable}

\clearpage
\subsection{Migration Matrices}

\begin{figure*}[th!]
  \includegraphics[width=0.45\textwidth]{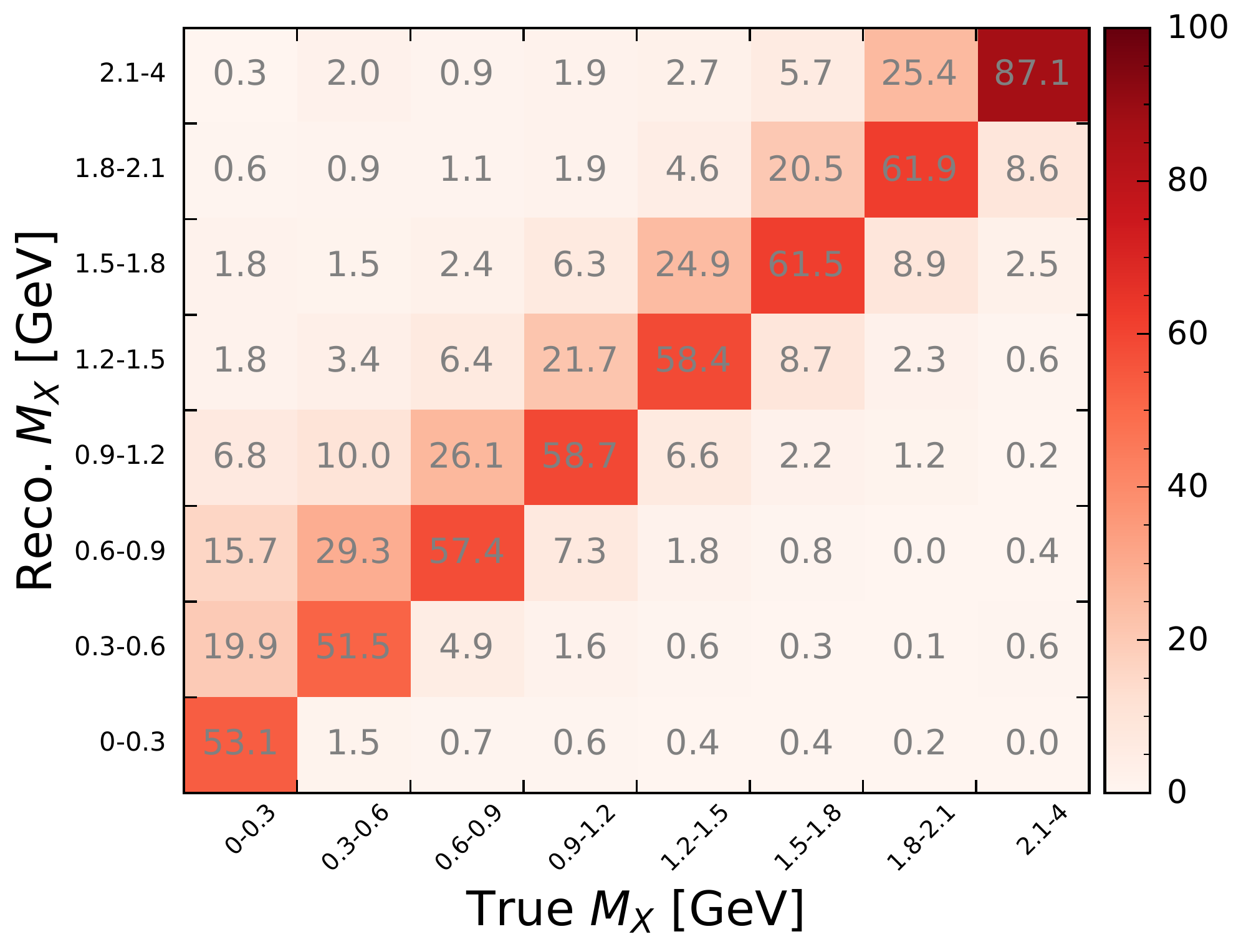} 
  \includegraphics[width=0.45\textwidth]{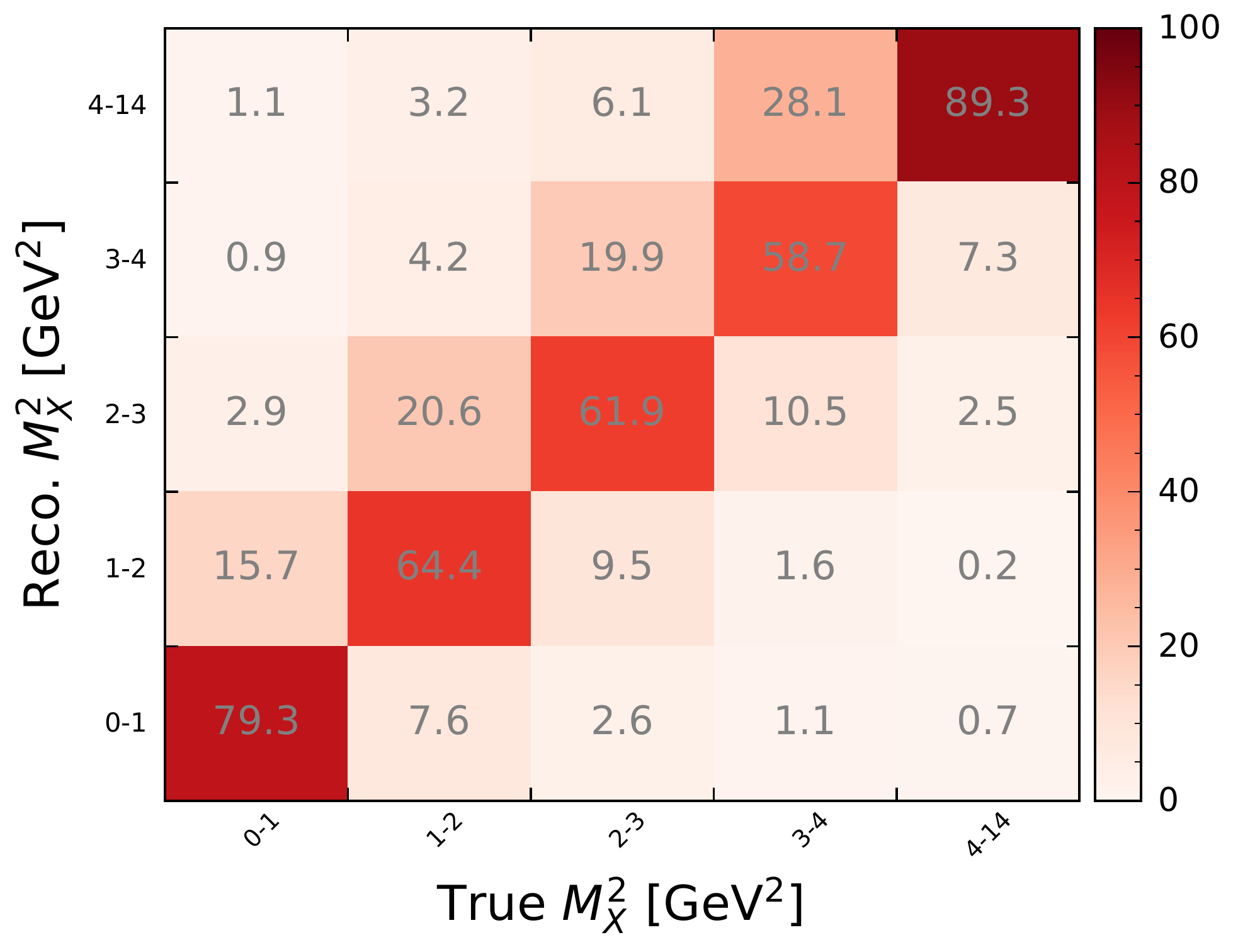} 
  \includegraphics[width=0.45\textwidth]{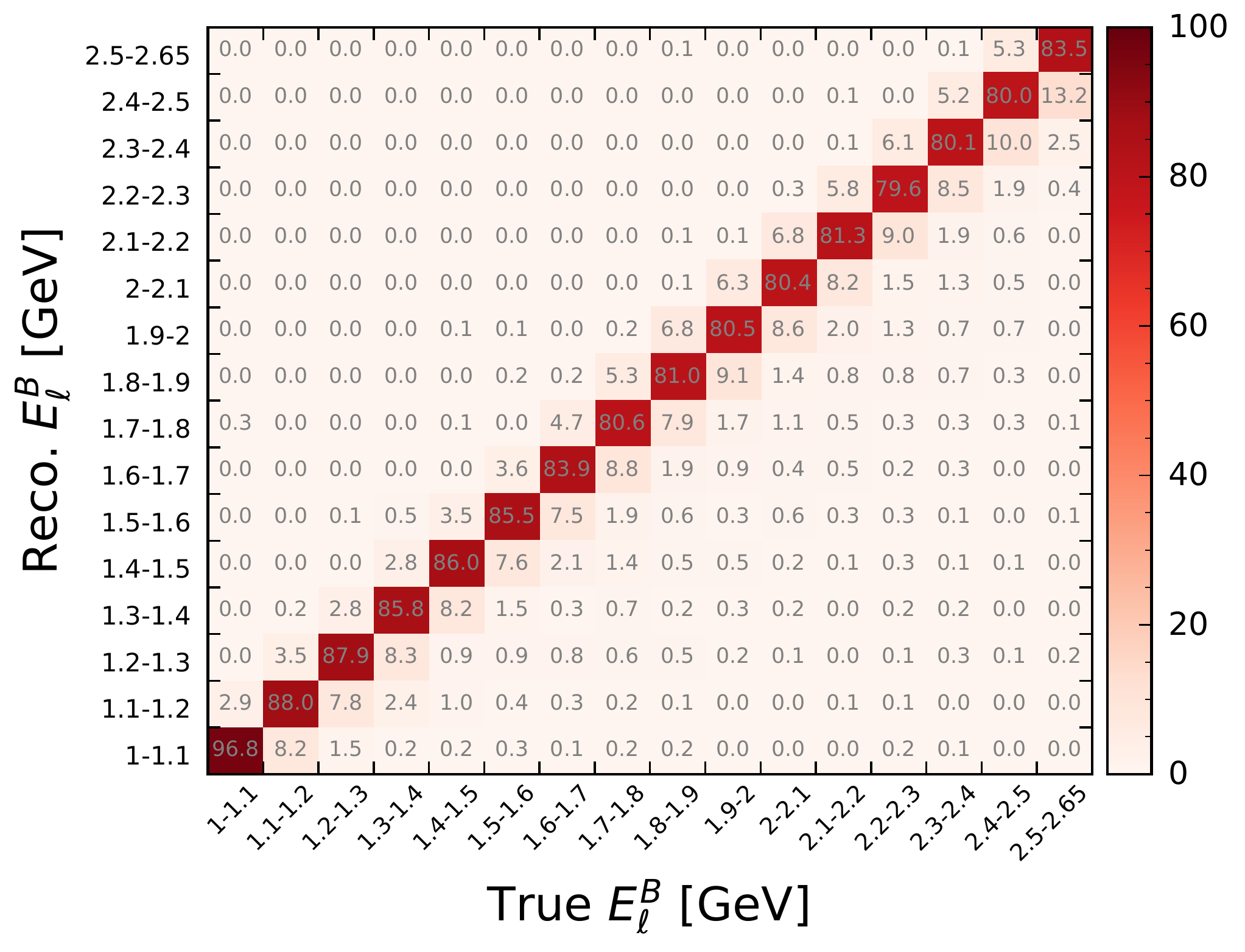} 
  \includegraphics[width=0.45\textwidth]{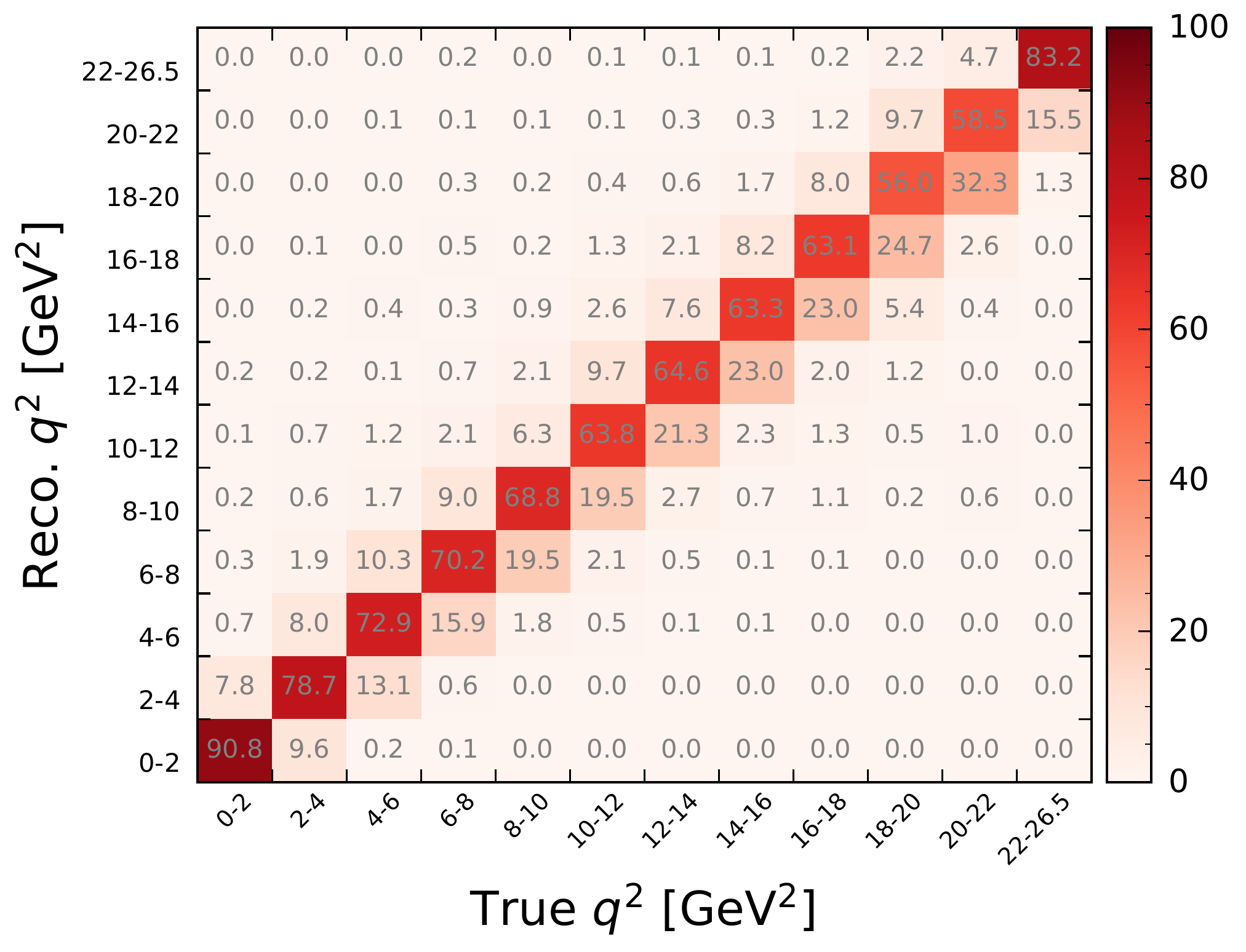} 
  \includegraphics[width=0.45\textwidth]{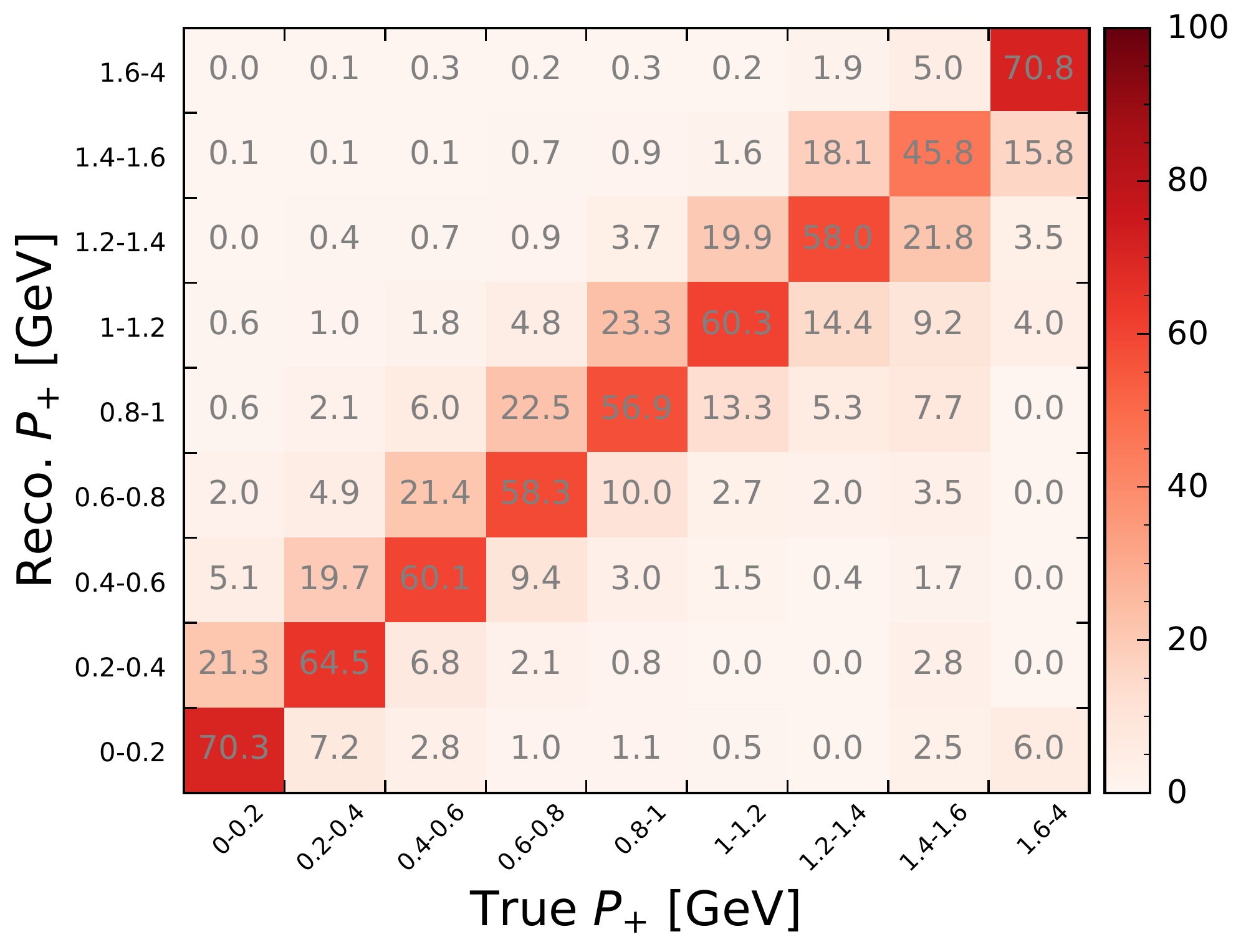} 
  \includegraphics[width=0.45\textwidth]{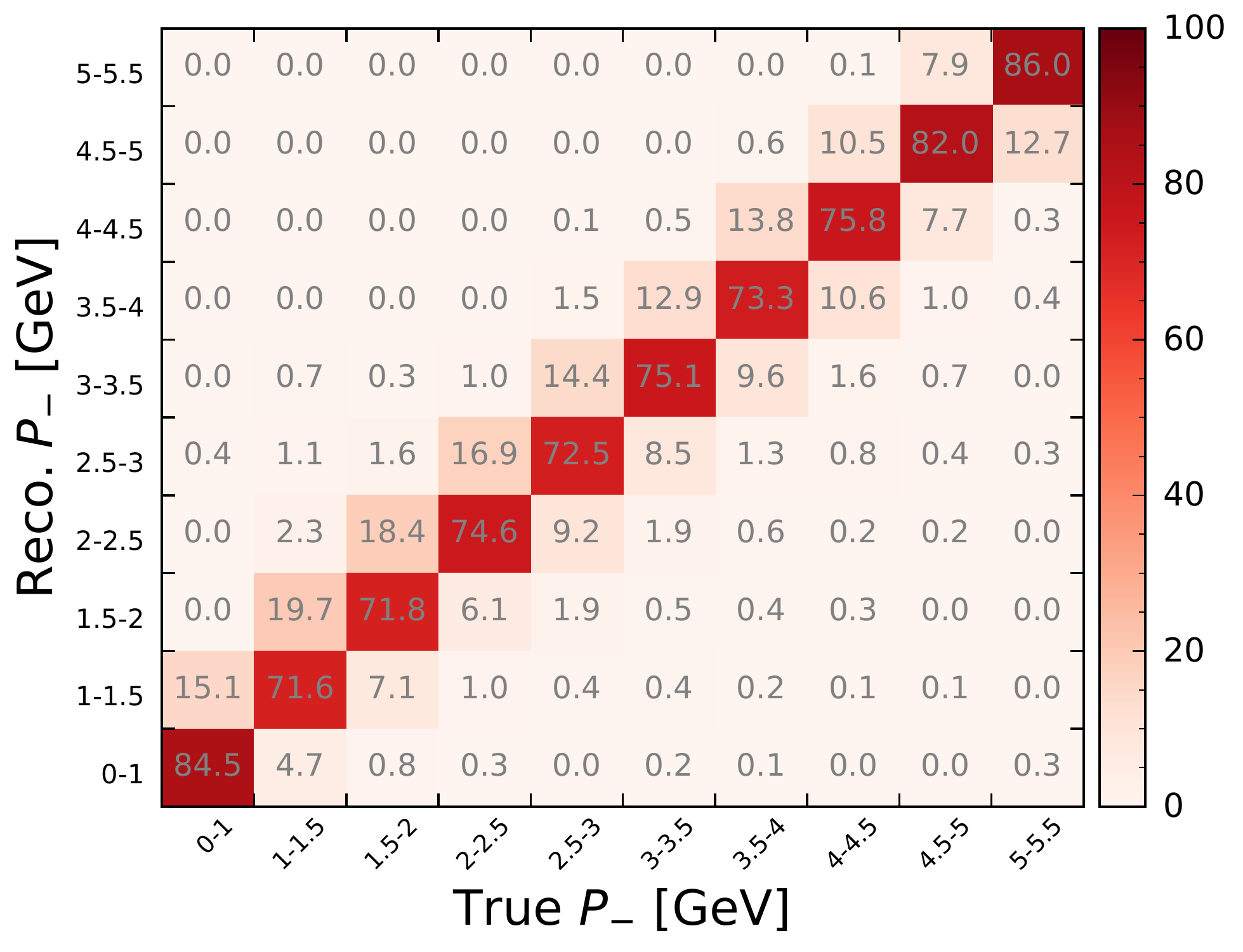} 
\caption{
The migration matrices for all studied variables are shown.
 }
\label{fig:mig}
\end{figure*}

The migration matrices of all studied distributions, defined as the conditional probability 
\begin{align}
 M_{ij} & = P( \text{event reconstructed in bin $i$} \, | \, \text{event generated in bin $j$}) \, ,
\end{align}
are shown in Fig.~\ref{fig:mig}. The binning of the measured distributions was chosen, such that the purity $M_{ii}$ is at least 50\%. The statistical uncertainty of the migration matrices due to the limited MC size is considered as a systematic uncertainty, cf. Tables~\ref{tab:data-dbf-sys-mx}-\ref{tab:data-dbf-sys-pminus}.

\subsection{Efficiency Correction Factors}

\begin{figure*}[th!]
  \includegraphics[width=0.45\textwidth]{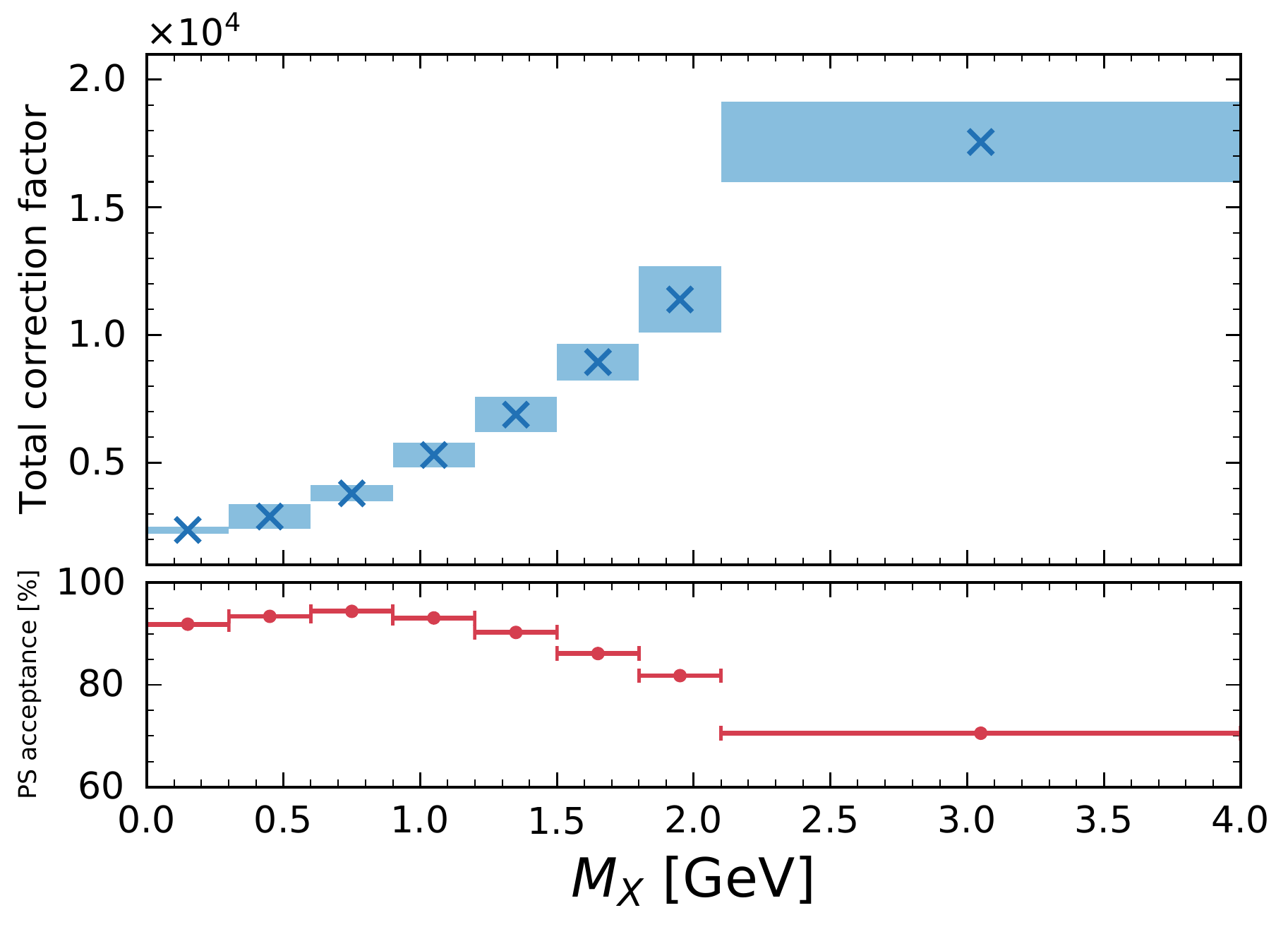} 
  \includegraphics[width=0.45\textwidth]{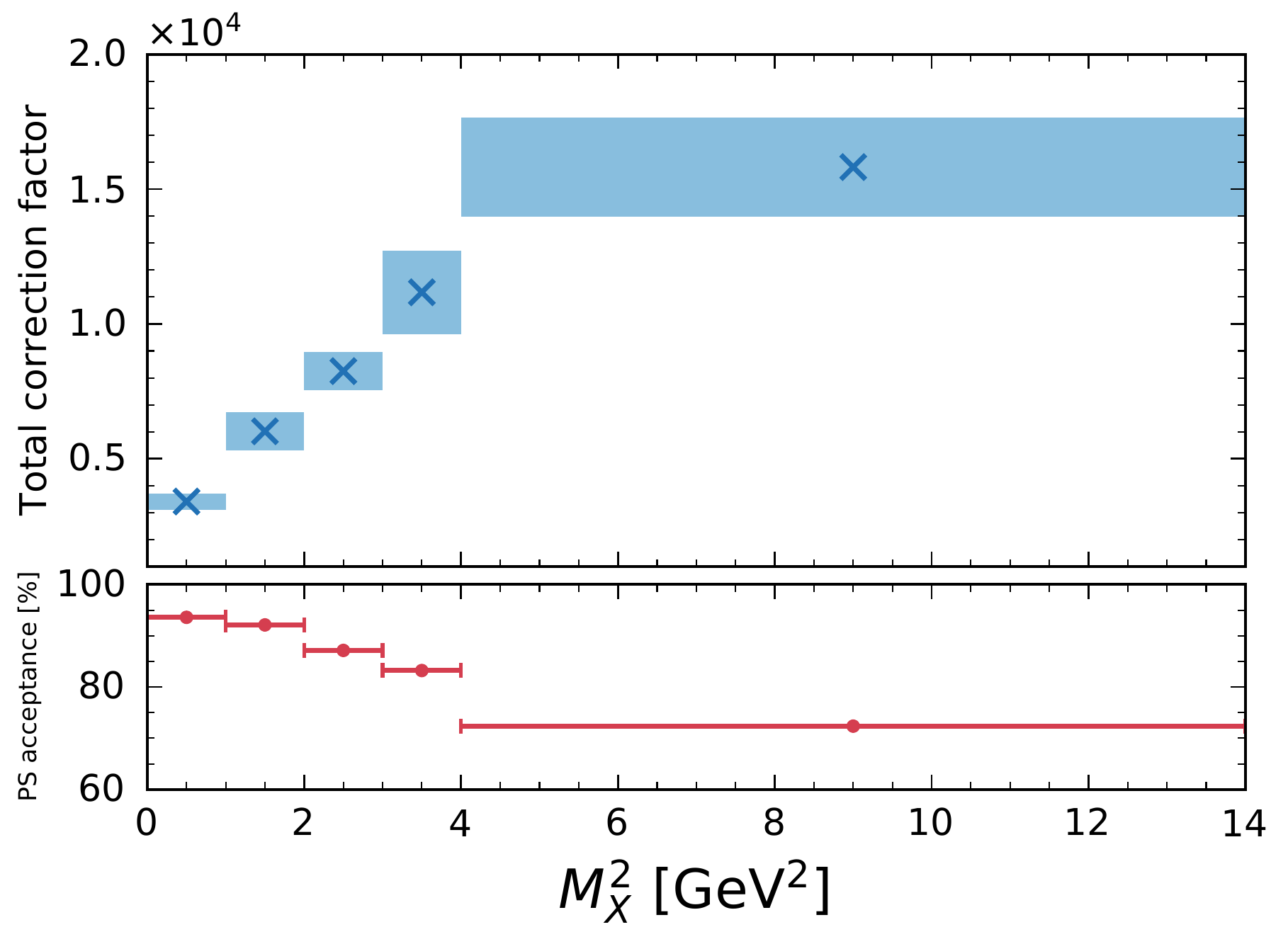} 
  \includegraphics[width=0.45\textwidth]{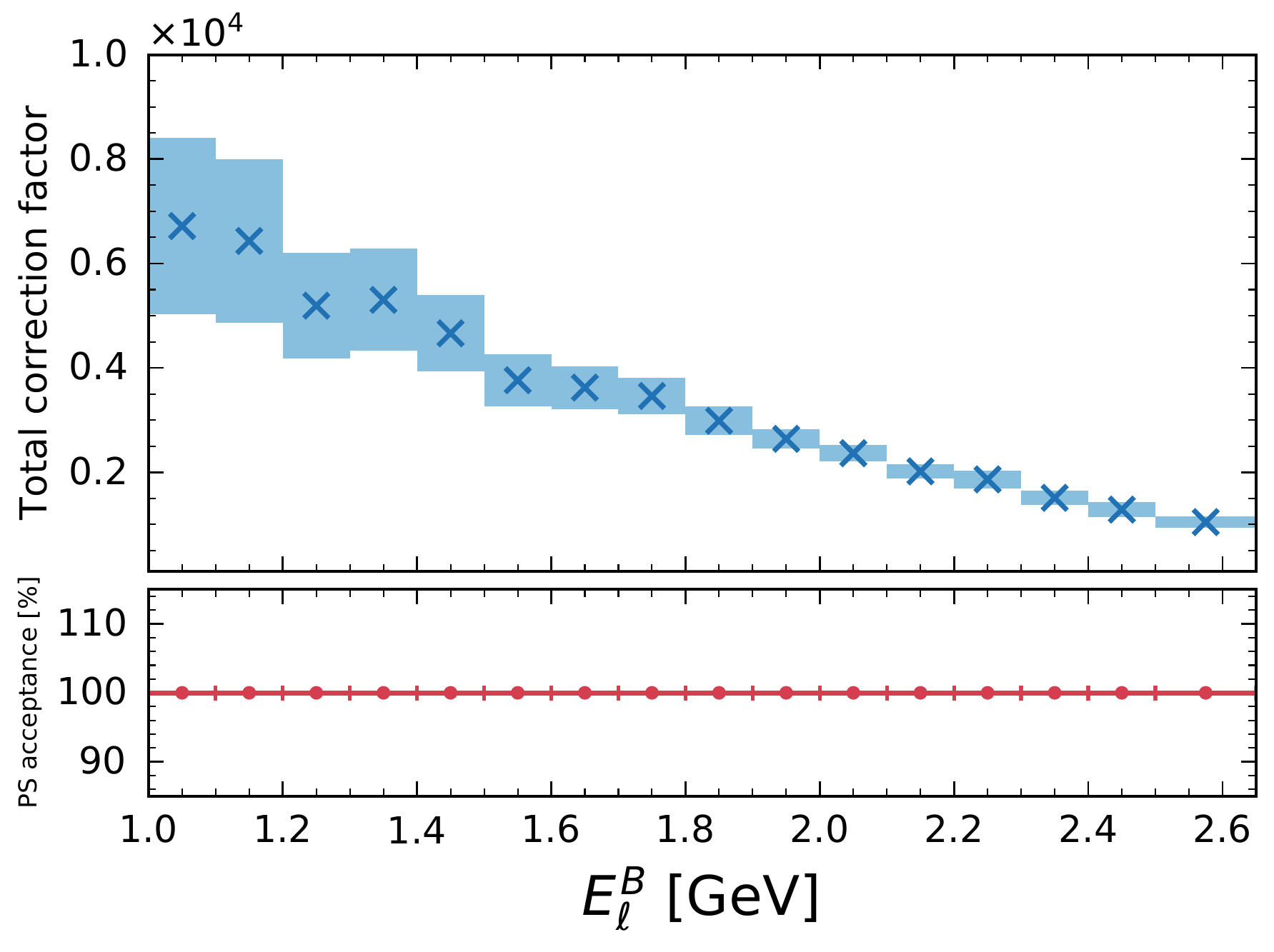} 
  \includegraphics[width=0.45\textwidth]{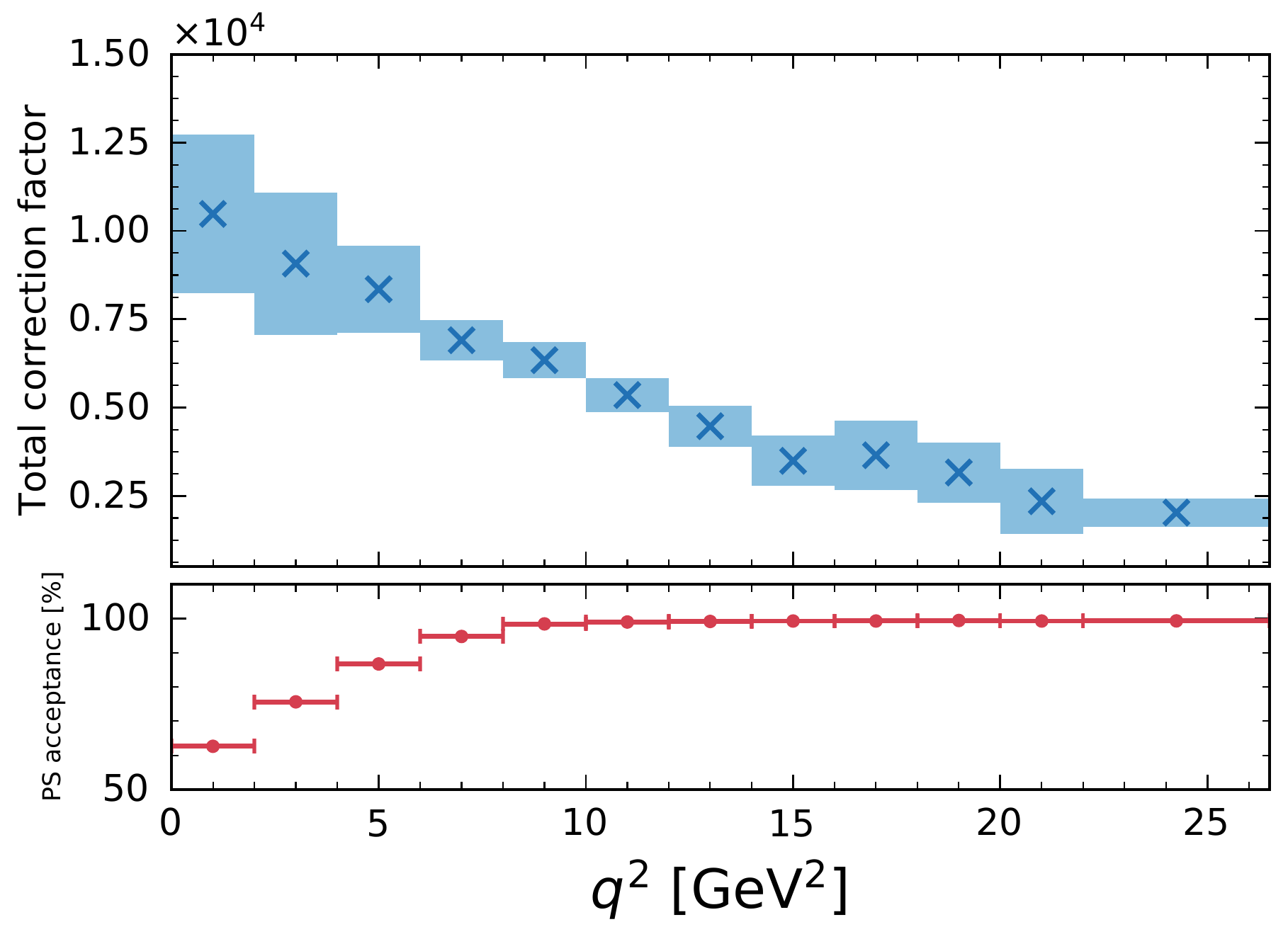} 
  \includegraphics[width=0.45\textwidth]{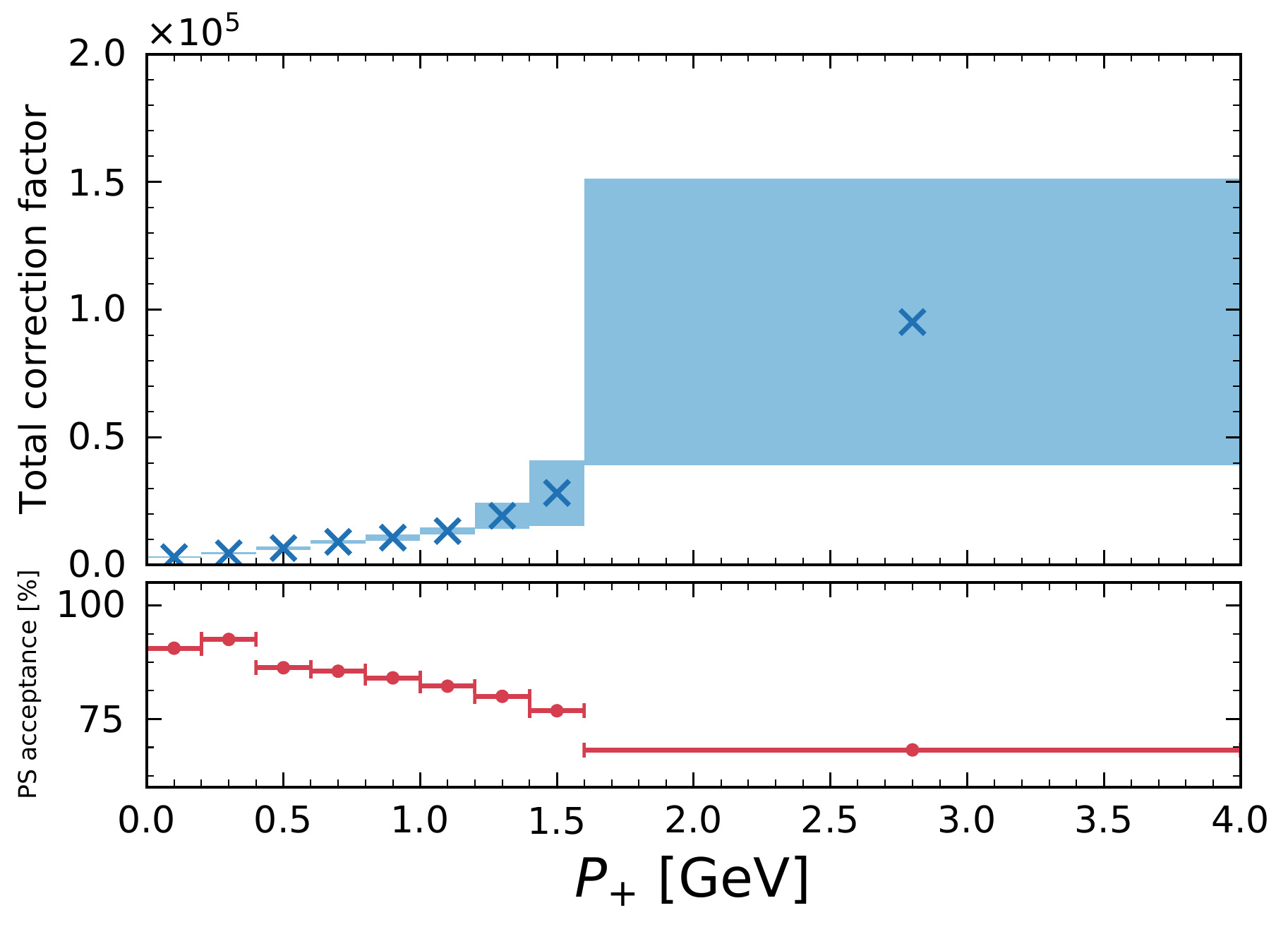} 
  \includegraphics[width=0.45\textwidth]{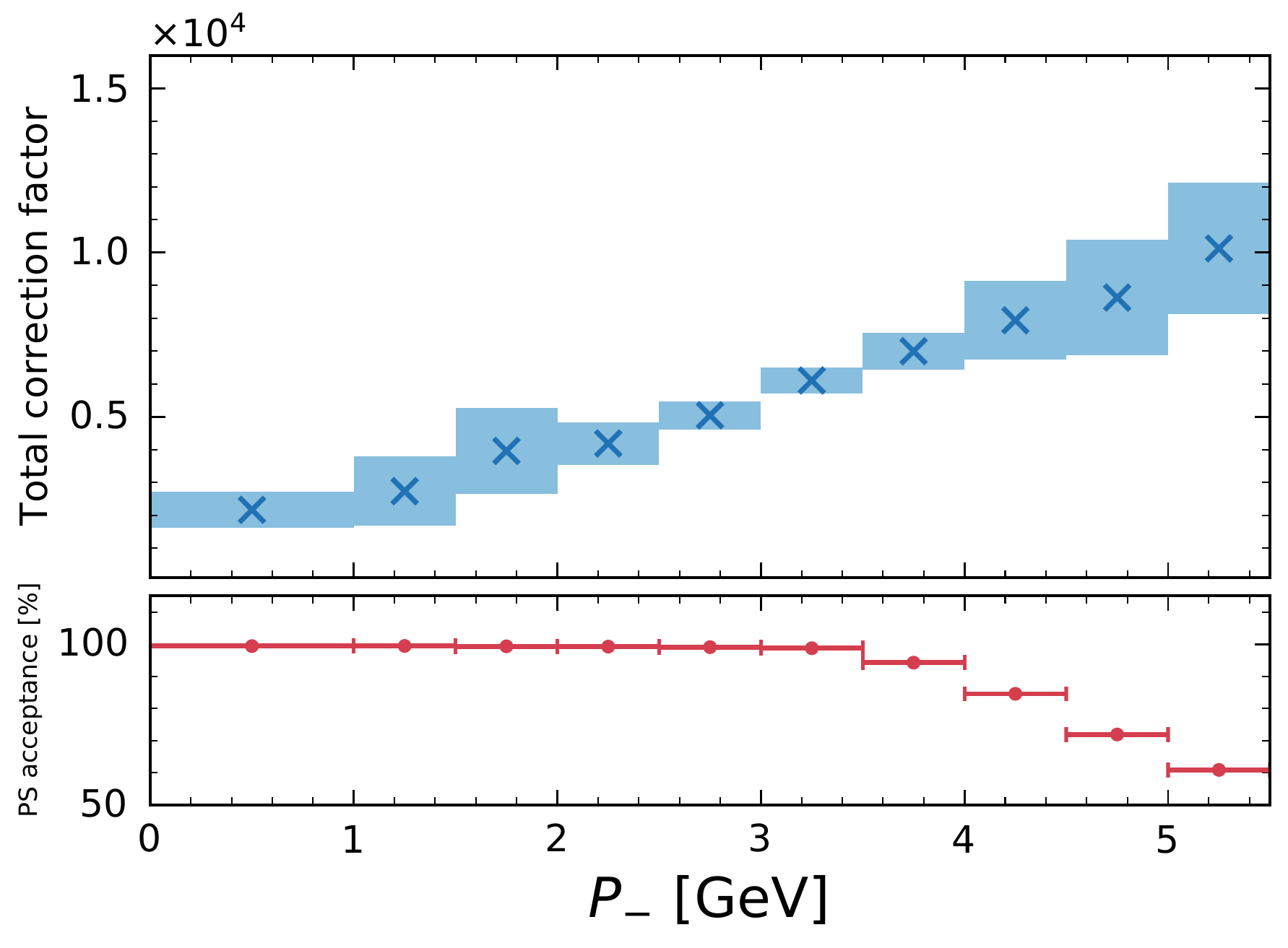} 
\caption{
The correction factors $\left( \epsilon_{\mathrm{tag}} \times \epsilon_{\mathrm{sel}} \right)^{-1} \times \epsilon_{ \Delta \mathcal{B}(E_\ell^B > 1 \, \mathrm{GeV})} $ (blue) and phase space acceptance factor $ \epsilon_{ \Delta \mathcal{B}(E_\ell^B > 1 \, \mathrm{GeV})}$ (red) are shown. The colored band of the total correction factor shows the full systematic uncertainty. 
 }
\label{fig:eps}
\end{figure*}

The unfolded signal yields $\nu_i$ of a given bin $i$ are corrected for selection efficiency, acceptance and phase-space effects, and normalized to the total number of recored $B$-meson pairs, $N_{\mathit{B\bar{B}}} = \left(771.58 \pm 9.78 \right) \times 10^6$, to obtain differential branching fractions: 
\begin{align}
 \Delta \mathcal{B}_i = \frac{1}{4 N_{\mathit{B\bar{B}}}} \times \nu_i \times \left( \epsilon_{\mathrm{tag}} \times \epsilon_{\mathrm{sel}} \right)^{-1} \times \epsilon_{ \Delta \mathcal{B}(E_\ell^B > 1 \, \mathrm{GeV})} \, .
\end{align}
The factor of 4 is due to $N_{\mathit{B\bar{B}}}$ and that we average over electron and muon final states. Further, $\epsilon_{\mathrm{tag}}$ and $  \epsilon_{\mathrm{sel}} $ denote the tagging and selection efficiencies, respectively, and $ \epsilon_{ \Delta \mathcal{B}(E_\ell^B > 1 \, \mathrm{GeV})}$ maps the branching fraction to the partial phase space with $E_\ell^B > 1 \, \mathrm{GeV}$ in the $B$ rest frame.  Figure~\ref{fig:eps} shows the product of $\left( \epsilon_{\mathrm{tag}} \times \epsilon_{\mathrm{sel}} \right)^{-1} \times \epsilon_{ \Delta \mathcal{B}(E_\ell^B > 1 \, \mathrm{GeV})} $ for all studied differential variables, including the full systematic uncertainties. The bottom panel shows the phase space acceptance $ \epsilon_{ \Delta \mathcal{B}(E_\ell^B > 1 \, \mathrm{GeV})}$. 

\clearpage
\subsection{Experimental Correlations of the Differential Branching Fractions}

\begin{figure}[h!]
  \includegraphics[width=0.95\textwidth]{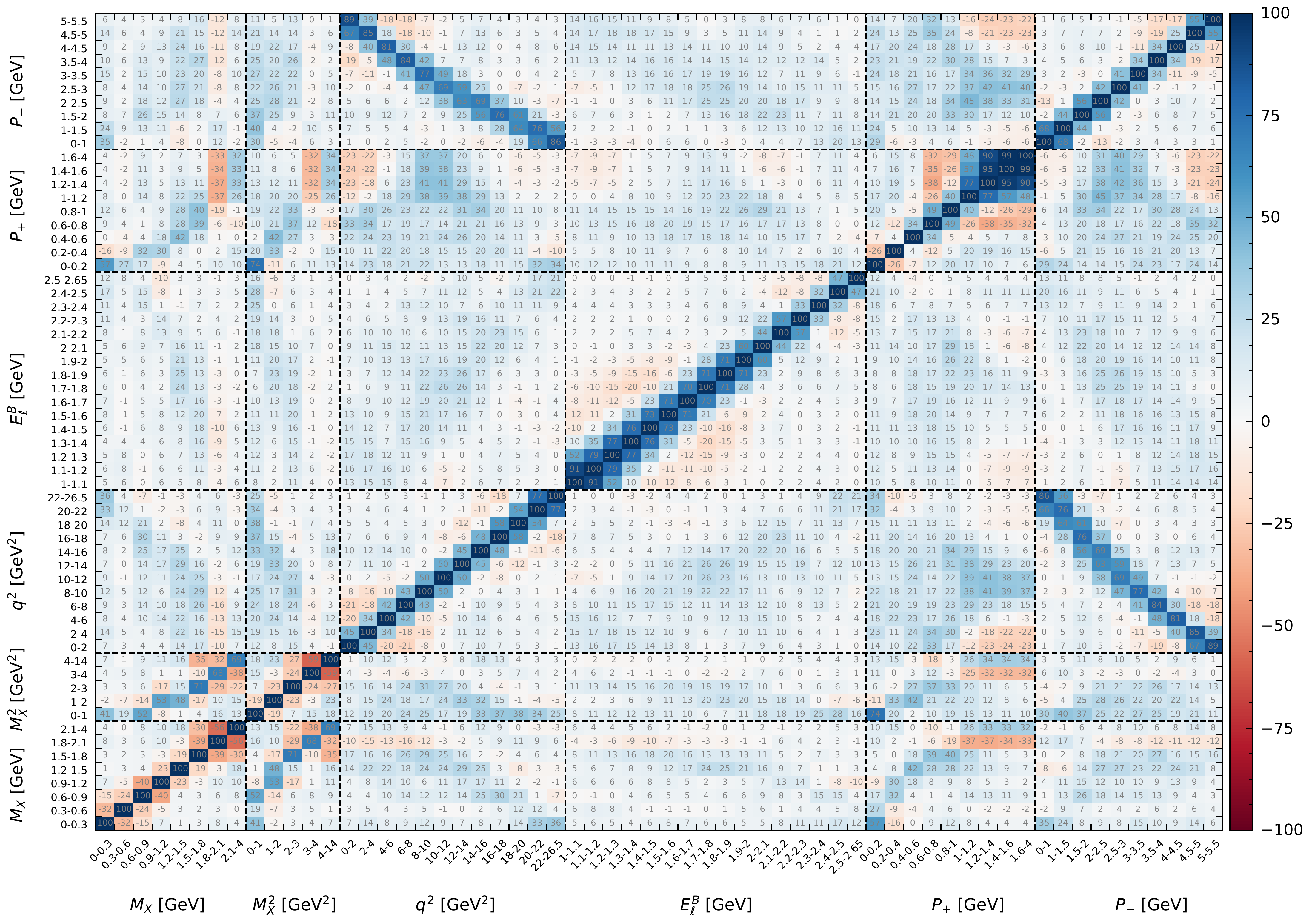} 
\caption{
The statistical correlations of the differential branching fractions are shown.
 }
\label{fig:cov_stat}
\end{figure}

\begin{figure}[t!]
  \includegraphics[width=0.95\textwidth]{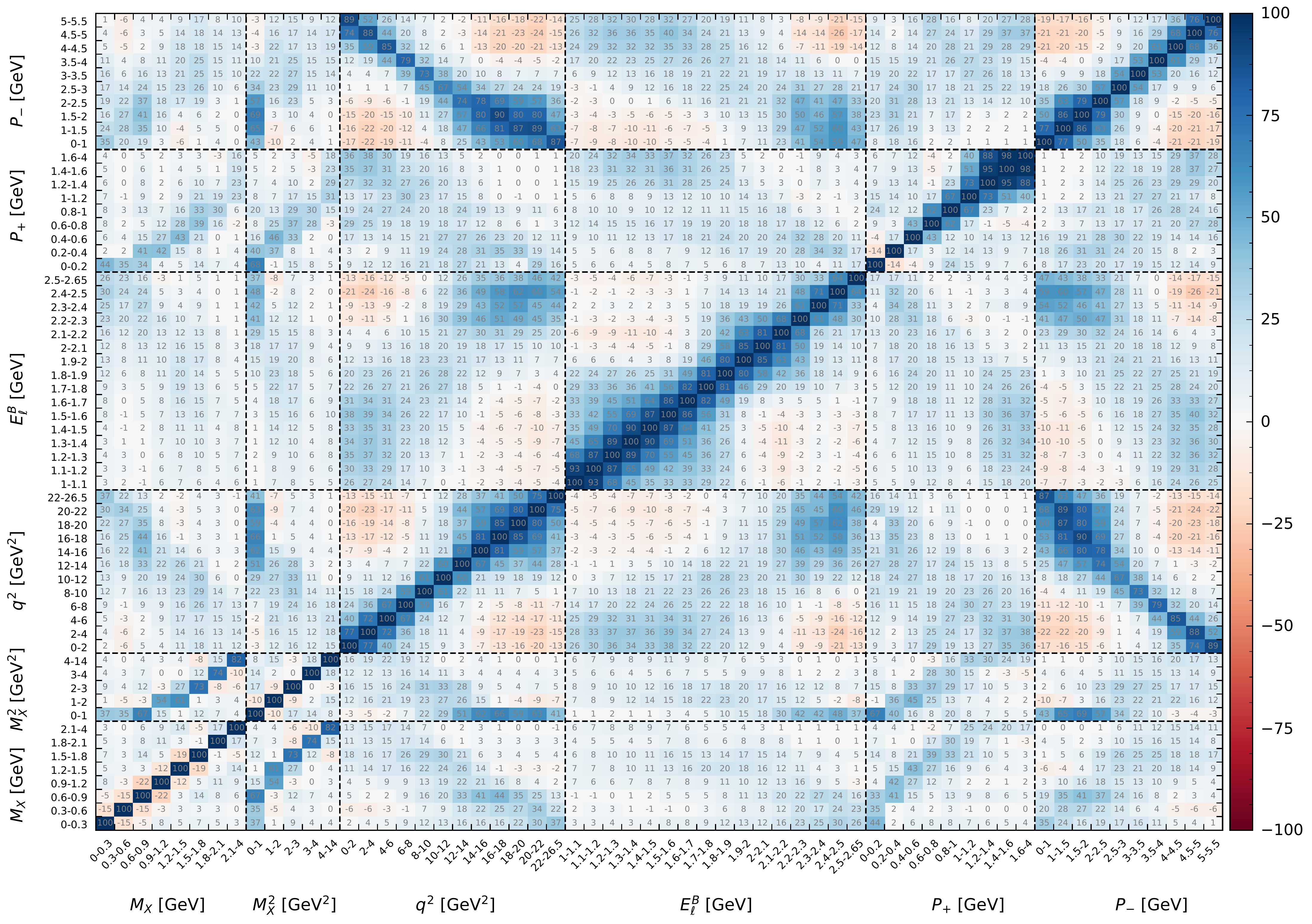} 
\caption{
The full experimental (statistical and systematical) correlations of the differential branching fractions are shown.
 }
\label{fig:cov_tot}
\end{figure}

The statistical and full experimental correlations of the measured distributions are shown in Fig.~\ref{fig:cov_stat} and \ref{fig:cov_tot}. The statistical correlations were obtained using a bootstrapping procedure: ensembles of the selected data events after the initial selection were created and the full analysis was repeated (binned likelihood fit in $M_X$, background subtraction in the variable of interest, unfolding, efficiency correction). From the obtained central values of these ensembles, the statistical correlations between observables were estimated using the Pearson correlation coefficient. 

\subsection{Total Partial Branching Fractions and Comparison to Ref.~\cite{Cao:2021xqf}}

\begin{figure}[h!]
  \includegraphics[width=0.45\textwidth]{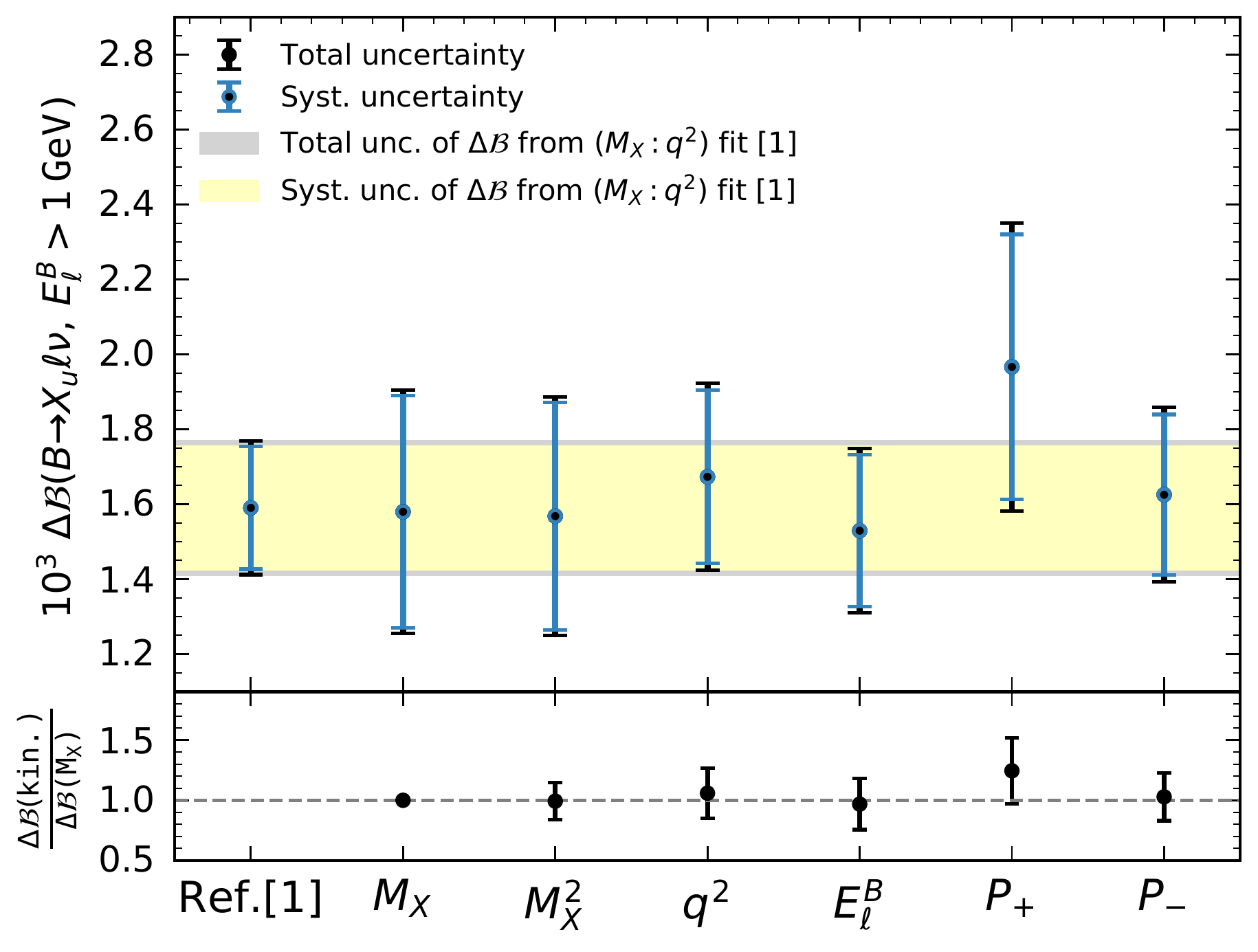} 
  \includegraphics[width=0.45\textwidth]{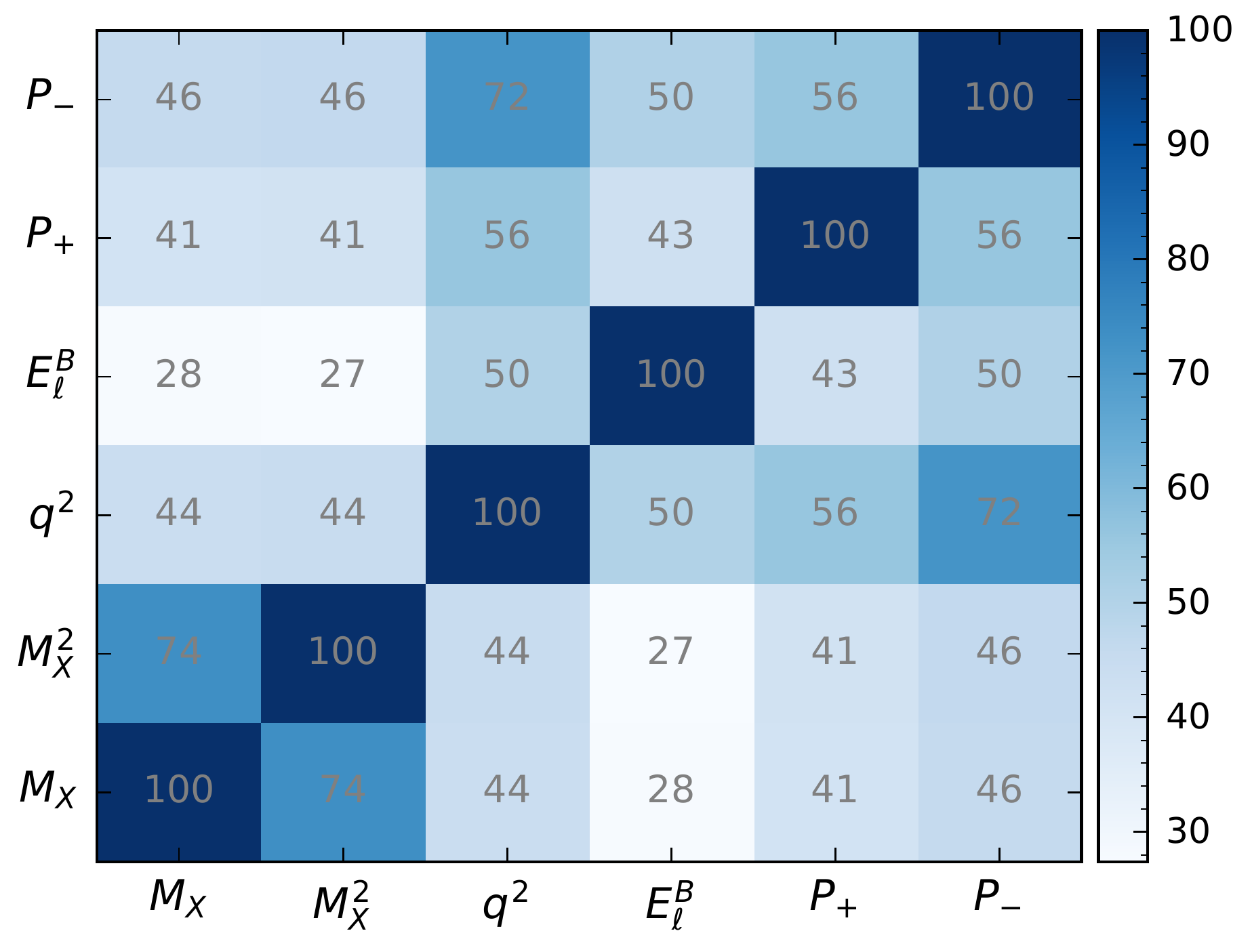} 
\caption{
Left: the total partial branching fraction with $E_\ell^B > 1 \, \mathrm{GeV}$ as calculated by each differential measurement is compared to the result of Ref.~\cite{Cao:2021xqf}, which is based on the 2D fit of $M_X:q^2$ and obtained with a looser selection. The ratio compares the total partial branching fractions to the result obtained by summing the measured $M_X$ distribution and the uncertainty takes into account the full statistical and systematic correlations between the different results. Right: the full experimental correlations between the total partial branching fractions from summing the individual bins are shown.
 }
\label{fig:tot_BF}
\end{figure}

The left panel of Fig.~\ref{fig:tot_BF} shows the partial branching fractions for $E_\ell^B > 1 \, \mathrm{GeV}$ as calculated when summing the individual bins of the differential measurements of $M_X$, $M_X^2$, $E_\ell^B$, $q^2$, and $P_\pm$. The values agree with the value reported in Ref.~\cite{Cao:2021xqf}, which uses the same analysis strategy but less strict selection criteria. To further study the compatibility, we evaluate the ratios of the partial branching fractions with respect to the partial branching fraction obtained from the $M_X$ distribution, taking into account the full statistical and systematic correlations. We find good overall agreement and the largest discrepancy is from the ratio of $P_+$, which agrees  with unity to within 0.9 standard deviation. The right panel of Fig.~\ref{fig:tot_BF} shows the experimental correlation of the partial branching fractions obtained by summing the individually measured bins. 

\clearpage

\subsection{Quantitative comparison between measured spectra and various modelings}

To quantify the agreement between the measured distributions and the three MC predictions (Hybrid, DFN~\cite{DeFazio:1999ptt}, BLNP~\cite{Lange:2005yw}), we carry out a $\chi^2$ test. For this test the full experimental correlations are taken into account and the obtained $\chi^2$ values are given in Table~\ref{tab:chi2}. Note that no theory uncertainties were included. Overall the agreement with the hybrid MC is fair for all measured distributions, but the comparisons in $M_{X}$, $M_{X}^{2}$ and $P_+$ show poor agreement for DFN and BLNP. This is due to that in these measurements the \bulnu resonance region is resolved, which is not adequately modelled by fully inclusive predictions. 

\begin{table}[h!]
\renewcommand\arraystretch{1.2}
\centering
\begin{tabular}{lllllll}
\hline
\hline
  $\chi^2$     & $E_{\ell}^{B}$ & $M_{X}$ & $M_{X}^{2}$ & $q^{2}$ & $P_{+}$ & $P_{-}$ \\
 \hline
n.d.f.    & 16             & 8       & 5           & 12      & 9       & 10      \\
\hline
Hybrid & 13.5          & 2.5    & 2.6        & 4.5    & 1.7    & 5.2    \\
DFN    & 16.2          & 63.2   & 13.1       & 18.5   & 29.3   & 6.1    \\
BLNP   & 16.5          & 61.0   & 6.3        & 20.6   & 23.6   & 13.7  \\
\hline
\hline
\end{tabular}
\caption{The $\chi^{2}$ of the measured differential branching fractions respect to various modelings. The number of degree of freedom (n.d.f.) is equal to the number of bins, which is also listed. }
\label{tab:chi2}
\end{table}

\subsection{The first three moments in the phase space region of $E_\ell^B > 1 \, \mathrm{GeV}$}

\begin{figure*}[b!]
  \includegraphics[width=0.32\textwidth]{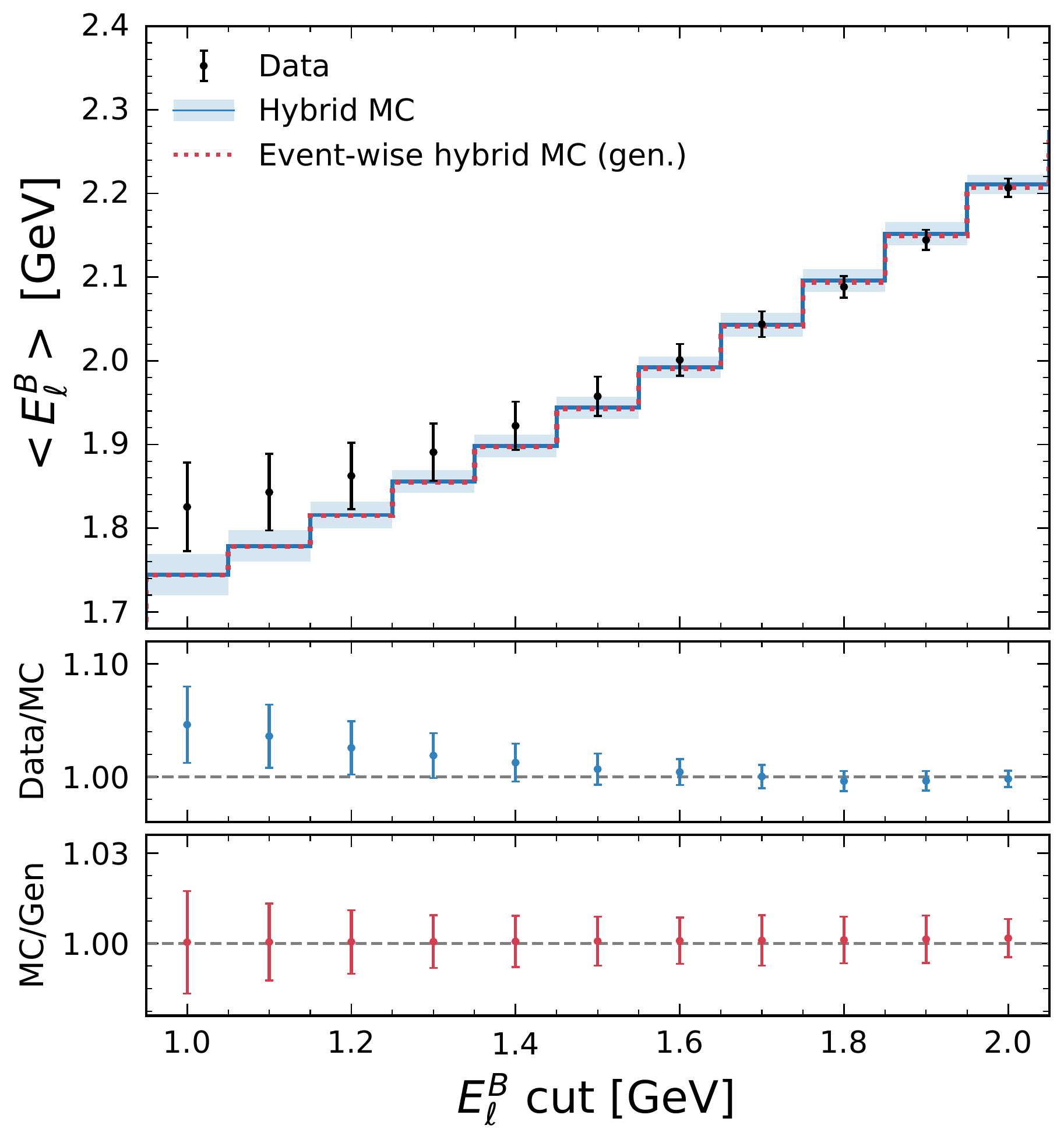} 
  \includegraphics[width=0.32\textwidth]{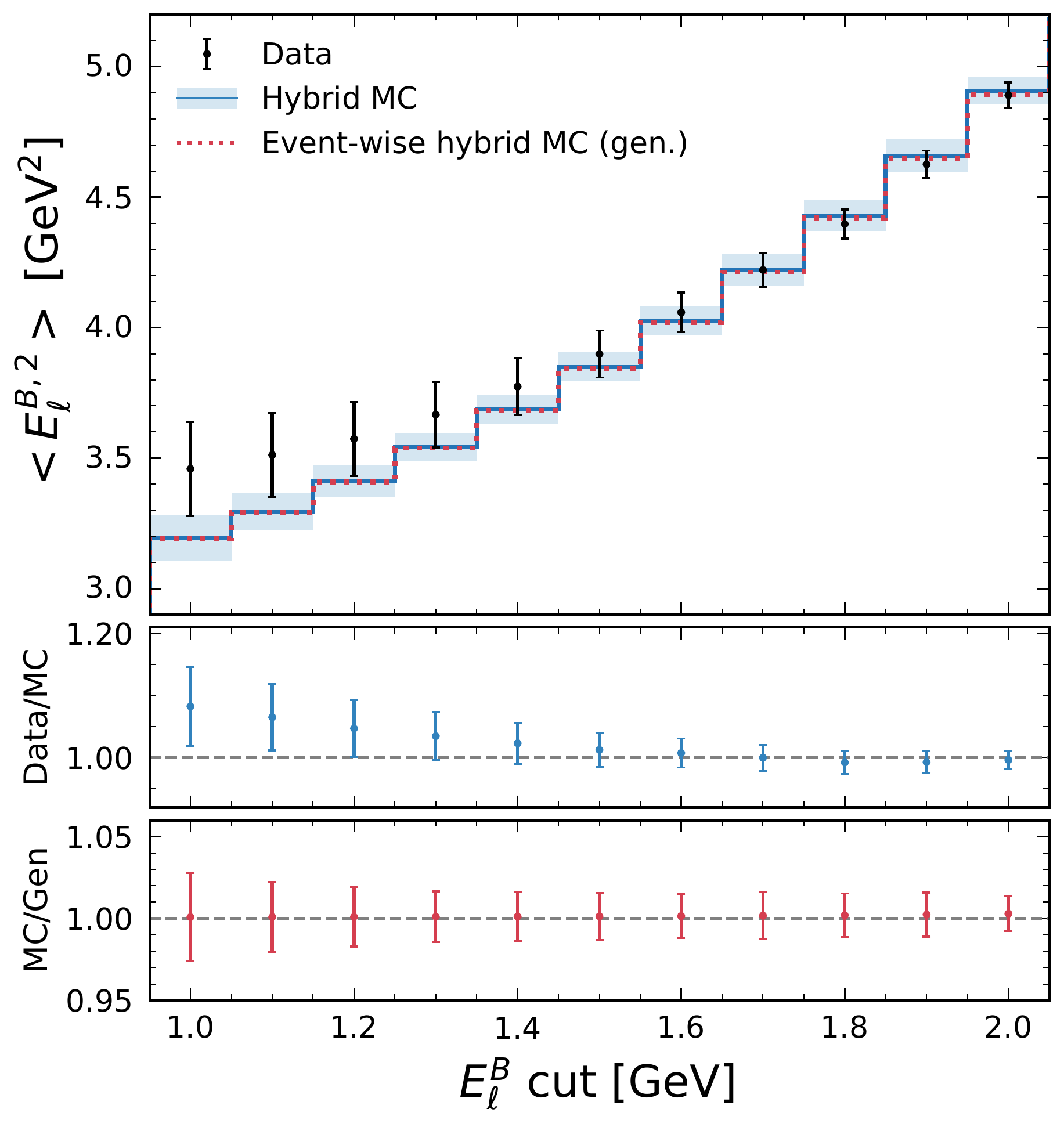} 
  \includegraphics[width=0.32\textwidth]{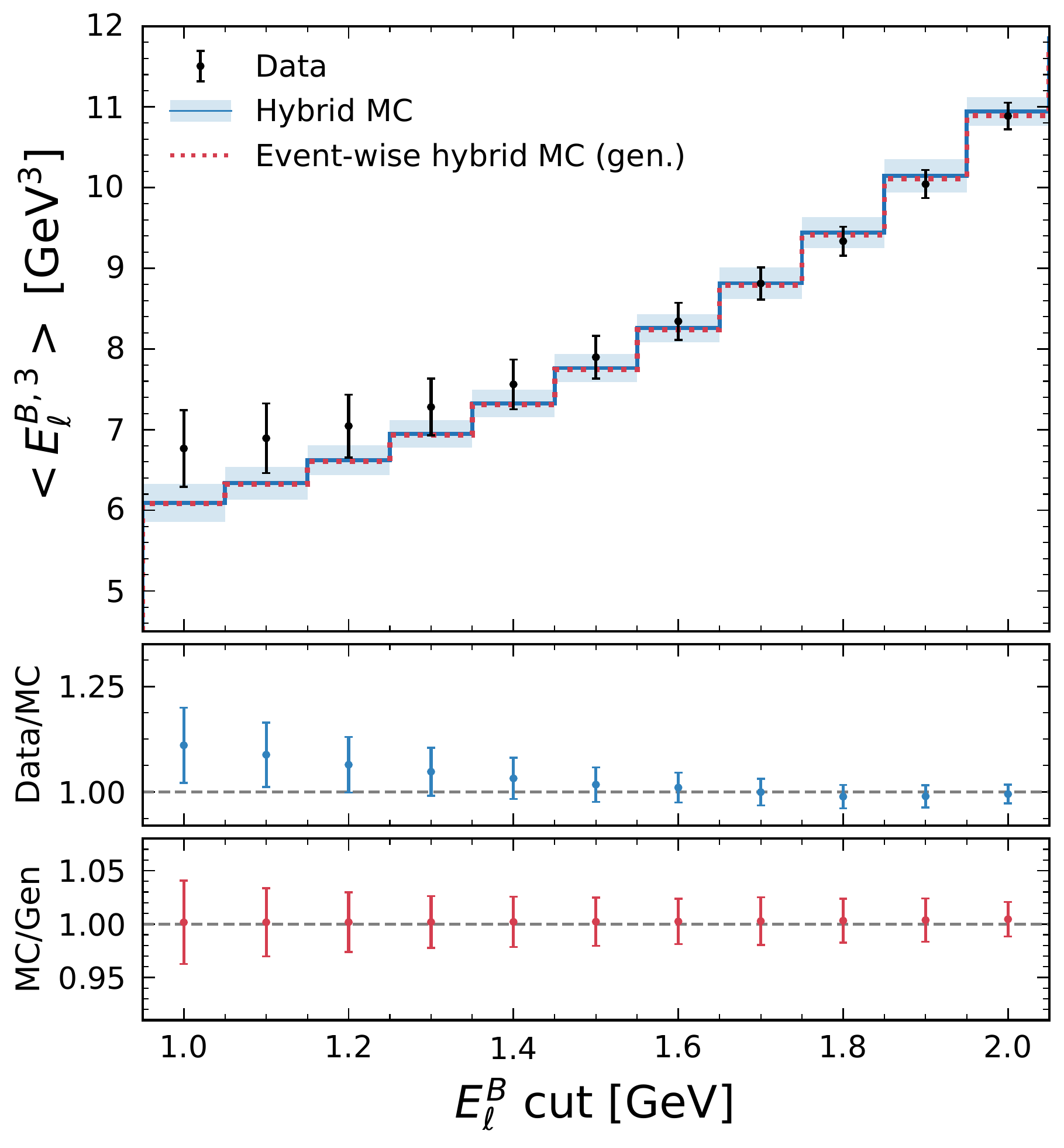} 
    
\caption{
The first (left), second (middle) and third (right) moment of the measured differential branching fraction of $E_\ell^B$. The full experimental uncertainty is included and shown for the extracted moments. The moments based on binned hybrid MC (blue and including full modelling uncertainty) are compared to measured data and the event-wise treatment of generator-level hybrid events (red dotted) in a ratio, respectively.
 }\label{fig:moments-1}
\end{figure*}

\begin{figure*}[t!]
     \includegraphics[width=0.32\textwidth]{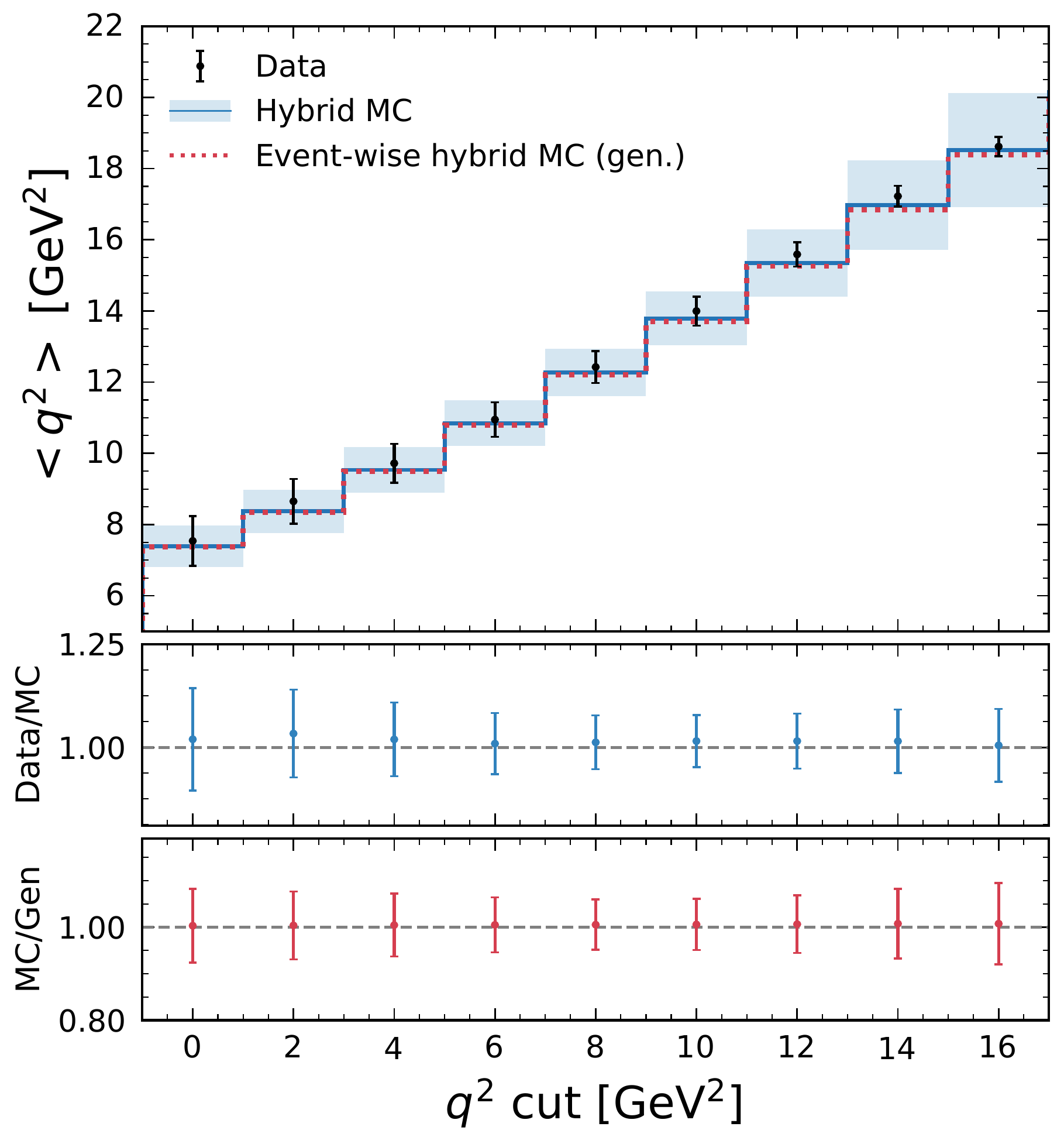} 
  \includegraphics[width=0.32\textwidth]{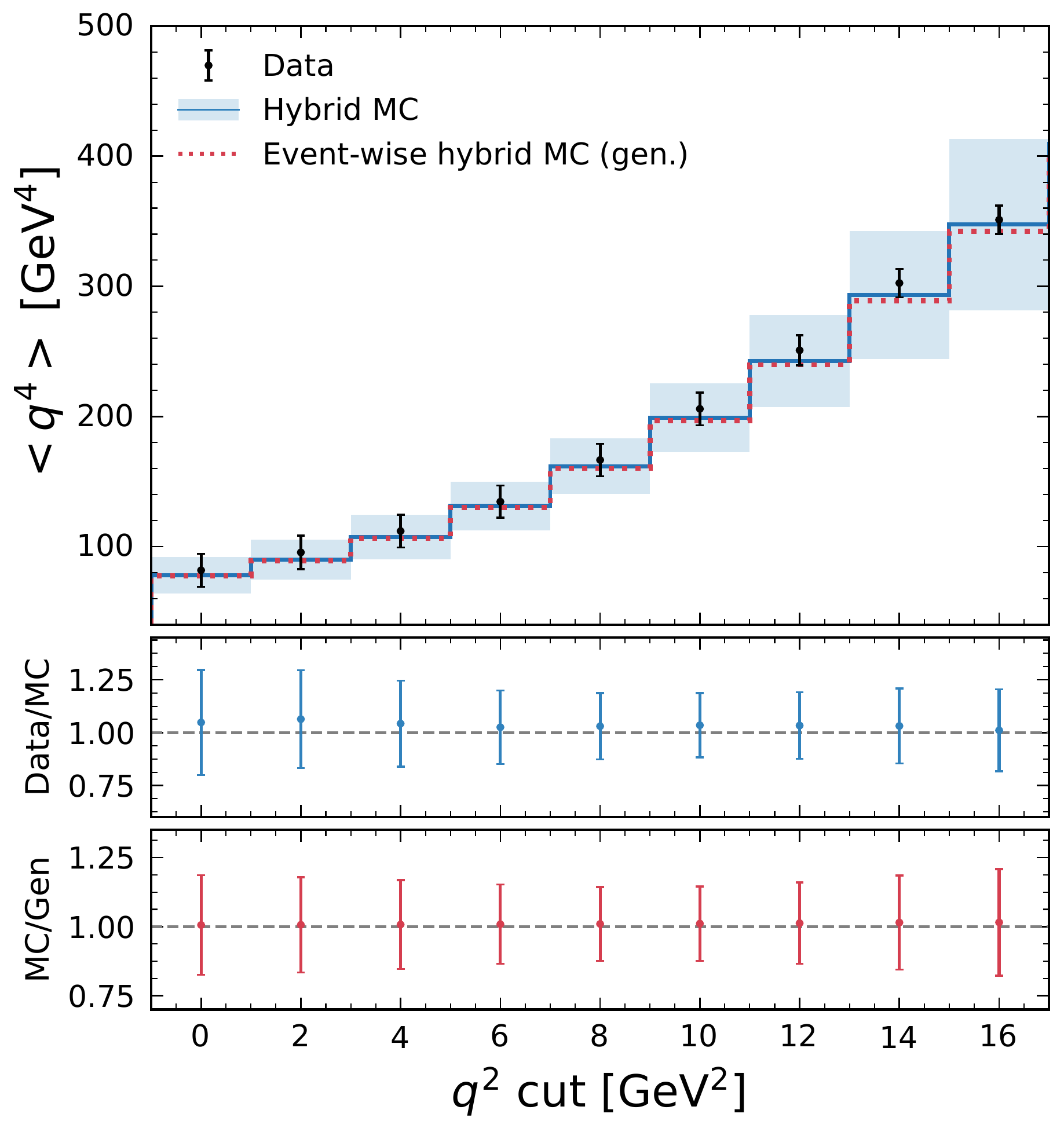} 
  \includegraphics[width=0.32\textwidth]{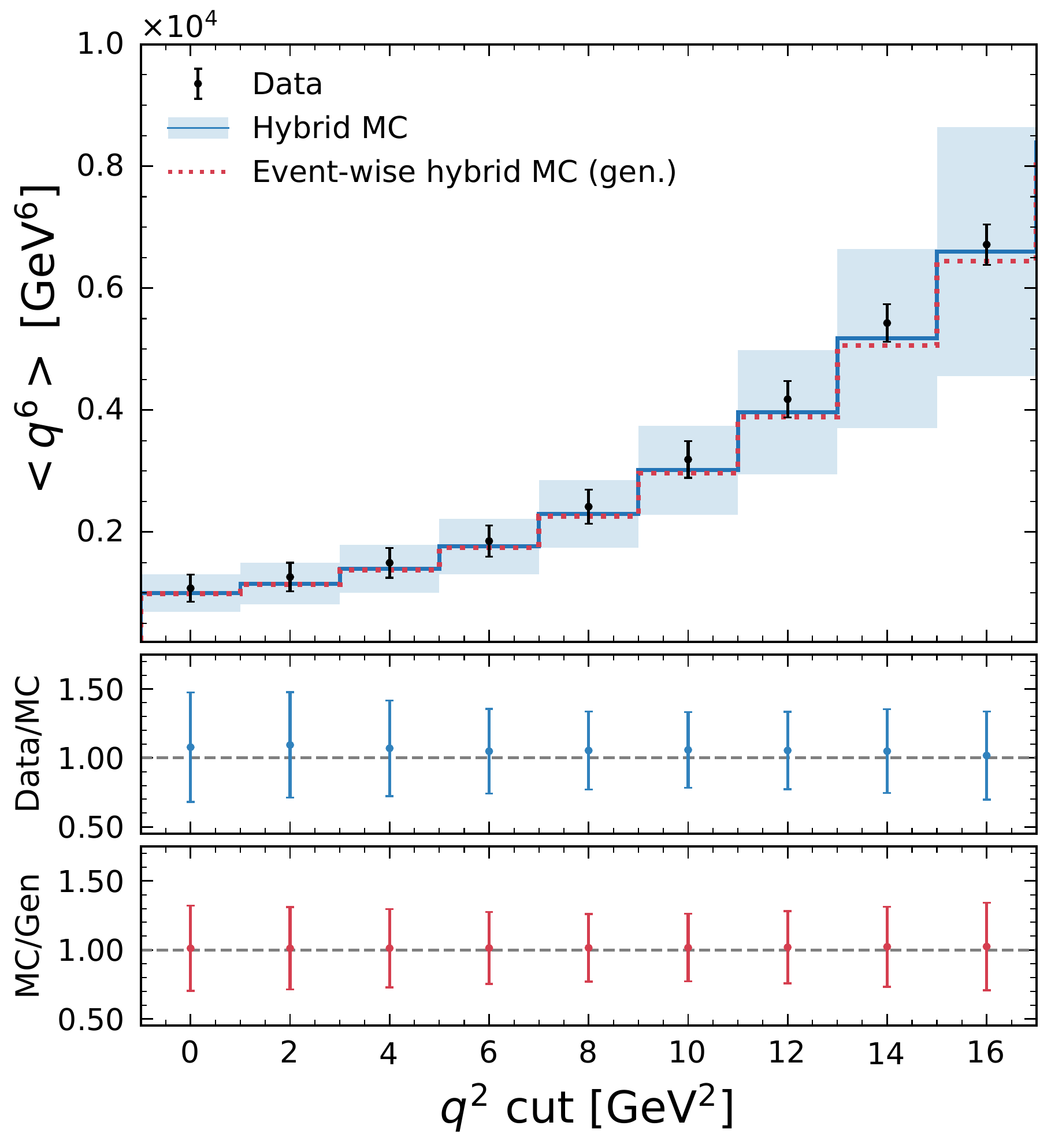} 
  
    \includegraphics[width=0.32\textwidth]{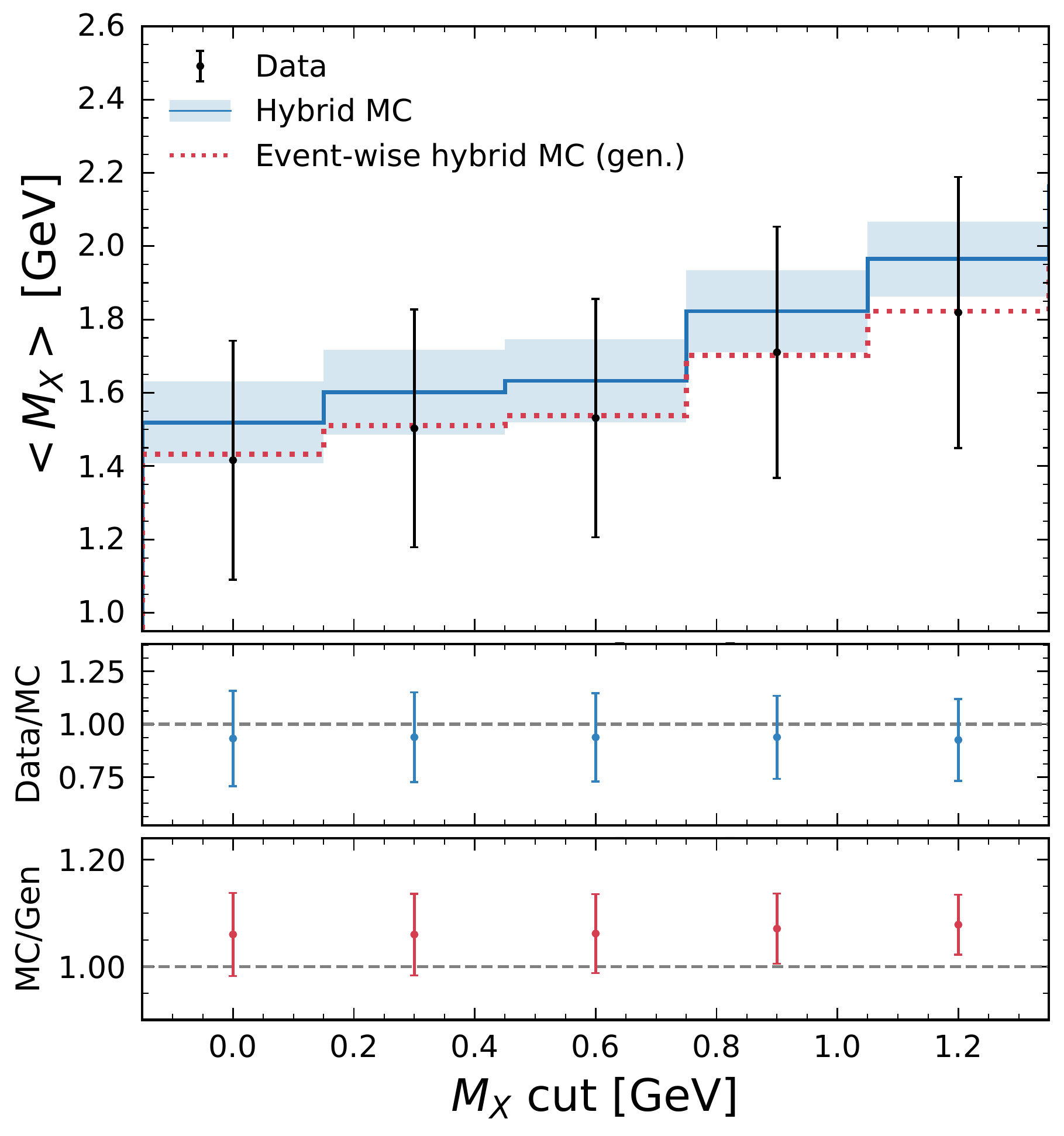} 
  \includegraphics[width=0.32\textwidth]{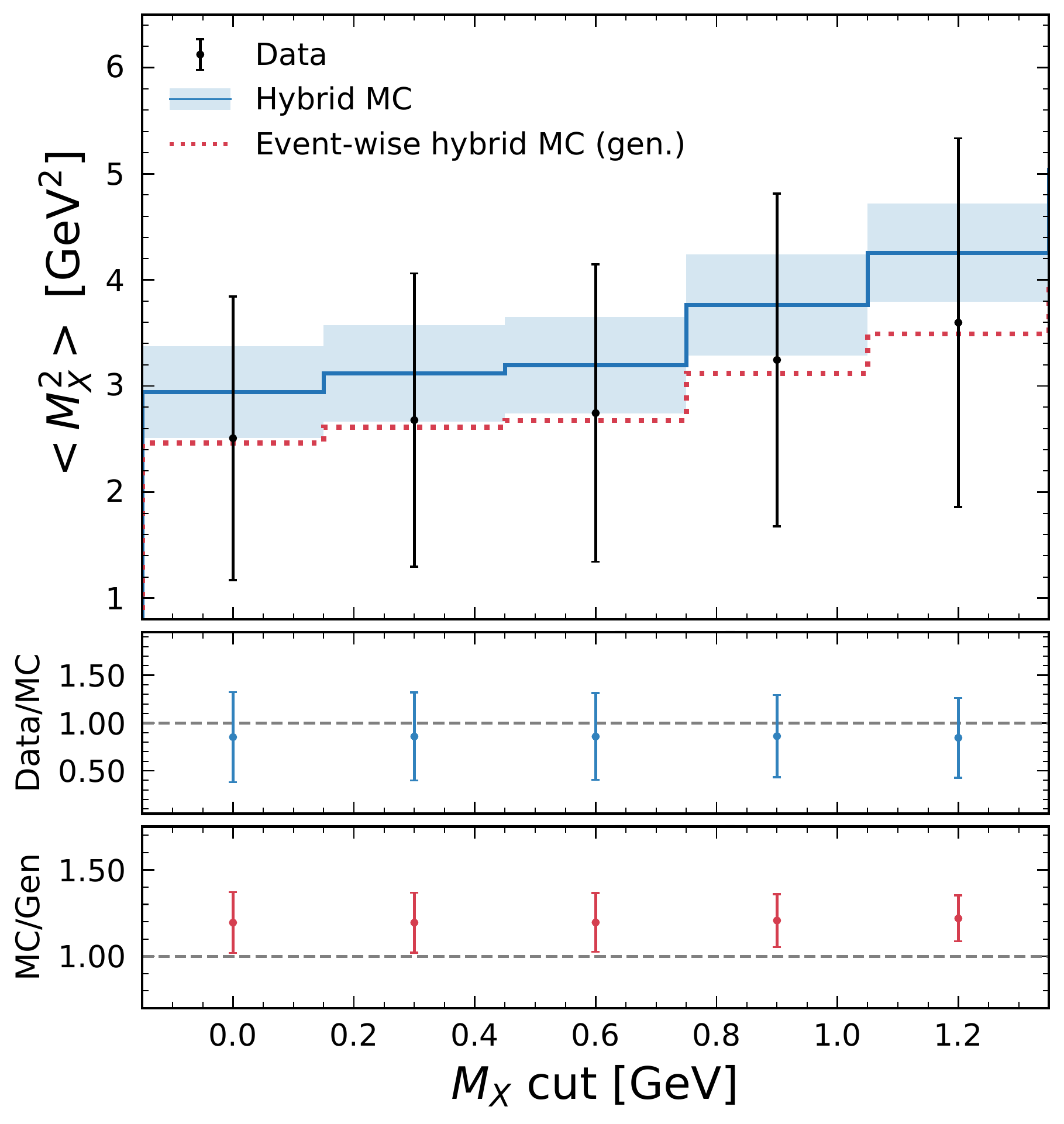} 
  \includegraphics[width=0.32\textwidth]{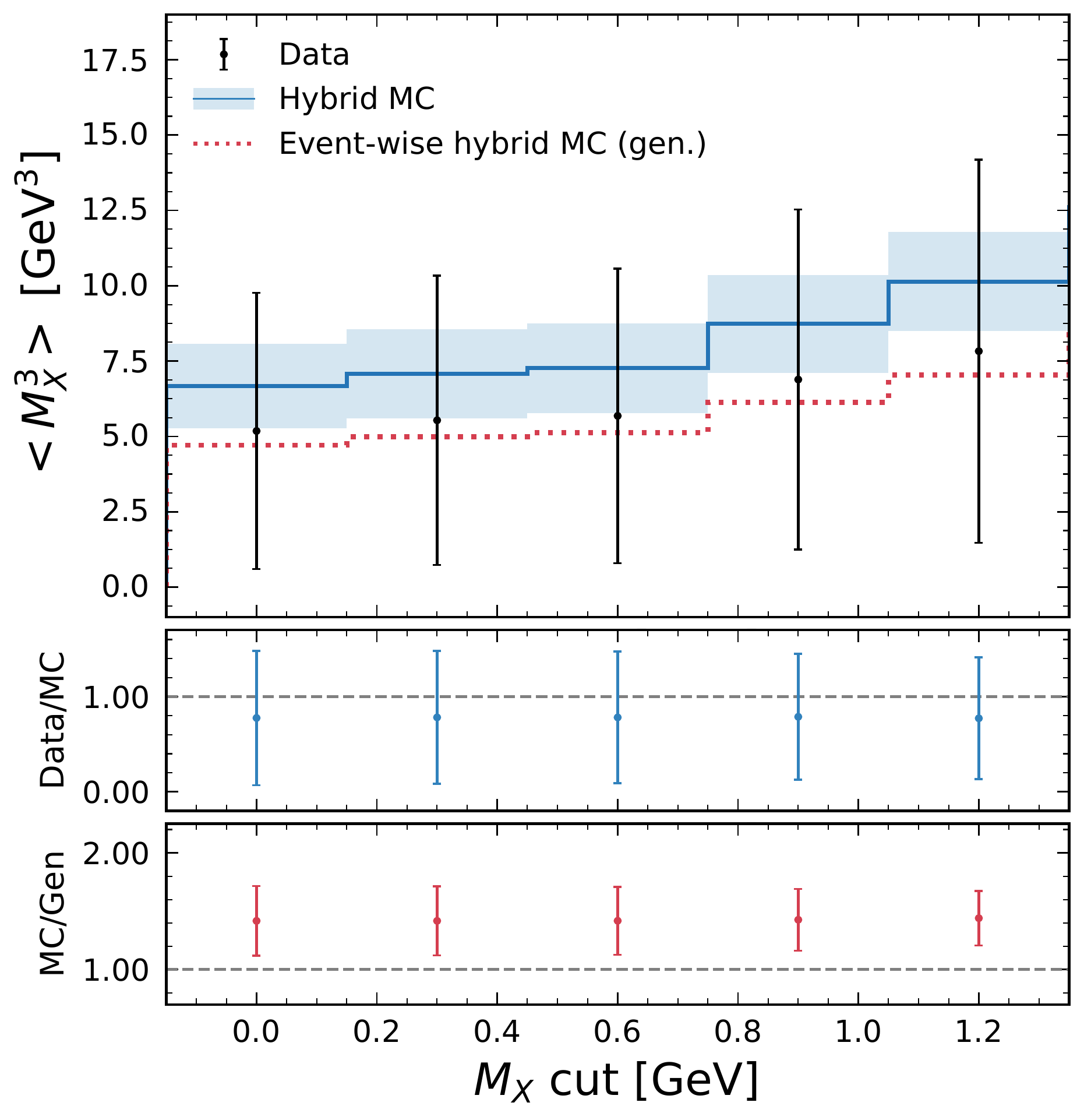} 
  
    \includegraphics[width=0.32\textwidth]{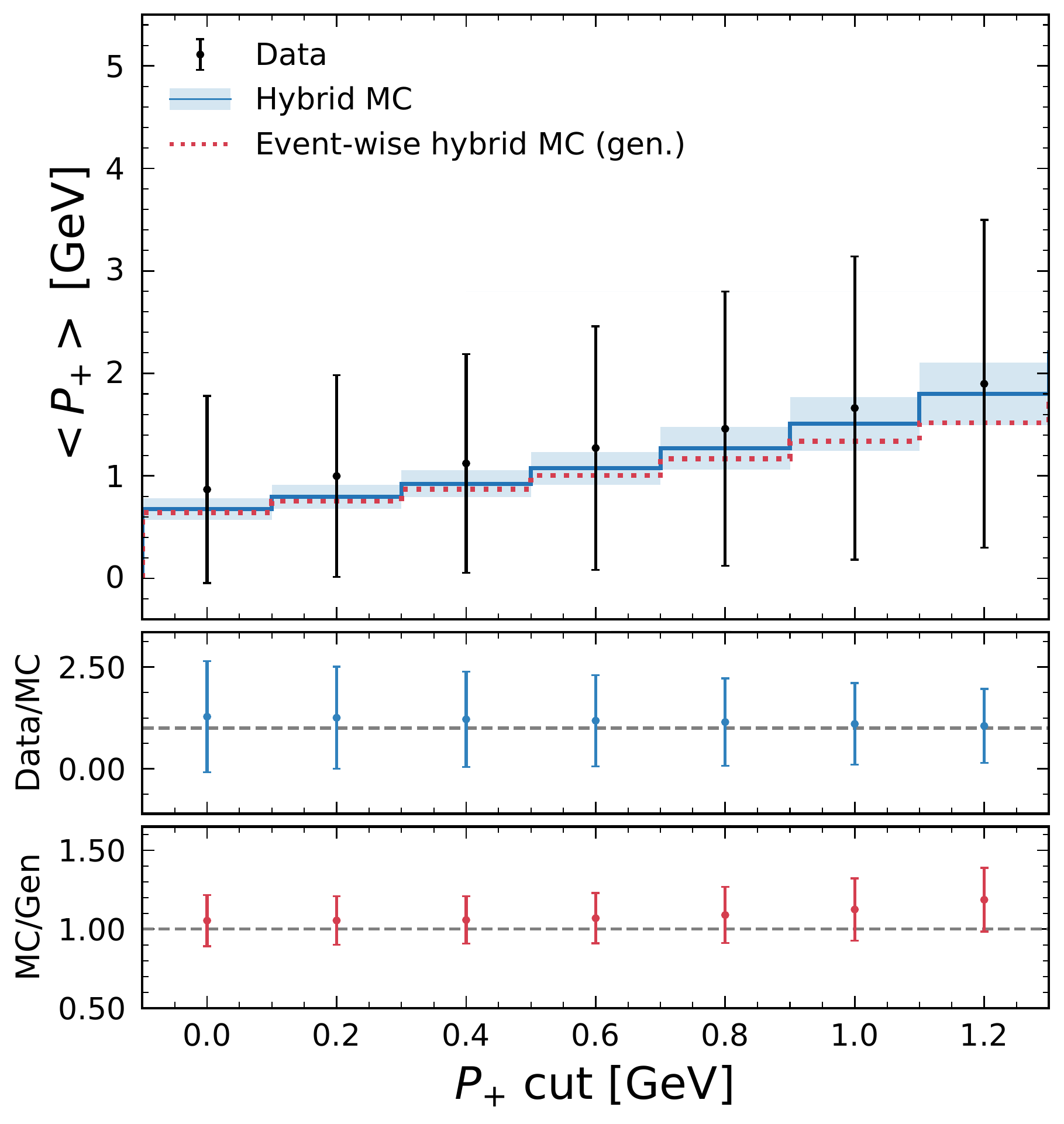} 
  \includegraphics[width=0.32\textwidth]{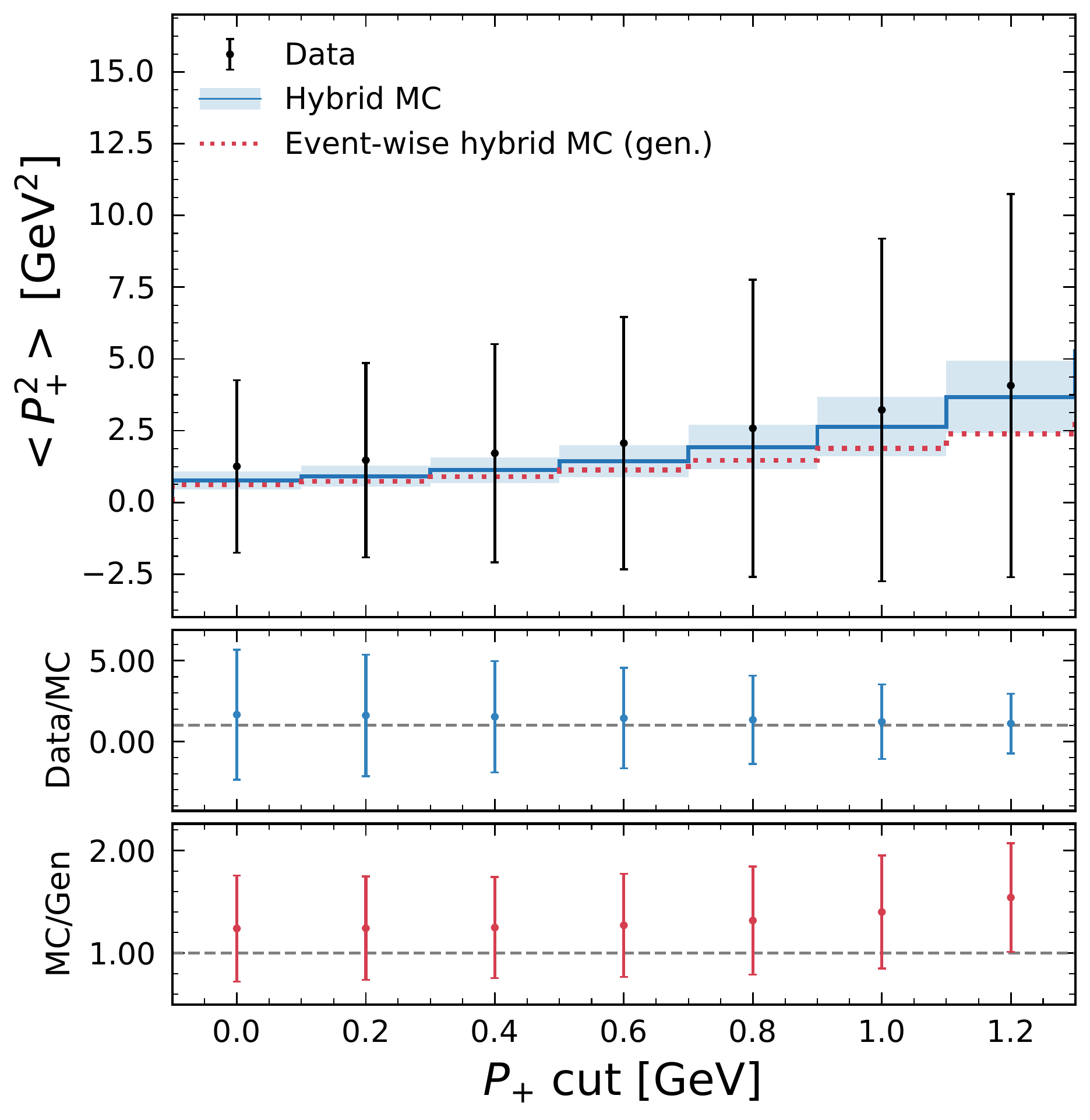} 
  \includegraphics[width=0.32\textwidth]{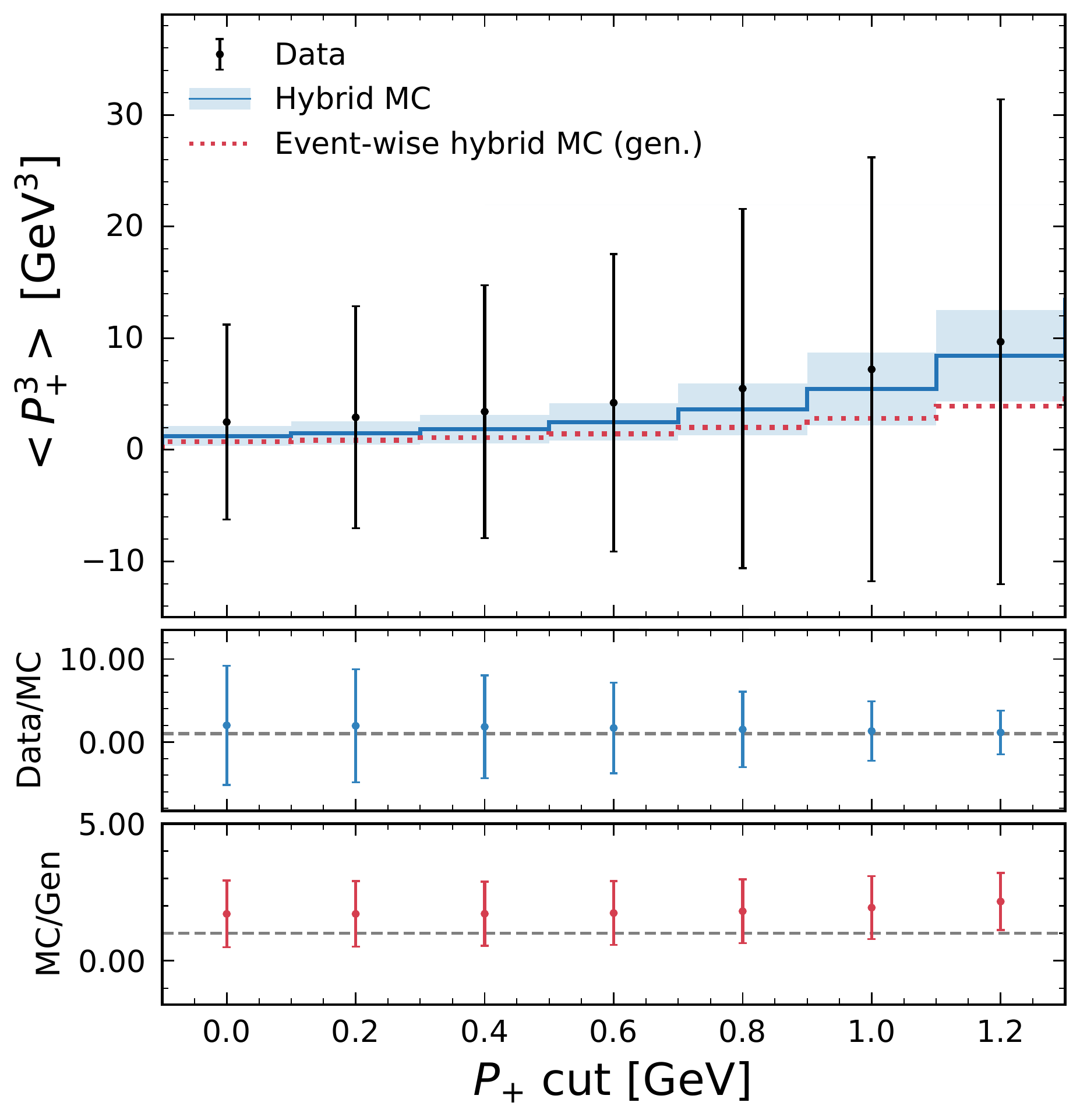}

\caption{
The first (left), second (middle) and third (right) moment of the measured differential branching fraction of $q^2$ (top row), $M_{X}$ (middle row) and $P_{+}$ (bottom row) in the phase space region of $E_\ell^B > 1 \, \mathrm{GeV}$. The full experimental uncertainty is included and shown for the extracted moments. The moments based on binned hybrid MC (blue and including full modelling uncertainty) are compared to measured data and the event-wise treatment of generator-level hybrid events (red dotted) in a ratio, respectively.
 }\label{fig:moments-2}
\end{figure*}

\begin{figure*}[]
   \includegraphics[width=0.32\textwidth]{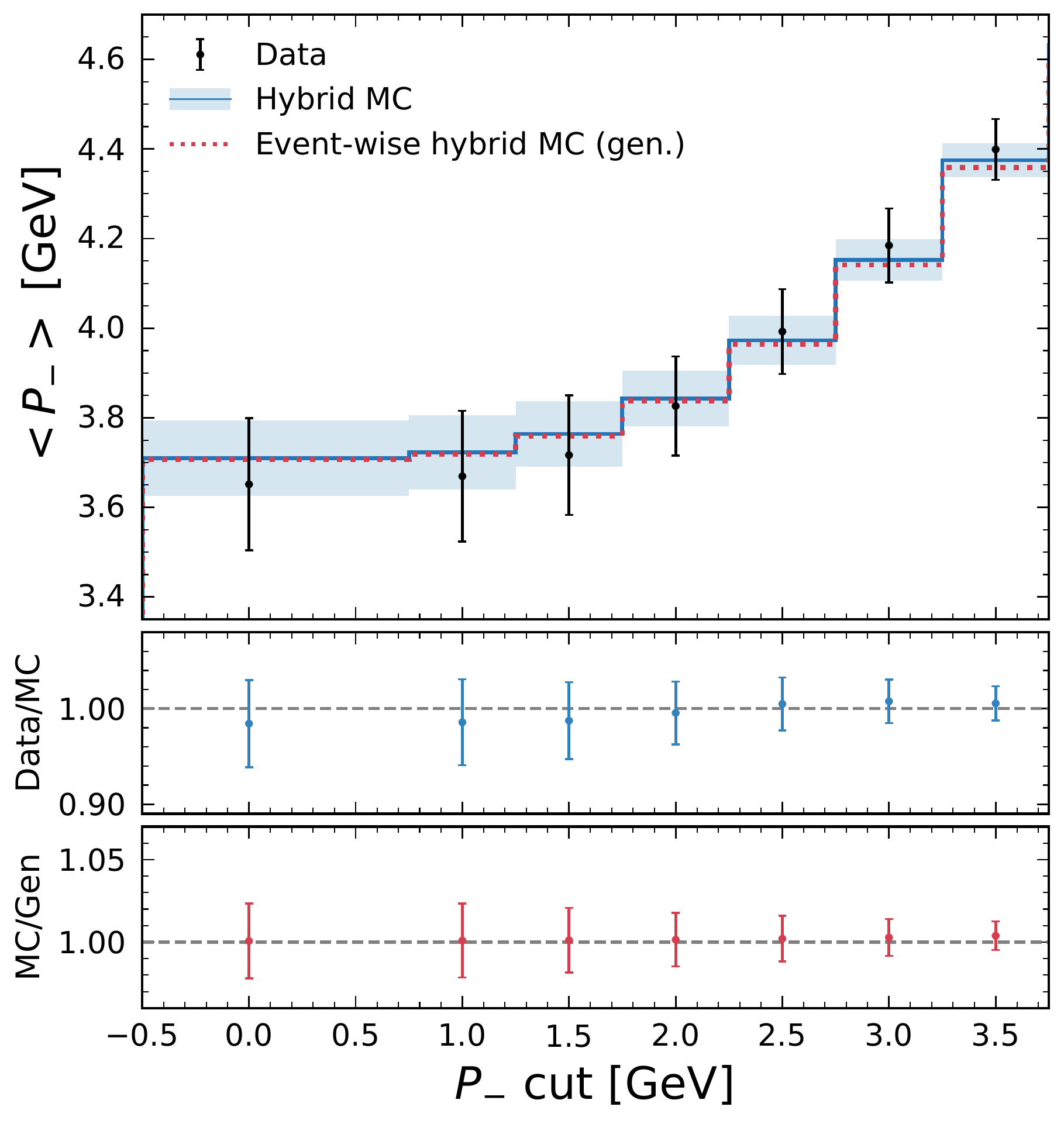} 
  \includegraphics[width=0.32\textwidth]{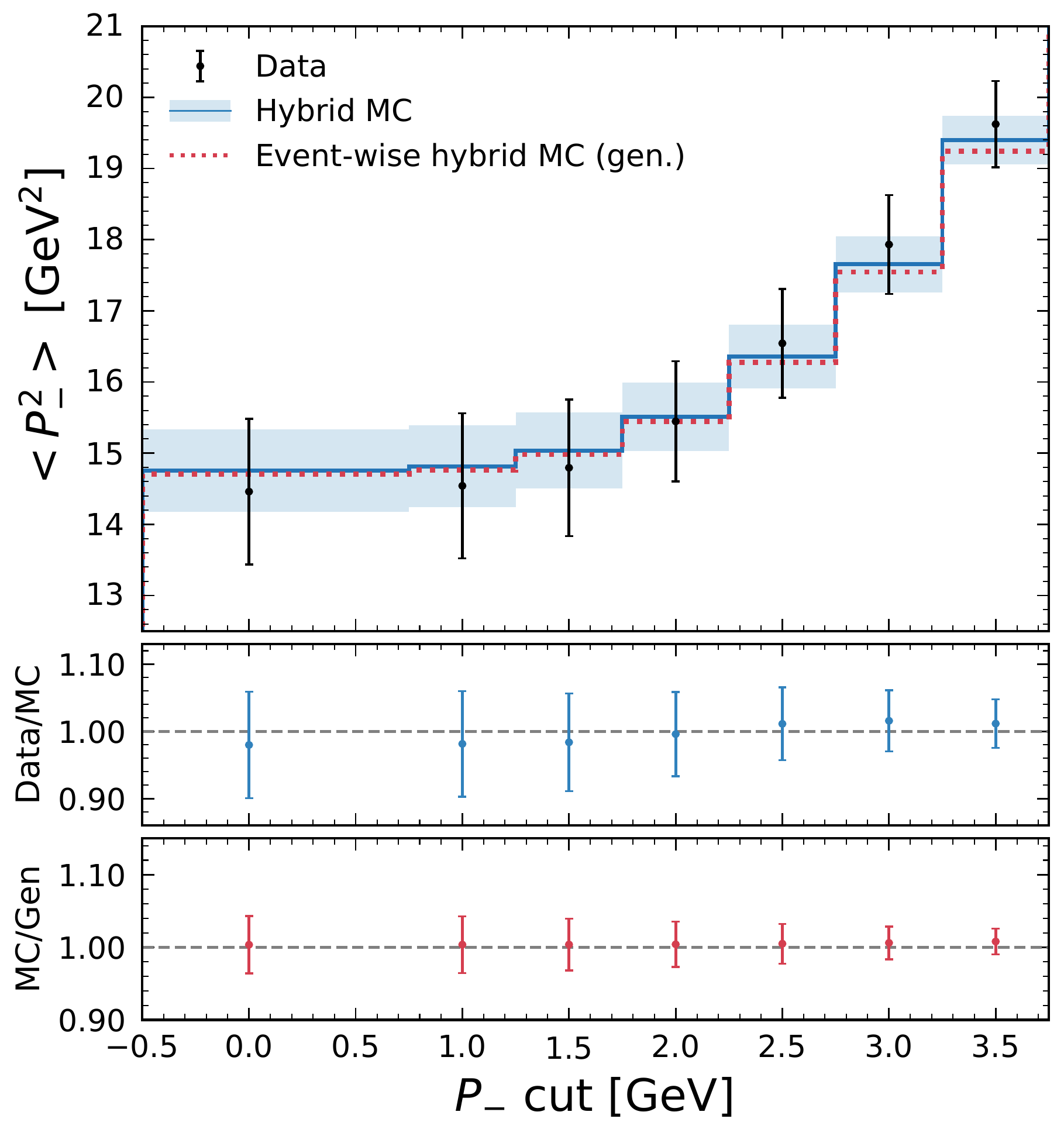} 
  \includegraphics[width=0.32\textwidth]{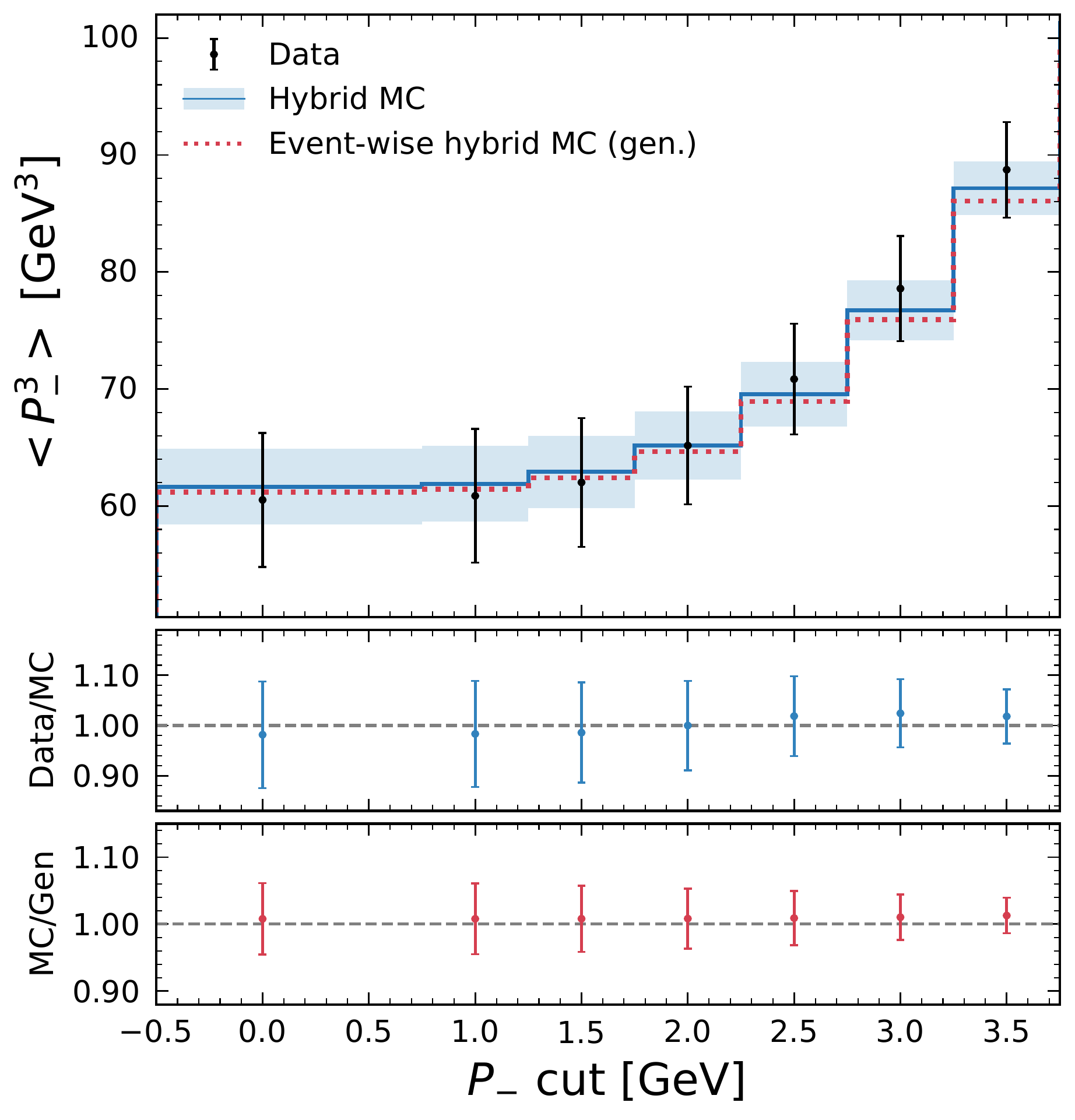} 
  
\caption{
The first (left), second (middle) and third (right) moment of the measured differential branching fraction of $P_{-}$ in the phase space region of $E_\ell^B > 1 \, \mathrm{GeV}$. The full experimental uncertainty is included and shown for the extracted moments. The moments based on binned hybrid MC (blue and including full modelling uncertainty) are compared to measured data and the event-wise treatment of generator-level hybrid events (red dotted) in a ratio, respectively.
 }\label{fig:moments-3}
\end{figure*}

Using the measured differential branching fractions, we determine the first to third moments of all measured kinematic observables. The moments are determined with a progression of the kinematic variable and defined for the partial phase-space with a selection of $E_\ell^B > 1 \, \mathrm{GeV}$ unless stated otherwise. As the moments are determined using binned information, we validate their accuracy using binned and unbinned \bulnu MC events. The resulting biases from using binned information is negligible for all distributions, expect for the moments of the hadronic mass spectrum. There, the resonance region leads to strong changes in the line-shape, which are not well captured by the utilized binning. The resulting biases are still small in comparison to the experimental errors and for the hadronic mass spectrum, we include them into the total experimental uncertainty. Figures~\ref{fig:moments-1}-\ref{fig:moments-3} shows the results for each measured kinematic variable, also showing the prediction from binned and unbinned \bulnu hybrid MC.

\end{document}